\documentclass[longauth,traditabstract,printer]{aa}  

\usepackage{graphicx}
\usepackage[varg]{txfonts}
\usepackage[final]{hyperref}
\usepackage{lscape}

\usepackage{url}
\usepackage{color}
\usepackage{natbib}
\usepackage{microtype}

\usepackage[utf8]{inputenc}
\usepackage[T1]{fontenc}

\makeatletter
\renewcommand*\aa@pageof{, page \thepage{} of \pageref*{LastPage}}
\makeatother

\makeatletter
 \def\@textbottom{\vskip \z@ \@plus 1pt}
 \let\@texttop\relax
\makeatother

\newcommand{\ha}{H$\alpha$}
\newcommand{\hb}{H$\beta$}
\newcommand{\heps}{H$\epsilon$}
\newcommand{\kms}{\,km\,s$^{-1}$}
\newcommand{\lsun}{\,L$_{\sun}$}
\newcommand{\msun}{\,M$_{\sun}$}
\newcommand{\ang}{\,\AA}
\newcommand{\ergs}{\,erg\,s$^{-1}$}
\renewcommand{\object}{SN~2015da}

\begin{document} 

\sloppy

\title{The long-lived Type IIn \object: Infrared echoes and strong interaction within an extended massive shell\thanks{Spectroscopic data and photometric tables are available through the Weizmann Interactive Supernova Data Repository (WISeREP) at the following address: \url{https://wiserep.weizmann.ac.il/object/13868}. Tables~A.1-A.4 are also available at the CDS via anonymous ftp to \url{cdsarc.u-strasbg.fr} (130.79.128.5) or at \url{http://cdsweb.u-strasbg.fr/cgi-bin/qcat?J/A+A/}}}
\titlerunning{The Type IIn SN~2015da}
\author{L.~Tartaglia\inst{1} 
\and A.~Pastorello\inst{2} 
\and J.~Sollerman\inst{1} 
\and C.~Fransson\inst{1}
\and S.~Mattila\inst{3}
\and M.~Fraser\inst{4} 
\and F.~Taddia\inst{1}
\and L.~Tomasella\inst{2}
\and M.~Turatto\inst{2}
\and A.~Morales-Garoffolo\inst{5}
\and N.~Elias-Rosa\inst{2,6,7} 
\and P.~Lundqvist\inst{1}
\and J.~Harmanen\inst{3} 
\and T.~Reynolds\inst{3}
\and E.~Cappellaro\inst{2}
\and C.~Barbarino\inst{1}
\and A.~Nyholm\inst{1}
\and E.~Kool\inst{1}
\and E.~Ofek \inst{8}
\and X.~Gao\inst{9}
\and Z.~Jin\inst{10}
\and H.~Tan\inst{11}
\and D.~J.~Sand \inst{12}
\and F.~Ciabattari \inst{13}
\and X.~Wang \inst{14}
\and J.~Zhang \inst{15,16}
\and F.~Huang \inst{17,14}
\and W.~Li \inst{14}
\and J.~Mo\inst{14}
\and L.~Rui \inst{14}
\and D.~Xiang \inst{14}
\and T.~Zhang \inst{18,19}
\and G.~Hosseinzadeh \inst{20}
\and D.~A.~Howell \inst{21,22}
\and C.~McCully\inst{21}
\and S.~Valenti\inst{23}
\and S.~Benetti\inst{2} 
\and E.~Callis\inst{4}
\and A.~S.~Carracedo\inst{24}
\and C.~Fremling\inst{25}
\and T.~Kangas\inst{26}
\and A.~Rubin\inst{27}
\and A.~Somero\inst{3} 
\and G.~Terreran\inst{28}
}

\institute{
Department of Astronomy and the Oskar Klein Centre, Stockholm University, AlbaNova, Roslagstullsbacken 21, SE 114 21 Stockholm, Sweden \\(\email{leonardo.tartaglia@astro.su.se}) 
\and INAF - Osservatorio Astronomico di Padova, Vicolo dell'Osservatorio 5, 35122 Padova, Italy
\and Tuorla Observatory, Department of Physics and Astronomy, FI 20014, University of Turku, Finland
\and School of Physics, O'Brien Centre for Science North, University College Dublin, Belfield Dublin 4, Ireland
\and Department of Applied Physics, Universidad de C\'{a}diz, campus of Puerto Real, 11510, C\'{a}diz, Spain
\and Institute of Space Sciences (ICE, CSIC), Campus UAB, Cam\'i de Can Magrans s/n, 08193, Cerdanyola del Vall\`es, Barcelona, Spain 
\and Institut d'Estudis Espacials de Catalunya (IEEC), c/Gran Capit\`a 2 -- 4, Edif. Nexus 201, 08034, Barcelona, Spain
\and Benoziyo Center for Astrophysics and the Helen Kimmel center for planetary science, Weizmann Institute of Science, 76100 Rehovot, Israel
\and Xinjiang Astronomical Observatory, 150 Science 1--Street, Urumqi, Xinjiang 830011, China
\and Xingming Observatory, Mountain Nanshan, Urumqi, Xinjiang 830011, China
\and Graduate Institute of Astronomy, National Central University, 300 Zhongda Rd., Zhongli District, Taoyuan City 32001, Taiwan, China
\and Department of Astronomy/Steward Observatory, 933 North Cherry Avenue, Rm. N204, Tucson, AZ 85721-0065, USA
\and Osservatorio Astronomico di Monte Agliale, Via Cune Motrone, 55023 Borgo a Mozzano, Lucca, Italy
\and Physics Department and Tsinghua Center for Astrophysics, Tsinghua University, Beijing, 100084, China
\and Yunnan Observatories, Chinese Academy of Sciences, Kunming 650216, China
\and Key Laboratory for the Structure and Evolution of Celestial Objects, Chinese Academy of Sciences, Kunming 650216, China
\and Department of Astronomy, School of Physics and Astronomy, Shanghai Jiaotong Univeristy, Shanghai, 200240, China
\and Key Laboratory of Optical Astronomy, National Astronomical Observatories, Chinese Academy of Sciences, 10101, Beijing, China
\and School of Astronomy and Space Science, University of Chinese Academy of Sciences, 101408, Beijing, China
\and Center for Astrophysics | Harvard \& Smithsonian, 60 Garden Street, Cambridge, MA 02138-1516, USA
\and Las Cumbres Observatory, 6740 Cortona Drive, Suite 102, Goleta, CA 93117-5575, USA
\and Department of Physics, University of California, Santa Barbara, CA 93106-9530, USA
\and Department of Physics, University of California, Davis, CA 95616, USA
\and The Oskar Klein Centre, Physics Department, Stockholm University, AlbaNova, Roslagstullsbacken 21, SE 114 21 Stockholm, Sweden
\and Division of Physics, Mathematics and Astronomy, California Institute of Technology, Pasadena, CA 91125, USA
\and Space Telescope Science Institute, 3700 San Martin Drive, Baltimore, MD 21218, USA
\and European Southern Observatory Karl -- Schwarzschild -- Str 2 85748, Garching bei M\"unchen, Germany
\and Center for Interdisciplinary Exploration and Research in Astrophysics (CIERA) and Department of Physics and Astronomy, Northwestern University, Evanston, IL 60208, USA}
\date{Received: 2019 August 22; Accepted: 2019 December 29.}
 
\abstract{
In this paper we report the results of the first $\sim$four years of spectroscopic and photometric monitoring of the Type IIn supernova \object~(also known as PSN~J13522411+3941286, or iPTF16tu). 
The supernova exploded in the nearby spiral galaxy NGC~5337 in a relatively highly extinguished environment. 
The transient showed prominent narrow Balmer lines in emission at all times and a slow rise to maximum in all bands.
In addition, early observations performed by amateur astronomers give a very well-constrained explosion epoch.
The observables are consistent with continuous interaction between the supernova ejecta and a dense and extended H-rich circumstellar medium.
The presence of such an extended and dense medium is difficult to reconcile with standard stellar evolution models, since the metallicity at the position of \object~seems to be slightly subsolar.
Interaction is likely the mechanism powering the light curve, as confirmed by the analysis of the pseudo bolometric light curve, which gives a total radiated energy $\gtrsim10^{51}\,\rm{erg}$.
Modeling the light curve in the context of a supernova shock breakout through a dense circumstellar medium allowed us to infer the mass of the prexisting gas to be $\simeq8$\msun, with an extreme mass-loss rate for the progenitor star $\simeq0.6\,\rm{M_{\odot}}\,\rm{yr^{-1}}$, suggesting that most of the circumstellar gas was produced during multiple eruptive events.
Near- and mid-infrared observations reveal a flux excess in these domains, similar to those observed in SN~2010jl and other interacting transients, likely due to preexisting radiatively heated dust surrounding the supernova. 
By modeling the infrared excess, we infer a mass $\gtrsim0.4\times10^{-3}$\msun~for the dust.}
\maketitle

\section{Introduction} \label{sec:intro}
Supernovae (SNe) interacting with a preexisting dense circumstellar medium (CSM) belong to an intriguing and not fully understood class of transients, including the Ibn \citep{2008MNRAS.389..113P} and IIn \citep{1990MNRAS.244..269S} classes.
Early optical spectra of Type IIn SNe show a blue continuum with narrow (from a few tens to $\simeq10^3$\kms) Balmer lines in emission, and broad wings resulting from electron scattering in an ionized, unshocked, H-rich CSM \citep[see, e.g.,][]{2001MNRAS.326.1448C}.
These lines, visible at all phases of the spectroscopic evolution, are the signature of underlying interaction between SN ejecta and slow moving CSM, as they are the result of recombination in the outer unshocked layers, ionized by photons emitted in the shocked regions.
When the fast-moving ejecta hit the preexisting CSM, forward/reverse shocks form at the interface between the two media \citep[also producing a contact discontinuity and a "cool dense shell" -- CDS;][]{1984A&A...133..264F,1994ApJ...420..268C}, and high energy photons propagate in both directions, either ionizing the freely expanding SN ejecta or the slow-moving CSM.

Under specific conditions (e.g., particular geometrical configurations), the resulting emission lines can be characterized by structured, asymmetric, and multicomponent profiles, produced by recombining gas shells moving at different velocities (see, e.g., the case of the prototypical SNe~1987F; \citealt{1996MNRAS.278...22W} and 1988Z; \citealt{1991MNRAS.250..786S,1993MNRAS.262..128T}), although in other cases, the overall profiles are characterized by pure Lorentzian profiles at all epochs \citep[see, e.g., the case of SN~2010jl;][]{2014ApJ...797..118F}.
Depending on the density of the CSM, the early light curve of IIn SNe might be dominated by photon diffusion rather than $^{56}$Ni decay \citep[e.g.,][]{2011MNRAS.414.1715B}, possibly extending the "shock breakout" signal \citep[e.g.,][]{2010ApJ...724.1396O,2011ApJ...729L...6C}.
Modeling the bolometric light curves of SNe exploding in a dense wind allows us to infer crucial information about the nature of the exploding star and its environment \citep[such as the mass-loss rate and the mass of the surrounding CSM; e.g.,][]{2011MNRAS.414.1715B,2012ApJ...759..108S}. 
Narrow lines, if resolved, may also be used to infer wind speeds directly from their P-Cygni absorption features.

Type IIn SNe are relatively rare \citep[$\sim9\%$ of all Type II, H-rich core-collapse SNe -- CC SNe;][]{2011yCat..74121441L}, although they are more common than Ibn SNe. 
They show a remarkable heterogeneity, with absolute peak magnitudes ranging from $M_r=-17$ to $-22\,\rm{mag}$ \citep{2012ApJ...744...10K,2013A&A...555A..10T}, with a mean rise time of $\simeq17\,\rm{d}$ \citep[based on the sample of 15 objects published in][]{2014ApJ...788..154O}.
On the other hand, this value might be strongly affected by the limited number of SNe IIn discovered soon after explosion and/or well-studied transients, whose publication is likely biased towards the most peculiar or luminous objects \citep[see, e.g., SN~2006gy;][and the "super-luminous" Type II SNe -- SLSN II; \citealt{2018arXiv181201428G}]{2007ApJ...659L..13O,2007ApJ...666.1116S}.
A recent analysis on a greater sample of 42 SNe IIn discovered by the Palomar Transient Factory (PTF) reveals an average peak luminosity of $M_r=-19.18\pm1.32\,\rm{mag}$ and a bimodal distribution of rise times, with $t_{rise}=19\pm8\,\rm{d}$ and $50\pm15\,\rm{d}$ \citep{2019arXiv190605812N}.
A wide range of photometric properties is also observed after peak, with a fraction of IIn SNe (e.g., SNe~1988Z; \citealt{1991MNRAS.250..786S,1993MNRAS.262..128T}, 2005ip and 2006jd; \citealt{2012ApJ...756..173S}) showing a very slow evolution, with the transient still visible several years after explosion \citep[see, e.g., SN~1995N;][]{2002ApJ...572..350F,2005ASPC..342..285P}. 
In other cases, light curves exhibit a plateau-like shape with steep post-plateau declines \citep[IIn-P;][]{2012MNRAS.424..855K,2013MNRAS.431.2599M} or otherwise almost "linear" declines \citep[see, e.g., the case of SN~1999el;][]{2002ApJ...573..144D}.
Long-lasting Type IIn SNe are generally brighter than IIn-P, although not as bright as SLSNe.
\begin{figure}
\begin{center}
\includegraphics[width=\linewidth]{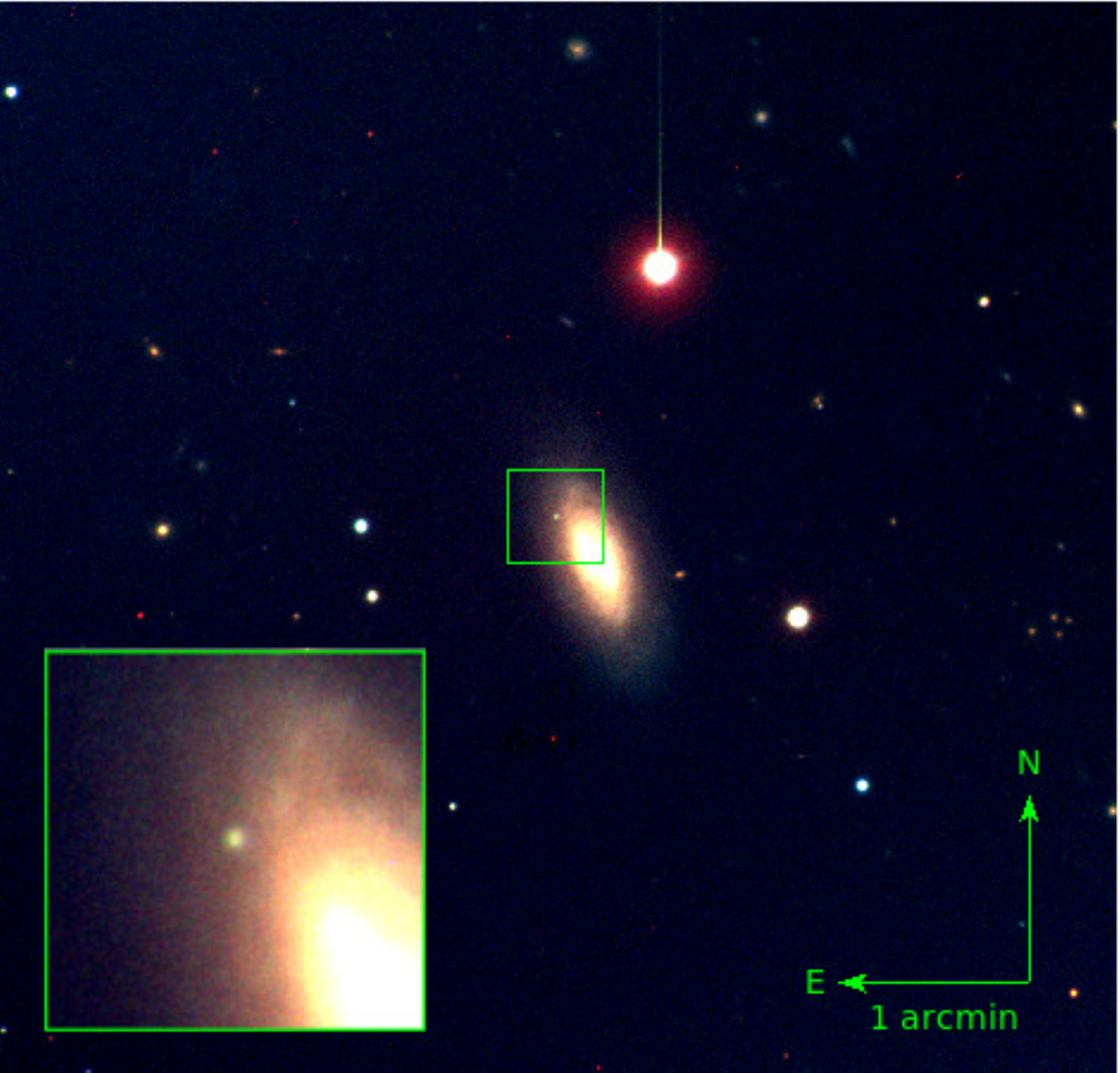}
\label{fig:fc}
\caption[]{Combined $gri$ images of the field of NGC~5337 obtained on 2019 January 23 with the Nordic Optical Telescope with ALFOSC.
The position of \object, still clearly visible, is highlighted within a green box. The insert shows \object, which~is the bright source within the zoomed-in region.}
\end{center}
\end{figure}

The controversy surrounding the nature of the progenitors of SNe IIn is still not completely solved.
The presence of an extended and dense CSM at the time of the explosion requires strong long-term winds, binary interactions, or giant eruptive events similar to those occasionally experienced by luminous blue variables \citep[LBVs;][and references therein]{2012ASSL..384..221V}.
Some observational evidence points to LBVs as progenitors of at least some Type IIn SNe.
The direct observation of the likely quiescent progenitors of SNe~2005gl \citep{2009Natur.458..865G} and 2009ip \citep{2010AJ....139.1451S} in deep archival {\it Hubble Space Telescope} (HST) images strongly supports this scenario, as well as the disappearance of the LBV candidate progenitor of SN~2005gl in a post-explosion HST image obtained when the SN had faded \citep{2009Natur.458..865G}. 
Nevertheless, pre-SN super-winds might not be rare in red supergiant (RSG) stars \citep[see, e.g., the case of the Galactic RSG VY Canis Majoris;][]{2009AJ....137.3558S} suggesting that they are viable progenitors for at least a fraction of SNe IIn \citep[as proposed for SN~1995N;][]{2002ApJ...572..350F,2004MNRAS.352..457P}.
In addition, \citet{2016ApJ...823L..23T} showed that also stars less massive than LBVs can experience major eruptive events.
Some observational evidences seem to suggest thermonuclear explosions of stars embedded in a dense and extended CSM as viable mechanisms to produce some Type IIn SNe.
These Ia-CSM SNe \citep[see, e.g.,][]{2013ApJS..207....3S,2015MNRAS.447..772F} show optical spectra dominated by prominent narrow Balmer lines in emission and luminosities usually exceeding those typically observed in normal SNe Ia.
A number of transients have been proposed to belong to this class \citep[see, e.g.,][]{2013ApJS..207....3S}, although their real nature is still a matter of discussion \citep[see, e.g.,][]{2006ApJ...653L.129B,2008A&A...483L..47T,2016MNRAS.459.2721I}.
On the other hand, the analysis performed on high-resolution optical spectra of PTF11kx \citep{2012Sci...337..942D} revealed features first resembling those of the overluminous Type Ia SN~1999aa \citep[see, e.g.,][and references therein]{2004AJ....128..387G}, but soon developing strong narrow \ha~features, therefore suggesting that a fraction of SNe showing narrow emission features arise from thermonuclear explosions.

More than $50\%$ of SNe showing late (or very late, see, e.g., SNe~1999el; \citealt{2002ApJ...573..144D} and 2005ip; \citealt{2009ApJ...691..650F,2010ApJ...725.1768F}, among others) infrared (IR) excess belong to the IIn class \citep[see][]{2002ApJ...575.1007G,2011ApJ...741....7F}, although this percentage might be biased by the still limited number of well studied SNe IIn in the IR domain. 
This suggests the presence of either preexisting or newly formed dust.
Dust can, in fact, form in the rapidly cooling post-shock layers \citep[e.g.,][]{2004MNRAS.352..457P,2008MNRAS.389..141M} or be located in the dense preexisting CSM, becoming visible after being shock- or radiatively heated, or because echoing and reprocessing some of the SN light into IR radiation \citep[e.g.,][]{2010ApJ...725.1768F,2012ApJ...756..173S,2014ApJ...797..118F}.
Modeling the IR emission of SNe IIn showing late-time excesses can independently infer crucial information on the environments of progenitor stars as well as their mass-loss rates \citep[e.g.,][]{2010ApJ...725.1768F}.

In this context, we report the results of our study of the long-lasting Type IIn \object.
The SN was discovered on 2015 January 9.90~UT in the nearby spiral galaxy NGC~5337 by Z.~Jin and X.~Gao\footnote{\url{http://www.cbat.pdf.harvard.edu/unconf/followups\\/J13522411+3941286.html}}, with an apparent magnitude $\simeq18\,\rm{mag,}$ and was later classified as a Type IIn SN \citep{2015ATel.6939....1Z}.
No source was detected at the position of the transient in an unfiltered image obtained on 2015 January 7 to a limiting magnitude $19.5\,\rm{mag}$, or in previous frames of the field.
\object~is located at RA=13:52:24.11, Dec.=+39:41:28.2 [J2000], 12\farcs54~E, 14\farcs04~N from the center of NGC~5337 \citep[assuming RA=13:52:23.024, Dec=$+39$:41:14.16 - J2000 - for the center of the host;][see also Figure~\ref{fig:fc}]{2006AJ....131.1163S}.
The SN was extensively imaged by amateur astronomers, who provided very good sampling of the rise to the maximum through unfiltered images, later calibrated to the $R$ band.
\object~was also followed by a number of facilities and collaborations, such as the Nordic Optical Telescope Un-biased Transient Survey (NUTS\footnote{\url{http://csp2.lco.cl/not/}}) and its extension NUTS2, and the intermediate Palomar Transient Facility\footnote{\url{https://www.ptf.caltech.edu/iptf}} (iPTF) under the designation of iPTF16tu.
The discovery of \object~was also reported by \citet{2017MNRAS.470.1881P}, who estimated a distance of $32.1\,\rm{Mpc}$ based on the redshift derived from the heliocentric recessional velocity of NGC~5337 \citep[$2165\pm17$\kms;][]{2001A&A...378..370V}, and a cosmology with $\Omega_{\rm{M}}=0.31$, $\Omega_{\Lambda}=0.69,$ and $H_0=69.9\,\rm{km}\,\rm{s^{-1}}\,\rm{Mpc^{-1}}$.

The paper is organized as follows: Section~\ref{sec:host} describes the host galaxy, metallicity, and star formation rate at the location of \object, including a discussion on the local reddening.
In Section~\ref{sec:analysis}, we report our analysis on data collected during the photometric (Section~\ref{sec:photometry}) and spectroscopic (Section~\ref{sec:spectroscopy}) follow-up campaigns, and give a qualitative interpretation of the observed quantities.
A summary of the main results of the paper is reported in Section~\ref{sec:conclusions}, while in the Appendix, we detail the observations and data-reduction techniques, provide the magnitude tables, and main information on the spectra.

\section{The host galaxy and local extinction} \label{sec:host}
\begin{table}
\centering
\caption[]{Summary of the main properties of \object~and its environment.}
\label{table:SNsum}
\begin{tabular}{cc}
\hline
\noalign{\smallskip}
$\alpha$ [J 2000] & $13^h52^m24^s.11$ \\
$\delta$ [J 2000] & $+39\degr41\arcmin28\farcs2$ \\
$M_{R,peak}$ & $-20.45\,\rm{mag}$ \\
Host & NGC~5337 \\
Distance modulus & $33.63\,\rm{mag}^a$ \\
Galactic reddening & $E(B-V)=0.01\,\rm{mag}^b$ \\
Host reddening & $E(B-V)=0.97\pm0.27\,\rm{mag}$ \\
Metallicity & $12+\log{[O/H]}=8.48\,\rm{dex}$ \\
\noalign{\smallskip}
\hline
\end{tabular}
\tablefoot{$^a$ \citet{2016AJ....152...50T}; $^b$ \citet{2011ApJ...737..103S}.}
\end{table}

NGC~5337 is a SBab galaxy, with a total corrected\footnote{Apparent magnitude corrected for foreground Galactic extinction in the direction of NGC~5337 \citep[$A_{B,MW}=0.052\,\rm{mag}$;][]{2011ApJ...737..103S}, internal extinction \citep[$A_{B,int}=0.35\,\rm{mag}$;][]{1995A&A...296...64B} and k--correction.} apparent $B-$band magnitude of $12.94\pm0.29\,\rm{mag}$ and mean heliocentric radial velocity ($cz$) $V_r=2127\pm2$\kms~(as reported in the HyperLeda database\footnote{\url{http://leda.univ-lyon1.fr/}}).
In the following, we assume a luminosity distance of $53.2\pm13.2\,\rm{Mpc}$ (corresponding to a distance modulus of $33.63\pm0.54\,\rm{mag}$), which is the most recent redshift--independent distance estimate for NGC~5337 \citep{2016AJ....152...50T}. 
We note, however, a large discrepancy among different distances reported in the literature for NGC~5337 (in particular among redshift-dependent distances reported in the NASA/IPAC Extragalactic Database -- NED\footnote{\url{https://ned.ipac.caltech.edu/}}). 
Assuming a distance modulus $\mu=33.63\pm0.54\,\rm{mag}$ implies a total absolute magnitude of $B=-20.69\pm0.61\,\rm{mag}$.
The main properties inferred for \object~and its environment described in this Section are reported in Table~\ref{table:SNsum}.

An estimate of the local metallicity in the environment of \object~can be inferred through the analysis of the spectral emission lines of a relatively nearby \ion{H}{II} region, SDSS~J135223.63+394136.2, located at RA=13:52:23.63, Dec.=+39:41:36.21 [J2000] (i.e., $5\farcs54\,\rm{W}$, $8\farcs01\,\rm{N}$ from the position of \object, corresponding to a projected distance of $2.51\,\rm{kpc}$ at $53.2\,\rm{Mpc}$).
The spectrum of the \ion{H}{II} region, available through the Sloan Digital Sky Survey Data Release 14 \citep[SDSS DR14;][]{2006ApJS..162...38A} was obtained through the SDSS Catalog Archive Server (CAS\footnote{\url{https://www.sdss.org/dr14/data\_access/tools/}}).
After correcting the spectrum for the foreground Galactic extinction, we estimated the internal reddening through the observed Balmer decrement, assuming an intrinsic ratio of 2.86 \citep[case B recombination scenario, see][]{2006agna.book.....O} and a standard extinction law with $R_V=3.1$ \citep{1989ApJ...345..245C}.
Following \cite{2012A&A...537A.132B}, we therefore corrected the spectrum of the \ion{H}{II} region by including an additional contribution of $E(B-V)=0.47\,\rm{mag}$ to the total color excess.
\begin{figure}
\begin{center}
\includegraphics[width=\linewidth]{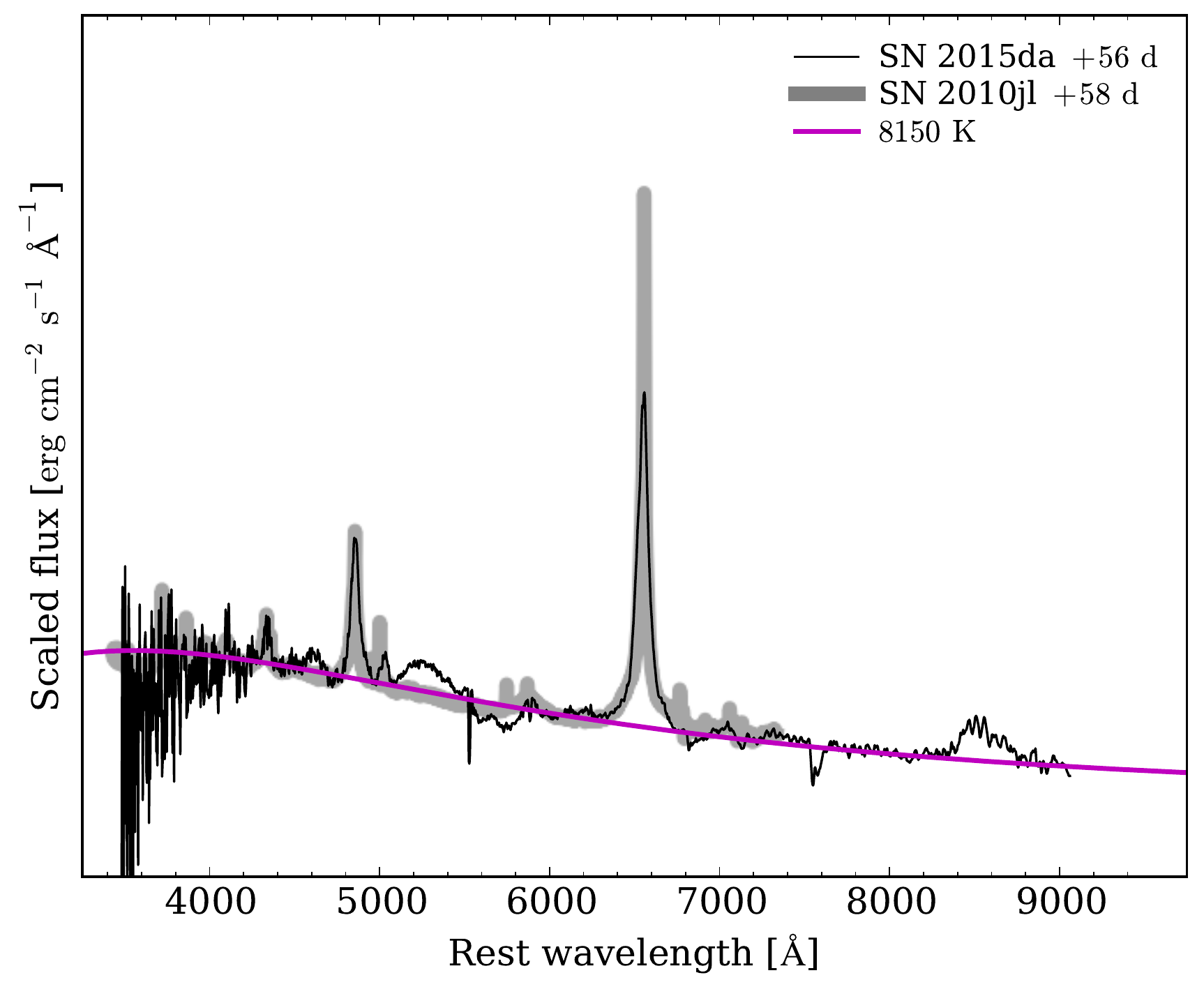}
\caption{Fit to the continuum of the optical spectrum obtained at $+56\,\rm{d}$ after $R-$band maximum compared to a spectrum of the Type IIn SN~2010jl at a similar phase. The fit was obtained by varying the amount of the extinction for \object~in order to have a BB temperature comparable with that computed for SN~2010jl. We first fixed $R_V=3.1$ and then let it vary within the $R_V=2-4$ range. \label{fig:deredSpec}}
\end{center}
\end{figure} 

The local metallicity in the environment of the \ion{H}{II} region was inferred from the integrated flux ratios of the diagnostic lines \ha~and \ion{[N}{II]}$ \lambda6583$. 
Following \citet{2004MNRAS.348L..59P} and using their recalibration of the N$2~\equiv~$log(\ion{[N}{II]}$\lambda6583/\rm{H\alpha}$) index \citep{2002MNRAS.330...69D}, we get $12+\log{\rm{[O/H]}}=8.45\,\rm{dex}$ from their linear relation, and $8.39\,\rm{dex}$ using their third-order polynomial fit (their Equations~1 and 2, respectively). 
We could not use other emission line diagnostics \citep[such as O3N2;][]{1979A&A....78..200A} since we did not detect the required lines for these methods (i.e., \ion{[O}{III]} $\lambda4363$ or $\lambda5007$). 
Since it is well established that peripheral \ion{H}{II} regions have lower metallicities than those closer to the galactic centers, we then used metallicity gradients available in the literature to infer an estimate of the local oxygen abundance at the position of \object~with respect to the value inferred at the position of the \ion{H}{II} region.
We computed the de-projected normalized distances ($r_{SN}/R_{25}$ and $r_{HII}/R_{25}$) from the center of NGC~5337 following \citet[][]{2009A&A...508.1259H} \citep[their Equations~1 and 2; see also][]{2013A&A...558A.143T}.
Using the coordinates of the SN and the host center, the inclination and major axis position angle (PA) reported in the HyperLeda database, we inferred de-projected distances $r_{SN}/R_{25}=0.35$ and $r_{HII}/R_{25}=0.47$ (17\farcs38 and 23\farcs65 offset from the center of NGC~5337) for the SN and the \ion{H}{II} region, respectively.
Assuming a metallicity gradient of $-0.47\,R/R_{25}$ \citep[i.e., the average of the metallicity gradients reported in][for a sample of galaxies]{2004A&A...425..849P}, we extrapolated the oxygen abundance at the galactocentric distance of \object, obtaining an averaged nearly solar value of $12+\log{\rm{[O/H]}}=8.48\,\rm{dex}$ \citep[assuming a solar metallicity of $12+\log{\rm{[O/H]}}=8.69\,\rm{dex}$;][]{2009ARA&A..47..481A}.
Although affected by the assumption on the metallicity gradient and by the fact that line diagnostic methods are generally believed to underestimate local abundances \citep[see, e.g.,][]{2012MNRAS.426.2630L}, this result is in agreement with the one reported by \citet{2015A&A...580A.131T} on a sample of interacting SNe, showing that long-lasting SNe IIn, like \object, seem to occur in marginally subsolar metallicity environments.

The host galaxy extinction was estimated as detailed below.
From the DEIMOS spectrum, we inferred an equivalent width (EW) of 1.3\ang~and 1.2\ang~for the D2 and D1 lines of the \ion{Na}{I\,D} doublet, respectively, which are both above the linearity range of the relation between the \ion{Na}{I\,D} EW and $E(B-V)$ \citep[$\simeq0.6$\ang; see][]{2012MNRAS.426.1465P}. 
This suggests a highly extinguished environment.
We therefore used different methods to estimate the local extinction.
Although Type IIn SNe show a remarkable heterogeneity, we notice a strong similarity between the spectroscopic and photometric evolutions of \object~and SN~2010jl \citep[e.g.,][see Section~\ref{sec:analysis}]{2011ApJ...730...34S,2012AJ....144..131Z,2014ApJ...797..118F,2015ApJ...801....7B}.
On the basis of this similarity, we used the available data to infer an estimate of the extinction, fitting the evolution of the spectral continuum of \object~to that of SN~2010jl.
Since Type IIn SNe typically show prominent Balmer lines in emission, we compared the evolution of the temperature of the pseudo photosphere estimated fitting black-body (BB) functions to selected regions of the spectral continuum (i.e., those not affected by the presence of strong emission lines, like \ha~or \hb).
We therefore fitted the temperatures using a standard extinction law \citep[assuming $R_V=3.1;$][]{1989ApJ...345..245C} within the first $\simeq200\,\rm{d}$ from the $R-$band maximum (in this context, we refer to the $R-$band maximum, since the explosion epoch of SN~2010jl is not well constrained).
For each epoch, we let the $V-$band extinction vary within the $A_V = 0-4\,\rm{mag}$ range with steps of $0.08\,\rm{mag}$, considering the one minimizing the $\chi^2$ distributions derived at each epoch as the best fit model.
With this method, we obtain $E(B-V)\simeq0.97\pm0.10\,\rm{mag}$ for $R_V=3.1$ (assuming a conservative 10\% error due to the uncertainty in the flux calibration of the spectra), whereas we do not notice a significant improvement (although we get a slightly higher extinction) when allowing $R_V$ to vary within the $2-4$ range (see Figure~\ref{fig:deredSpec}).
A similar approach is to consider the evolution of the spectral energy distribution (SED) computed using those bands typically unaffected by the presence of strong emission lines, which, therefore, map the shape of the spectral continuum ($BVIHK$) more closely.
Following the prescriptions of \citet{2018ApJ...853...62T} \citep[see also][for an equivalent approach]{2006MNRAS.369.1880E}, we fitted the evolution of the SED of \object~to that of SN~2010jl (computed using the same bands), finding a good agreement with the previous result, obtaining $E(B-V)\simeq0.92\pm0.10\,\rm{mag}$. 
\begin{figure}
\begin{center}
\includegraphics[width=\linewidth]{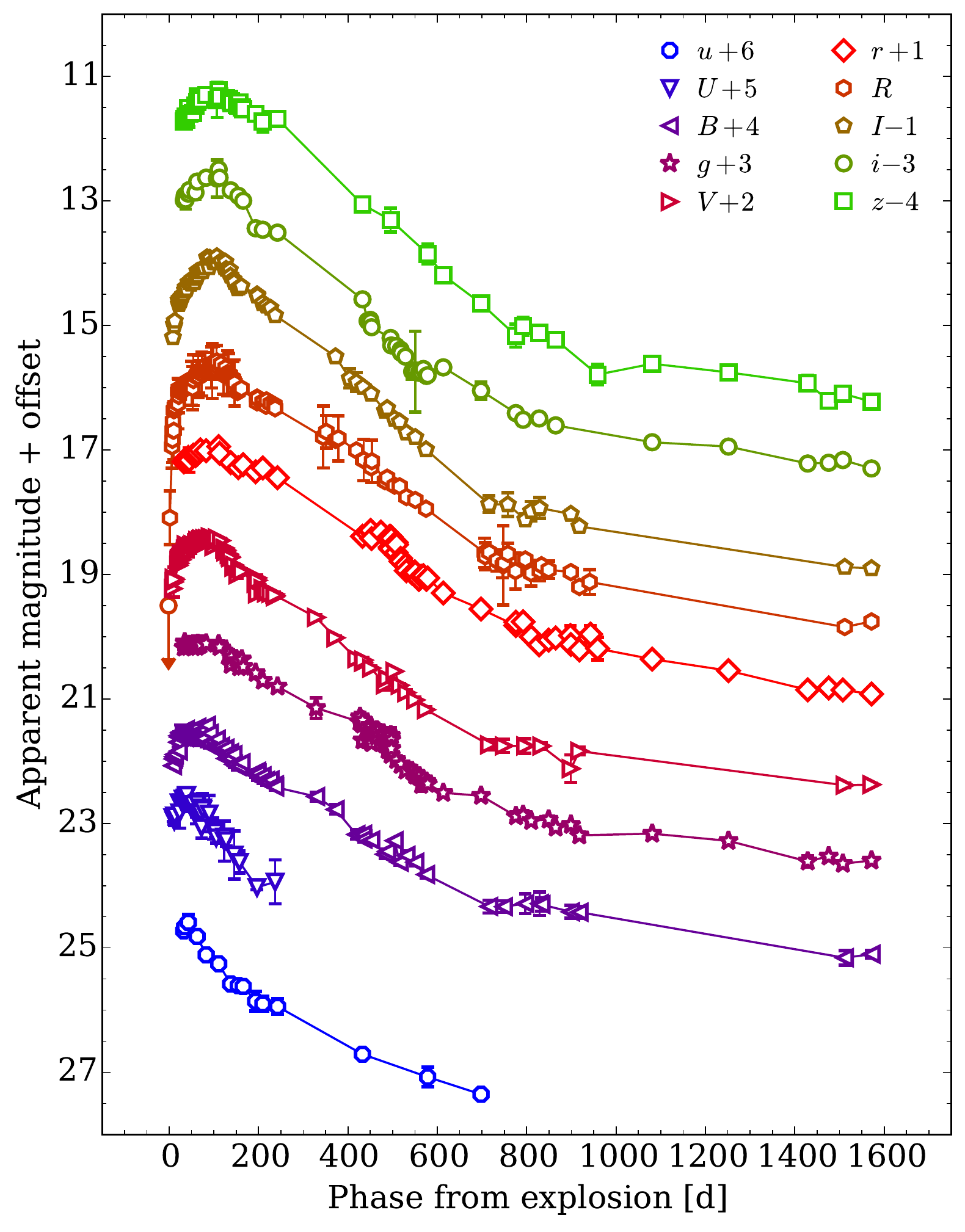}
\caption{Optical light curves of \object~during the first $\simeq4\,\rm{yr}$ after the explosion. Magnitudes are calibrated in the Vega ($UBVRI$) and AB systems ($ugriz$) and are not corrected for the MW or host galaxy extinction. \label{fig:optlc}}
\end{center}
\end{figure}

In order to test the robustness of these estimates, we compared this value with that obtained using an alternative method, based on the presence of particular features in sufficiently high-resolution spectra of highly extinguished SNe.
While we do not notice any strong evidence of diffuse interstellar bands \citep[DIBs; see, e.g.,][]{1995ARA&A..33...19H,2005A&A...429..559S} in the DEIMOS spectrum, we inferred an independent estimate of the local extinction by fitting the strong \ion{Na}{I\,D} features using the Voigt profile fitting {\sc vpfit}\footnote{\url{https://www.ast.cam.ac.uk/\~rfc/vpfit.html}} code (and $R_V=3.1$), convolving theoretical Voigt line profiles with the spectral resolution to fit the observed features, therefore obtaining an \ion{Na}{I} column density of $\log{N_{Na}}=13.53\pm0.08\,\rm{dex}$.
Following \citet{1985ApJ...298..838F} we used the relation: 
\begin{equation}
\log{N(Na\,I)=1.04[\log{N(H\,I+H_2)}]-9.09},
\end{equation}
(with $N$(\ion{Na}{I}) and $N(H)$ in $cm^{-2}$) to infer the neutral hydrogen column density and thus compute an estimate of the color excess through the relation given by \citet{1978ApJ...224..132B}:
\begin{equation}
\frac{N(\ion{H}{I}+\mathrm{H_2})}{E(B-V)}=5.8\times10^{21}\,\rm{cm^{-2}}\,\rm{mag^{-1}}.
\end{equation}
Using these prescriptions, we get an additional color excess due to the extinction in the SN environment of $E(B-V)=0.96\pm0.27\,\rm{mag}$, which is in agreement with the estimates reported above.
Therefore, in the following, we use $E(B-V)\simeq0.97\,\rm{mag}$ (assuming a standard extinction law with $R_V=3.1$) as the host galaxy reddening, implying a total color excess of $E(B-V)=0.98\pm0.30\,\rm{mag}$ (accounting for the uncertainties of the different methods used above) in the direction of \object.

\section{Analysis and discussion} \label{sec:analysis}
In the following, we qualitatively discuss the main results of the analysis on the available photometric and spectroscopic data.
In-depth modeling and focused analysis are to be the subject of a forthcoming paper.
\subsection{Photometry} \label{sec:photometry}
\begin{figure}
\begin{center}
\includegraphics[width=\linewidth]{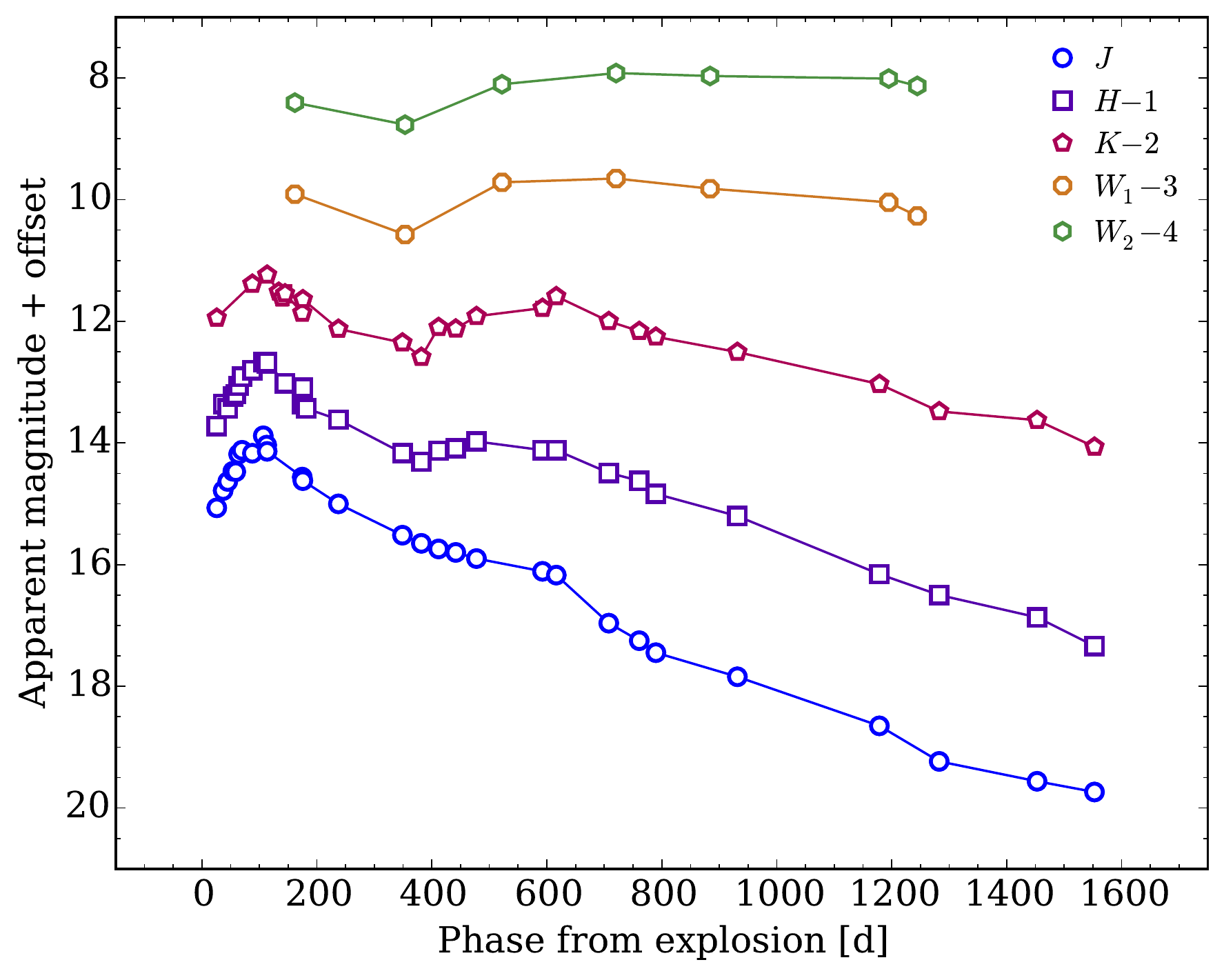}
\caption{IR light curves of \object~during the first $\simeq4\,\rm{yr}$ after the explosion. Magnitudes are calibrated in the Vega system. \label{fig:irlc}}
\end{center}
\end{figure}

Optical, near- and mid-IR (NIR and MIR, respectively) light curves of \object~are shown in Figures~\ref{fig:optlc} and \ref{fig:irlc}, respectively, and the apparent magnitudes are reported in Tables~A.1, A.2, A.3, and A.4 (Appendix~\ref{sec:obsredu}, available through the CDS), along with apparent magnitudes of the local standard stars used for the photometric calibration (Table~\ref{table:localstandards}).
Optical and NIR photometric data were mainly obtained using the telescopes of the Las Cumbres Observatory\footnote{\url{https://lco.global/}} network \citep{2013PASP..125.1031B} within the Supernova Key Project and by the NUTS collaboration\footnote{\url{http://csp2.lco.cl/not/}}.
\begin{figure*}
\begin{center}
\includegraphics[width=0.48\linewidth]{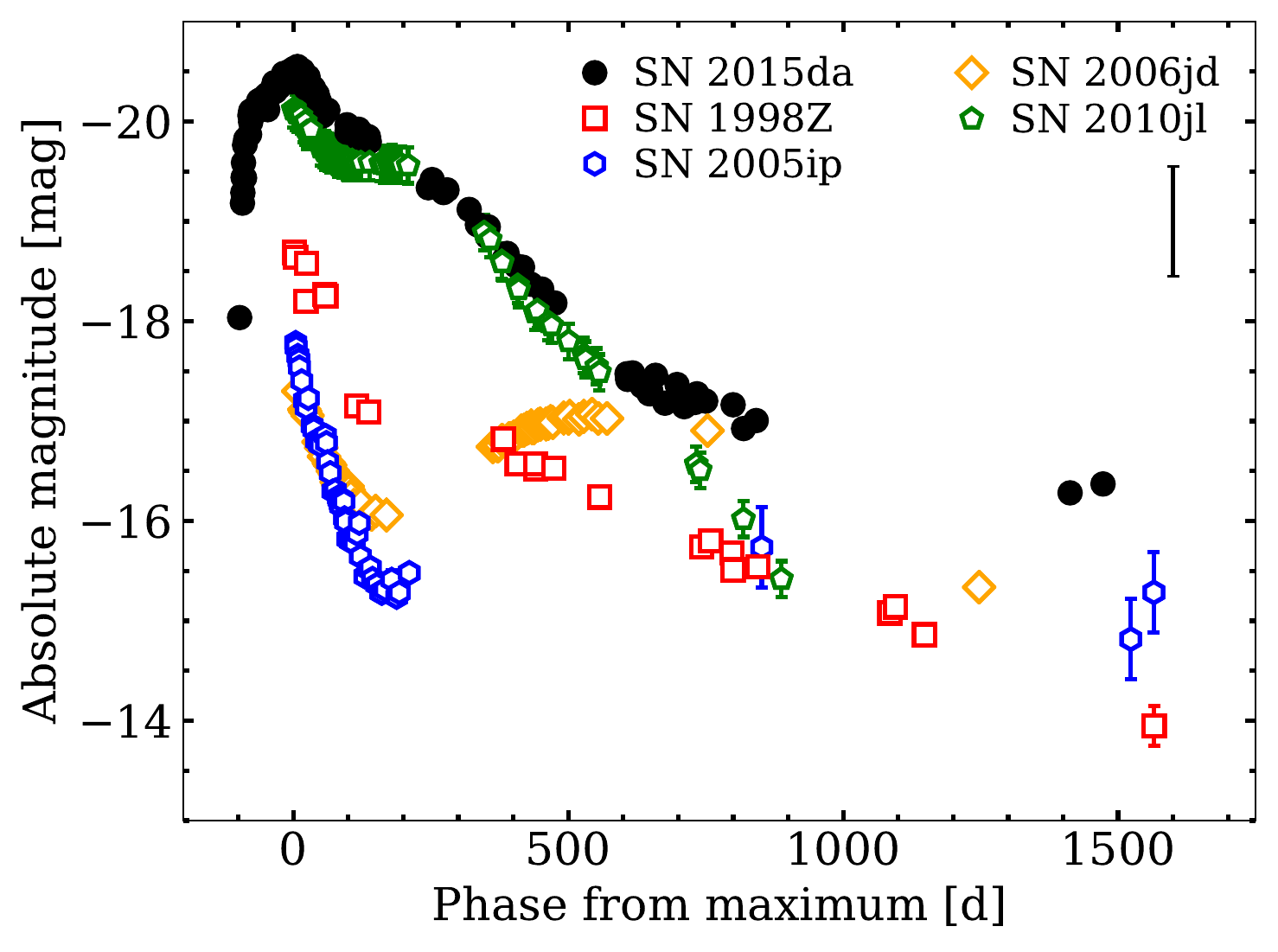}
\includegraphics[width=0.48\linewidth]{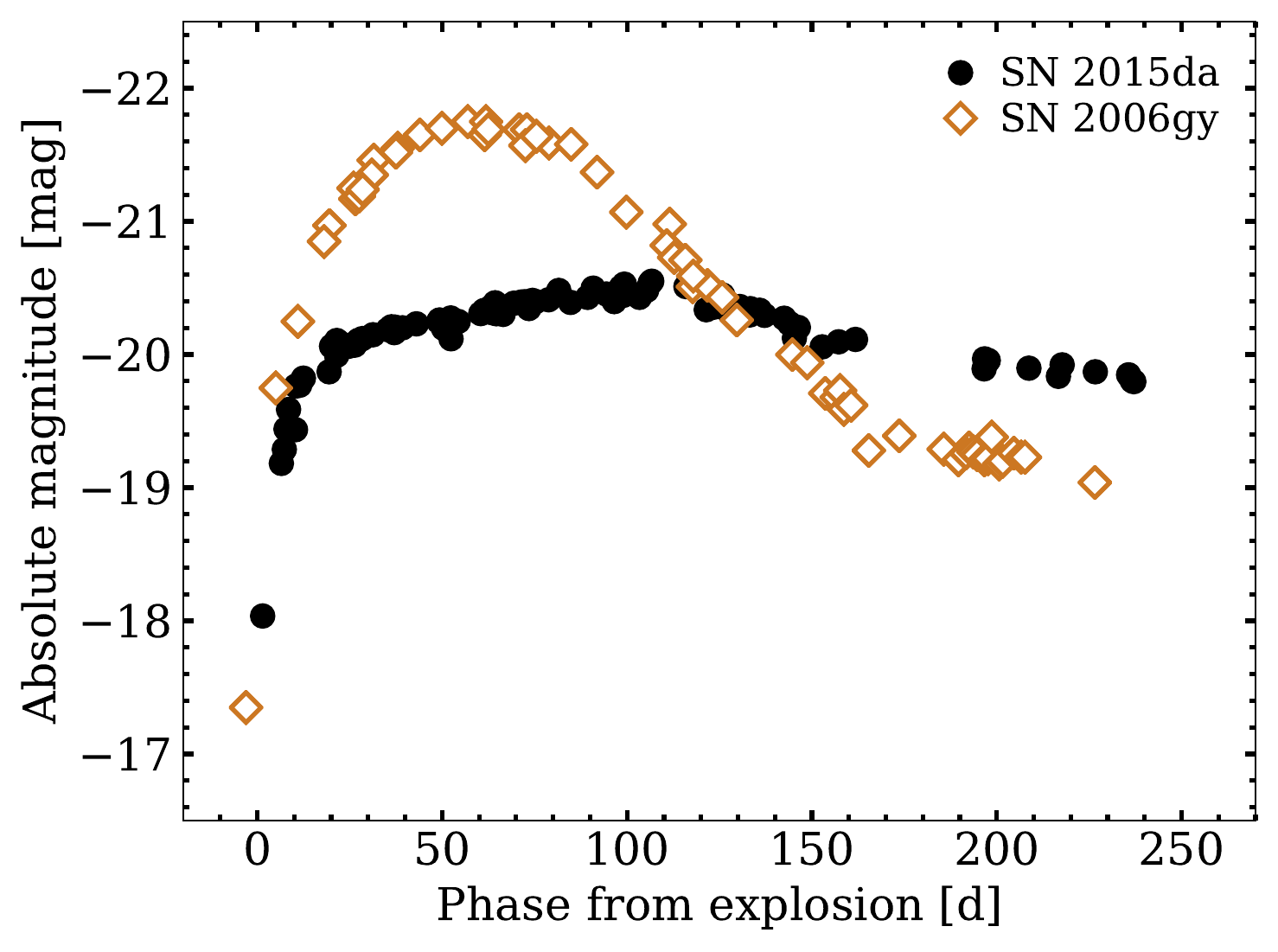}
\caption{{\bf Left:} Comparison of the $R-$band light curve of \object~with those of SNe~1988Z \citep{1993MNRAS.262..128T}, 2005ip, 2006jl \citep{2012ApJ...756..173S} and 2010jl \citep{2014ApJ...797..118F}. Absolute magnitudes were obtained using Milky Way and host galaxy extinctions and adopting distances reported in the literature. The representative error bar corresponds to the uncertainty on the distance of NGC~5337. {\bf Right:} Comparison of the $R-$band light curve of \object~with that of the overluminous SN~2006gy up to $\simeq250\,\rm{d}$. \label{fig:absRr}}
\end{center}
\end{figure*}

To estimate the explosion epoch, we first fitted a power law to the early $R-$band light curve (the one with the best coverage of the rise time including the latest nondetection limit) at $t_{\rm{MJD}}\le57122$ (i.e. $\simeq90\,\rm{d}$ after the discovery). 
This choice is based on the shape of the $R-$band light curve during the rise, which is well-reproduced by a power law at early phases only.
Performing $10^4$ Monte Carlo (MC) simulations randomly shifting the data points within their errors, we find that the early $R-$band light curve of \object~evolves as $L\propto(t-57031.8893\pm0.0007)^{0.283\pm0.006}$, suggesting an explosion epoch roughly coincident with the discovery.
On the other hand, given the large uncertainties surrounding the early points, we prefer to take the midpoint between the discovery and the last nondetection as a more conservative estimate, therefore assuming 2015 January 8.45~UT ($\rm{JD=2457030.945}$, $\simeq1.45\,\rm{d}$ before the first detection) as the explosion epoch of \object. 
Rise times in different bands were computed in the same way, fitting high-order polynomials to the light curves through MC simulations.
We find slow rise times in all bands, increasing from the blue ($t_{rise,U}=35\pm10\,\rm{d}$) to the red bands ($t_{rise,R,I}=100\pm5\,\rm{d}$), with the longest period found in the NIR bands ($106\pm10$, $110\pm10,$ and $126\pm10\,\rm{d}$ in $J$, $H,$ and $K-$band, respectively), resulting in bright absolute magnitudes at peak (e.g., $M_R=-20.45\pm0.55$ and $M_I=-20.39\pm0.55\,\rm{mag}$; see Table~\ref{table:risePeaks}).
Errors on the absolute magnitudes are dominated by the large uncertainty in the distance to NGC~5337.
In Figure~\ref{fig:absRr}, we show that the resulting $R-$band peak magnitude is consistent with that of SN~2010jl \citep[][]{2014ApJ...797..118F}, also showing many other features comparable to those observed in \object~(see below and Section~\ref{sec:spectroscopy}).
\begin{table}
\centering
\caption[]{Rise times and absolute peak magnitudes of \object~in optical and NIR bands.}
\smallskip
\label{table:risePeaks}
\begin{tabular}{cccc}
\hline\hline
\noalign{\smallskip}
Band & $t_{max}-t_{expl}$ & Error & $M_{peak}$ \\
         & (d)                           & (d) & (mag)             \\
\noalign{\smallskip}
\hline
\noalign{\smallskip}
$u$ & 33 & 15 & $-19.61$ \\
$U$ & 35 & 10 & $-20.62$ \\
$B$ & 52 & 5 & $-20.08$ \\
$g$ & 70 & 10 & $-20.18$ \\
$V$ & 87 & 10 & $-20.19$ \\
$R$ & 100 & 3 & $-20.45$ \\
$i$ & 107 & 10 & $-20.16$ \\
$I$ & 100 & 5 & $-20.39$ \\
$z$ & 107 & 10 & $-19.85$ \\
$J$ & 106 & 6 & $-20.45$ \\
$H$ & 110 & 5 & $-20.48$ \\
$K$ & 126 & 6 & $-20.74$ \\
\noalign{\smallskip}
\hline
\end{tabular}
\tablefoot{Errors on the absolute magnitudes ($\sim0.55\,\rm{mag}$ in all bands) are dominated by the uncertainties on the distance modulus and the extinction.}
\end{table}

The $R-$band light curve is the one with the best coverage thanks to the well-cadenced early observations provided by the amateurs.
During the first $\sim26\,\rm{d}$, it shows a relatively fast rise at a rate of $0.2\,\rm{mag}\,\rm{d^{-1}}$, followed by a flattening in the light curve with a rate decreasing to $\simeq0.06\,\rm{mag}\,\rm{d^{-1}}$ in the remaining $74\,\rm{d}$ before maximum.
Only a handful of Type IIn SNe have rises with such good coverage and well-estimated explosion epochs (see, e.g., the samples discussed in \citealt{2014ApJ...788..154O} and \citealt{2019arXiv190605812N}).
A good sampling of the peak was also obtained in $V$ and $I$, where we notice similar slow rises ($0.005\,\rm{mag}\,\rm{d^{-1}}$ and $0.009\,\rm{mag}\,\rm{d^{-1}}$) at $+21\lesssim t\lesssim+100\,\rm{d}$.
After peak, the $R-$band light curve shows a first rapid decline, with the luminosity getting $\simeq0.5\,\rm{mag}$ fainter in $\sim65\,\rm{d,}$ and then it flattens at a rate of $0.004\,\rm{mag}\,\rm{d^{-1}}$.
At $\simeq+400\,\rm{d}$, the slope slightly increases to $0.005\,\rm{mag}\,\rm{d^{-1}}$ until $\simeq+700\,\rm{d}$, when we note a further flattening, with the exception of the $U/u$ bands, where the light curves show linear declines after peak ($0.003\,\rm{mag}\,\rm{d^{-1}}$ and $0.008\,\rm{mag}\,\rm{d^{-1}}$ in $u$ and $U$, respectively, where the discrepancy can be attributed to the larger scatter in the $U-$band light curve).
We notice a similar behavior in the other optical bands.

The relatively bright peak magnitudes of \object~are comparable to the brighter end of the distribution presented in \citet{2019arXiv190605812N} \citep[see also][for a preliminary analysis on PTF SNe IIn]{2013ApJS..207....3S}, extending the range of luminosities inferred by \citet{2012ApJ...744...10K} to slightly higher values.
The rise times observed in \object, on the other hand, seem to fall way outside the range found for their sample of 42 Type IIn SNe, although a similar (but still $\sim40\,\rm{d}$ shorter) rise was also observed in the overluminous Type IIn SN~2006gy \citep[][see Figure~\ref{fig:absRr}, right panel]{2007ApJ...659L..13O,2007ApJ...666.1116S,2015MNRAS.454.4366F,2009ApJ...691.1348A}. 
An explanation for the lack of very long-lasting SNe IIn might be the still low number of early discoveries for this type of transient, for which some progress has been made to drastically increase the quality of the data in samples, yet it might still significantly affect statistical studies.

NIR light curves evolve in a similar way up to $\simeq+400\,\rm{d}$, where we notice a rebrightening, more pronounced in the $K-$band, lasting $\simeq240\,\rm{d}$ with an increase of $\simeq1\,\rm{mag}$.
In order to collect the most complete information possible, we searched for NEOWISE Reactivation Survey detections of \object~in the $W_1$ ($3.4\,\rm{\mu m}$) and $W_2$ ($4.6\,\rm{\mu m}$) bands. 
The first detection of a source at the SN position occured during the pass of June 2015 (18 to 25), $162\,\rm{d}$ after the explosion (see Table A.4).
No detections were recorded during the previous pass occurred a few days before the SN explosion (25 to 31 Dec. 2014), for which we only obtained upper limits  ($W1>15.7\,\rm{mag}$ and $W2>15.3\,\rm{mag}$).
Since then, the SN was detected at all passings, with a cadence of approximately six months in both bands (Table A.4) showing a similar evolution with respect to the NIR bands (see Figure~\ref{fig:irlc}).

\subsubsection{Evolution of the spectral energy distribution and analysis of the near infrared excess} \label{sec:NIRexcess}
\begin{figure*}
\begin{center}
\includegraphics[width=\linewidth]{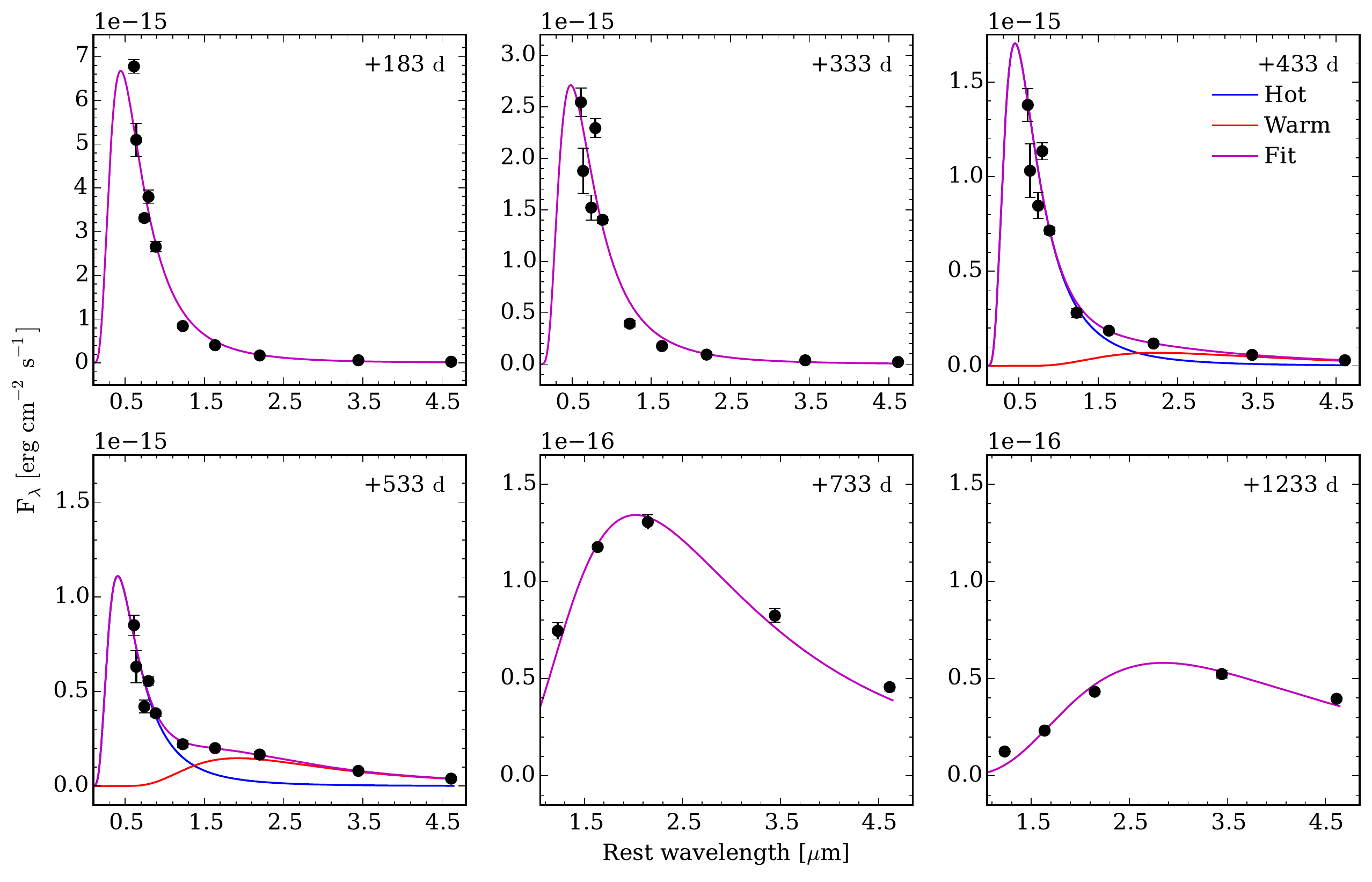}
\caption{Fit of a hot and a warm BB component to the evolution of the SED of \object. A warm component dominating at NIR wavelengths is clearly visible from $t\gtrsim+430\,\rm{d}$, increasing its strength with time. At $t\gtrsim+730\,\rm{d}$ a single BB is sufficient to reproduce the SED at IR wavelengths ($\lambda\gtrsim1.2\,\mu m$). \label{fig:2bbody}}
\end{center}
\end{figure*}

In order to model the excess in the IR luminosity observed at $t\gtrsim400\,\rm{d}$, we analyzed the evolution of the SED of \object~from $t=+33\,\rm{d}$ to $+1233\,\rm{d}$ with a regular interval of $50\,\rm{d}$.
This choice is based on the available photometric points (e.g., the first available $u-$band point is at $+33\,\rm{d}$).
Apparent magnitudes were interpolated at each epochs, corrected for the total extinction as estimated in Section~\ref{sec:host}, and converted in flux using the appropriate effective wavelengths and zero points available in the literature for each band.
Effective wavelengths were corrected for the heliocentric recessional velocity inferred from the narrow \ion{Na}{I\,D} features in absorption (see Section~\ref{sec:spectroscopy}).
Figure~\ref{fig:2bbody} shows the results of the fit at selected epochs, reported in Table~\ref{table:SEDfit}.

We used BB functions (using a single, or a combination of two BBs, when appropriate) to fit the SED evolution, excluding $u-$band fluxes at $t\ge+83\,\rm{d}$ and bands bluer than $R$ at $t\ge+183\,\rm{d}$, since strong blends of \ion{Fe}{II} lines start to dominate the blue spectral continuum around this epoch, and the SED can no longer be approximated by a BB at $\lambda\lesssim5500$\ang.
On the other hand, we estimated the contributions of \ha~and $\rm{Pa}\gamma$ lines in order to remove them from the integrated fluxes in the $r$, $R$ and $J$ bands.
Since the integrated flux of $\rm{Pa}\gamma$ could be directly estimated from the spectra only at two epochs (i.e., $+26\,\rm{d}$ and $+606\,\rm{d}$), we linearly interpolated its contribution to the $J-$band at the desired epochs, while the evolution of \ha~was estimated with higher accuracy (see Section~\ref{sec:HEv}).
Similarly, to avoid extrapolation, we did not include MIR fluxes at $t<+183\,\rm{d}$.

The SEDs were modeled following \citet[][see their Equations~1 and 2]{2018AJ....156..219S} through a Markov chain Monte Carlo (MCMC\footnote{\url{http://dfm.io/emcee/current/}}) routine based on the {\sc python} {\sc emcee} package, using the following expression:
\begin{equation}
f_{\lambda}(R,T)=\frac{2\pi hc^2}{D^2\lambda^5} \left(\frac{R^2_h}{e^{hc/\lambda k_BT_{h}}-1}+\frac{R^2_w}{e^{hc/\lambda k_BT_w}-1} \right),
\label{eq:2BBfit}
\end{equation}
where $R_h$, $T_h$, $R_w,$ and $T_w$ are the radii and temperatures of the hot and warm components, respectively, and $D$ is the distance to \object.
Adopting a luminosity distance of $53.2\pm13.2\,\rm{Mpc}$ (see Sections~\ref{sec:intro} and \ref{sec:host}), we could therefore infer the radii of the two emitting regions, assuming pure thermal emission at the equilibrium temperatures $T_h$ and $T_w$ and spherical symmetry.
A more detailed analysis, including geometrical considerations \citep[such as the one performed by][for SN~2010jl]{2011AJ....142...45A}, is beyond the scope of this paper and is to be presented in a forthcoming paper.
Up to $+433\,\rm{d}$, the SED is well-reproduced using a single BB with a temperature decreasing from $12700\pm150\,\rm{K}$ to $5850\pm90\,\rm{K}$.
During the same period, the radius of the photospheric component increases from $(1.81\pm0.02)\times10^{15}$ to $(3.49\pm0.07)\times10^{15}\,\rm{cm}$ in the first $150\,\rm{d}$, and then decreases during the following $500\,\rm{d}$ (see Table~\ref{table:SEDfit}).

From $+433\,\rm{d}$, we need a second BB to properly reproduce the SED, showing a slightly decreasing temperature (after an initial increase from $1280\pm{40}\,\rm{K}$ to $1510\pm20\,\rm{K}$, probably due to a residual contamination from the photospheric component) and a BB radius increasing from $(2.06\pm0.10)\times10^{16}$ to $(2.60\pm0.05)\times10^{16}\,\rm{cm}$.
At later times ($t\ge+733\,\rm{d}$), the fit to the IR flux is well-reproduced by a single warm component with slightly decreasing temperatures (from 1400 to $1000\rm{K}$, see Table~\ref{table:SEDfit}).
This suggests a negligible contribution of the photospheric component in the IR domain and the absence of an additional cooler component at these epochs.
\begin{table}
\centering
\caption[]{Main parameters of the BB fit to the observed SED evolution of \object.}
\smallskip
\label{table:SEDfit}
\scriptsize
\begin{tabular}{cccccc}
\hline\hline
\noalign{\smallskip}
Phase & $T_{h}$ (err) & $R_{h}$ (err) & $T_{w}$ (err) & $R_{w}$ (err) & $\log{L_{NIR}} (err)$\\
(d)       & (K)                & ($10^{15}\,\rm{cm}$) & (K) & ($10^{16}\,\rm{cm}$) & (\ergs) \\
\noalign{\smallskip}
\hline
\noalign{\smallskip}
$+$33 & 12700(150)   & 1.81(0.02) & --        & --         & --           \\
$+$83 & 9650(110)    & 2.84(0.04) & --        & --         & --           \\
$+$133 & 8280(180)   & 2.93(0.08) & --        & --         & --           \\
$+$183 & 6500(90)    & 3.49(0.07) & --        & --         & --           \\
$+$233 & 6820(90)    & 3.01(0.05) & --        & --         & --           \\
$+$283 & 6410(100)   & 2.85(0.06) & --        & --         & --           \\
$+$333 & 5990(100)   & 2.74(0.06) & --        & --         & --           \\
$+$383 & 5850(90)    & 2.42(0.05) & --        & --         & --           \\
$+$433 & 6450(260)   & 1.81(0.11) & 1280(40)  & 2.06(0.10) & 41.91(0.07)  \\
$+$483 & 6420(330)   & 1.56(0.12) & 1430(30)  & 1.98(0.07) & 42.07(0.05)  \\
$+$533 & 7300(520)   & 1.10(0.11) & 1510(20)  & 2.01(0.06) & 42.18(0.03)  \\
$+$583 & 8260(710)   & 0.81(0.09) & 1490(20)  & 2.13(0.05) & 42.20(0.03)  \\
$+$633 & 8060(660)   & 0.74(0.08) & 1430(10)  & 2.30(0.05) & 42.20(0.02)  \\
$+$683 & 12620(1890) & 0.41(0.06) & 1400(10)  & 2.41(0.05) & 42.20(0.02)  \\
$+$733 & --          & --         & 1400(10)  & 2.32(0.05) & 42.17(0.02)  \\ 
$+$783 & --          & --         & 1340(10)  & 2.42(0.05) & 42.13(0.02)  \\ 
$+$833 & --          & --         & 1300(10)  & 2.50(0.05) & 42.11(0.02)  \\ 
$+$883 & --          & --         & 1250(10)  & 2.60(0.05) & 42.07(0.02)  \\ 
$+$933 & --          & --         & 1190(5)   & 2.89(0.02) & 42.07(0.02)  \\
$+$983 & --          & --         & 1153(5)   & 2.97(0.02) & 42.04(0.02)  \\
$+$1033 & --         & --         & 1120(5)   & 3.05(0.02) & 42.02(0.02)  \\
$+$1083 & --         & --         & 1089(5)   & 3.14(0.03) & 41.99(0.02)  \\
$+$1133 & --         & --         & 1058(5)   & 3.25(0.02) & 41.97(0.02)  \\
$+$1183 & --         & --         & 1043(5)   & 3.21(0.05) & 41.93(0.04)  \\
$+$1233 & --         & --         & 1011(5)   & 3.39(0.05) & 41.93(0.03)  \\
\noalign{\smallskip}
\hline
\end{tabular}
\tablefoot{Phases refer to the epoch of the explosion.}
\end{table}

IR excesses are more common in long-lasting Type IIn SNe at late times ($t\gtrsim100\,\rm{d}$) than in other SN types \citep[e.g.,][]{2011ApJ...741....7F} and are generally interpreted as thermal emission from either preexisting or newly formed dust \citep[see, e.g.,][]{2002ApJ...575.1007G,2008MNRAS.389..141M}.
To get a rough estimate of the mass of dust needed to produce the observed excess, we followed the prescriptions of \citet{2010ApJ...725.1768F}, fitting the following expression to the observed SEDs:
\begin{equation}
f(\lambda)=B_\lambda(T_h)+\frac{M_{dust}\,B_\lambda(T_{dust})\,\kappa_\lambda(a)}{D^2},
\label{eq:modifiedBB}
\end{equation}
where $B_\lambda(T_h)$ is the BB resulting from the hot component (from Equation~\ref{eq:2BBfit}), $M_{dust}$ is the total mass of the dust, $B_\lambda(T_{dust})$ the BB at the temperature of the dust, $T_{dust}$, $\kappa_\lambda(a)$ the mass absorption coefficient for a dust particle of radius $a$ as derived from Mie theory, and $D$ is the distance of the emitting source to the observer. 
This formalism assumes only thermal emission at the equilibrium temperature $T_d$ and dust composed entirely either of carbon- or oxygen-rich grains (graphite and silicates, respectively) of a single size.
Following \citet{2012ApJ...756..173S}, we fitted the model to the observed SEDs using dust grains of three sizes (0.01, 0.1, and $1\,\rm{\mu m}$, see Table~\ref{table:DustMass}), fixing $T_h$, $R_h$ to the values derived from Equation~\ref{eq:2BBfit} (Table~\ref{table:SEDfit}).
For $s=0.1\,\rm{\mu m}$, at $+1233\,\rm{d}$ we get $\simeq5$ and $10\times10^{-3}$\msun~of preexisting dust in the CSM for graphite and silicates, respectively.
As a comparison, for the same size of graphite grains, \citet{2013AJ....146....2F} found masses in the $2\times10^{-5}-3\times10^{-2}$\msun~range for their sample of SNe IIn with available Spitzer data, so the masses inferred using Equation~\ref{eq:modifiedBB} for \object~are compatible with their derived values.
Assuming a standard gas-to-dust ratio of $1:100$, this indicates a CSM mass of $0.5-1$\msun~for \object, which is similar to that inferred by \citet{2013AJ....146....2F} for SN~2010jl \citep[$1-2$\msun~taking the gas-to-dust ratio computed by][]{2011A&A...526A.156M}.
On the other hand, in Section~\ref{sec:boloLC} we find a much higher value of swept up CSM ($\simeq8$\msun).
This difference could be due to the assumptions made when estimating the dust mass (such as grain sizes), composition, and geometrical configuration.
More accurate models are necessary to infer the physical properties of the dust producing the IR excess observed in \object.

We note that BB radii inferred for the IR component (Table~\ref{table:SEDfit}) are similar to those derived for the dust evaporation radius in SN~2010jl by \citet{2014ApJ...797..118F} for large graphite grains.
This would seem to favor large graphite grains as the main components of the preexisting dust.
On the other hand, since we do not have access to MIR spectra, we cannot rule out silicates or graphite as the main composition for the dust grains.
The highest $T_d$ we infer fitting Equation~\ref{eq:2BBfit} ($\simeq1510\,\rm{K}$ at $+533\,\rm{d}$, see Table~\ref{table:SEDfit}) is roughly coincident with the typical evaporation temperature of silicates \citep[$1500\,\rm{K}$; see, e.g.,][]{1984ApJ...285...89D}, although at those epochs, the photospheric component might still contribute to the IR SED, and, at later times, the temperature decreases well below the evaporation value for both types of grains.
In addition, we do not have measurements at $\lambda>4.6\,\rm{\mu m}$, the region where graphite and silicates show the most divergent behavior in their emission efficiencies \citep[see, e.g., Figure~4 in][]{2010ApJ...725.1768F}.

We therefore derived the dust mass from Equation~\ref{eq:modifiedBB}, using $\kappa_\lambda(a)$ for both graphite and silicates as given in \citet{1984ApJ...285...89D} and \citet{1993ApJ...402..441L}.
From the results reported in Table~\ref{table:DustMass}, we note that graphite grains with a radius of $1\,\rm{\mu m}$ give the closest $T_d$ values with respect to those inferred fitting Equation~\ref{eq:2BBfit} to the SED evolution (see Table~\ref{table:SEDfit}), although comparable values are also obtained for smaller silicate grains ($a=0.01\,\rm{\mu m}$).
The discrepancies among the results obtained using the two models might be a result of the assumptions on the geometrical distribution, composition and grain sizes, as well as the dust covering factor (a covering factor $f<1$ would give larger radii), which do not allow us to favor carbon-- or oxygen--rich grains as the main components of the dust shell.
\begin{table*}
\centering
\caption{Mass and temperature evolution of the dust shell for grains of different composition (graphite/silicates) and sizes.}
\smallskip
\label{table:DustMass}
\tiny
\begin{tabular}{ccccccc}
\hline\hline
\noalign{\smallskip}
Phase & $M_{dust}$ (err) & $T_{dust}$ (err) & $M_{dust}$ (err) & $T_{dust}$ (err) & $M_{dust}$ (err) & $T_{d}$ (err) \\
(d)       & ($\times10^{-3}\,M_{\odot}$) & (K) & ($\times10^{-3}\,M_{\odot}$) & (K) & ($\times10^{-3}\,M_{\odot}$) & (K) \\
\hline
\noalign{\smallskip}
           & \multicolumn{2}{c}{$a=0.01\,\rm{\mu m}$} & \multicolumn{2}{c}{$a=0.1\,\rm{\mu m}$} & \multicolumn{2}{c}{$a=1\,\rm{\mu m}$} \\
\noalign{\smallskip}
\hline
\noalign{\smallskip}
$+433$ & 2.43(0.25)/3.78(0.37) &  930(20)/1100(25) & 2.25(0.23)/3.95(0.38) & 875(20)/1080(25) & 0.42(0.04)/2.14(0.22) & 1225(30)/1025(25) \\
 $+483$ & 2.13(0.16)/3.43(0.25) & 1020(15)/1215(20) & 1.97(0.15)/3.60(0.27) & 955(15)/1190(20) & 0.38(0.03)/1.84(0.14) & 1370(25)/1140(20) \\
 $+533$ & 2.16(0.14)/3.52(0.23) & 1065(15)/1270(20) & 2.01(0.13)/3.70(0.24) & 995(10)/1245(15) & 0.39(0.03)/1.87(0.13) & 1440(20)/1195(15) \\
 $+583$ & 2.41(0.15)/3.92(0.24) & 1060(11)/1265(15) & 2.23(0.14)/4.12(0.26) & 990(10)/1240(15) & 0.44(0.03)/2.08(0.13) & 1430(20)/1190(15) \\
 $+633$ & 2.65(0.15)/4.51(0.25) & 1035(10)/1225(10) & 2.46(0.14)/4.74(0.26) & 970(10)/1200(10) & 0.51(0.03)/2.40(0.13) & 1380(15)/1150(10) \\
 $+683$ & 2.82(0.15)/4.88(0.26) & 1020(10)/1200(10) & 2.61(0.14)/5.13(0.27) & 955(10)/1175(10) & 0.55(0.03)/2.56(0.14) & 1350(15)/1130(10) \\
 $+733$ & 2.27(0.13)/4.36(0.23) & 1040(10)/1210(10) & 2.09(0.12)/4.61(0.24) & 970(10)/1180(10) & 0.50(0.03)/2.23(0.12) & 1350(10)/1140(10) \\
 $+783$ & 2.47(0.14)/4.72(0.25) & 1010(10)/1170(10) & 2.28(0.13)/4.99(0.26) & 945(10)/1145(10) & 0.54(0.03)/2.41(0.13) & 1300(10)/1110(10) \\
 $+833$ & 2.67(0.15)/5.08(0.26) &  980(10)/1135(10) & 2.46(0.13)/5.37(0.28) & 920(10)/1110(10) & 0.59(0.03)/2.60(0.14) & 1260(10)/1080(10) \\
 $+883$ & 2.90(0.16)/5.50(0.29) &  955(10)/1100(10) & 2.67(0.15)/5.80(0.30) & 900(10)/1080(10) & 0.63(0.03)/2.83(0.15) & 1220(10)/1045(10) \\
 $+933$ & 4.83(0.88)/7.32(0.13) &  890(10)/1040(10) & 4.47(0.81)/7.64(0.14) & 840(10)/1020(10) & 0.81(0.01)/4.60(0.85) & 1150(10)/965(10) \\
 $+983$ & 5.06(0.93)/7.69(0.14) &  870(10)/1010(10) & 4.67(0.85)/8.02(0.15) & 820(10)/1000(10) & 0.85(0.02)/4.81(0.88) & 1120(10)/945(10) \\
$+1033$ & 5.31(0.97)/8.09(0.15) &  850(10)/990(10) & 4.90(0.89)/8.43(0.15) & 805(10)/970(10) & 0.90(0.02)/5.05(0.93) & 1090(10)/920(10) \\
$+1083$ & 5.60(0.10)/8.56(0.16) &  830(10)/965(10) & 5.17(0.94)/8.92(0.16) & 790(10)/950(10) & 0.96(0.02)/5.34(0.98) & 1060(10)/900(10) \\
$+1133$ & 5.99(0.11)/9.17(0.17) &  815(10)/940(10) & 5.52(0.10)/9.56(0.17) & 775(10)/925(10) & 0.10(0.02)/5.71(0.10) & 1030(10)/880(10) \\
$+1183$ & 4.80(0.20)/8.46(0.34) &  820(10)/930(10) & 4.42(0.18)/8.88(0.35) & 780(10)/920(10) & 0.97(0.04)/4.57(0.19) & 1020(10)/890(10) \\
$+1233$ & 5.36(0.22)/9.47(0.38) &  800(10)/910(10) & 4.93(0.20)/9.93(0.39) & 760(10)/895(10) & 0.11(0.04)/5.15(0.21) &  990(10)/865(10) \\
\noalign{\smallskip}
\hline
\end{tabular}
\tablefoot{Phases refer to the estimated epoch of the explosion.}
\end{table*}

The analysis of the IR luminosity evolution may give additional information about the nature of the dust grains responsible for the excess of radiation observed at $t\ge433\,\rm{d}$.
IR bolometric luminosities were computed using the $R_w$ and $T_w$ obtained fitting Equation~\ref{eq:2BBfit} to the observed SEDs (Table~\ref{table:SEDfit}).
The resulting light curve shows a slow rise during the first $100\,\rm{d}$ (after $+433\,\rm{d}$, the onset of the IR excess), a peak at $\simeq1.6\times10^{42}$\ergs,~and subsequently a settling onto a relatively long "plateau".
At $+1233\,\rm{d}$, the last epoch at which we could interpolate available NIR and MIR photometry, we inferred a luminosity of $\simeq8.5\times10^{41}$\ergs, suggesting a very slow decline after peak (see Table~\ref{table:SEDfit}).

According to the discussion on SN~2010jl \citep[see, e.g.,][]{2011AJ....142...45A,2012AJ....143...17S,2014ApJ...797..118F,2014Natur.511..326G}, dust responsible for the IR excess can originate from different mechanisms, including rapid formation of large dust grains in the post-shocked regions at the interface between forward and reverse shock, heating and evaporation of preexisting dust by the SN shock, or an "echo" from a preexisting outer dust shell.
While emission from relatively large dust grains ($\simeq1\,\rm{\mu m}$) is able to reproduce the temperature of the dust inferred from broad-band photometry (see Tables~\ref{table:SEDfit} and \ref{table:DustMass}), in the following, we show that an IR echo is the most plausible interpretation for the observed late excess in the IR luminosity of \object.

Although the formation and survival of dust grains in post-shocked regions behind a radiative shock proved to be a viable explanation for the IR excess observed in a few Type IIn SNe \citep[see, e.g., the cases of SN~2005ip and 2006jd;][]{2012ApJ...756..173S}, \citet{2014ApJ...797..118F} showed that emission from newly formed dust would not be sufficient to explain the large IR luminosity observed at late time in SN~2010jl, which is comparable to that observed in \object.
In addition, condensation of dust either within the SN ejecta or in post-shocked regions is expected to scatter and absorb light emitted by the underlying onrushing ejecta, causing the attenuation of the red wings of the most prominent lines (e.g., \ha) and steeper decline rates, in particular in the bluer optical light curves.
We do not see such features in \object: \ha~always shows symmetric profiles with respect to its centroid, and the optical light curves do not show steeper declines. 
We therefore do not consider newly formed dust in the CDS of \object~as a plausible interpretation for the IR excess observed at $t\ge+433\,\rm{d}$.

A useful probe to infer the origin of the dust emission is to compare the BB radius of the dust component obtained from Equation~\ref{eq:2BBfit} to the radius of the shocked region inferred from the maximum velocity of the shock, as estimated in Section~\ref{sec:boloLC} ($R_s\simeq V_{s}t$, with $V_s\simeq3000$\kms), that is $\simeq1.1\times10^{15}\,\rm{cm}$ at $+433\,\rm{d}$.
This value is a factor of two smaller than that inferred for $R_{w}$ (see Table~\ref{table:SEDfit}).
For the shock to reach the dust shell at $+433\,\rm{d,}$ we would therefore need significantly higher values for $V_s$.
We stress, however, that the value of the dust shell radius has to be considered a lower limit, since it is computed assuming spherical, symmetric geometrical configurations, and a dust covering factor $f=1$. 

Another useful quantity is the evaporation radius for a given grain size, composition and bolometric luminosity at peak, namely the radius of the dust--free cavity produced by an SN outburst around the progenitor star.
According to \citet{2014ApJ...797..118F}, an SN outburst with a peak luminosity of $10^{43}$\ergs~produces a dust-free cavity with a radius $\gtrsim3.5\times10^{16}\,\rm{cm}$, depending on the dust composition, the grain size, and the effective temperature of the SN, with the lower limit reached for $T_{SN}=6000\,\rm{K}$ and carbon-rich dust grains with $a=1\,\rm{\mu m}$.
The radius of the dust-free cavity is larger than the BB radii inferred from the IR SEDs of \object~at all epochs, although, as discussed above, these have to be considered as lower limits.
The evaporation radius, on the other hand, also depends on the assumed evaporation temperature, which may be lower than the typical value for graphite/silicates if a more luminous outburst precedes the SN explosion, leading to a larger radius of the dust-free cavity \citep{1983ApJ...274..175D}. 

These considerations lead us to the conclusion that collisional heating and evaporation of preexisting dust is not likely to be the mechanism responsible for the IR excess in \object,~and that an IR echo from heated dust is a more promising alternative.
A simple approach to modeling IR echoes from preexisting radiatively heated dust is to consider a spherically symmetric shell around the SN composed solely of spherical grains of a single size and composition, optically thin (at IR wavelengths) to the SN radiation \citep[see][]{1983ApJ...274..175D,1986MNRAS.221..789G}.
The observed IR radiation therefore arises from a paraboloidal surface and is significantly delayed with respect to the SN light due to light-travel time effects.
In this context, the observed dust temperature is a function of the angle between the vector radius from the SN to the emitting shell element and the line of sight (i.e., shell elements closer to the SN are observed at higher temperatures) as well as the distance to the SN and the time from explosion.

One piece of evidence favoring this interpretation is the long plateau and very slow decline after peak observed in the IR bolometric luminosity of \object.
The IR light curve of an SN embedded in a dusty circumstellar shell depends on the chemical composition of the dust grains and the size of the dust-free cavity produced by the explosion.
\citet{1980ApJ...242L..23W} proposed an analytical solution for the bolometric light curve of a IR light echo showing a "flat top" lasting $\sim2*R_{ev}/c$, where $R_{ev}$ is the radius up to which dust is vaporized by the SN explosion.
Assuming a plateau length of $200\,\rm{d,}$ the dust-free cavity produced by \object~is $\simeq2.6\times10^{14}\,\rm{cm}$.
Following the same prescriptions, \citet{1985ApJ...297..719D} showed that the IR luminosity produced by a carbon-rich shell is expected to be brighter and shorter in time with respect to an oxygen-rich one (see his Figure~1).
In this context, the IR light curve observed in \object~would be explained by an extended shell of dust, mainly composed of silicates.

Despite the two different methods used to compute luminosities, we compared the total radiated energies obtained integrating the photospheric part of the pseudo-bolometric light curve up to $+433\,\rm{d}$ ($9.1\times10^{50}\,\rm{erg}$) and the IR bolometric light curve up to $+1233\,\rm{d}$ estimated by the BB fit ($8.5\times10^{49}\,\rm{erg}$).
While up to $+1233\,\rm{d,}$ we still do not see a clear fall from the plateau in the IR light curve: the fact that only $\sim10\%$ of the integrated IR luminosity is reprocessed by the dust suggests that it is confined in a preexisting optically thin shell.

As the NIR and MIR follow-up campaigns of \object~are still ongoing, it is likely that further photometric epochs at these wavelengths, as well as additional observations at $\lambda>4.6\,\rm{\mu m}$, could help to unveil the nature of the IR excess.

\subsubsection{Evolution of the pseudo-bolometric light curve} \label{sec:boloLC} 
The evolution of the bolometric luminosity of \object~was computed from the available photometric data in \citet{2016ApJ...823L..23T}, integrating the observed SEDs without considering those regions not covered by the observations, hence only from the $u$ to the $W_2$ bands.
The resulting "pseudo-bolometric" light curve shows a slow rise ($\simeq100\,\rm{d}$) with a bright peak luminosity $\simeq3.0\times10^{43}$\ergs.
However, the early SN bolometric light curves can be significantly underestimated without taking into account the UV contribution \citep[see also the discussion in][]{2018MNRAS.475.1937T}.
We therefore took the temperatures obtained fitting Equation~\ref{eq:2BBfit} to the observed SEDs to infer the scaling factor to account for the lack of observations at $\lambda<3500$\ang, assuming a blackbody form. 
Due to the strong contamination of \ion{Fe}{II} lines at wavelengths bluer than $\simeq5000$\ang~(see Sections~\ref{sec:spectroscopy} and \ref{sec:NIRexcess}), at $t>400\,\rm{d}$ we fixed the temperature of the pseudo continuum to $5850\,\rm{K}$, although we note that the correction at these epochs is small (less than 10\%).
We note, however, that this might still underestimate the UV contribution at early epochs, as shown by \citet[][their Figure~13]{2015MNRAS.449.4304D}.

\begin{figure}
\begin{center}
\includegraphics[width=\linewidth]{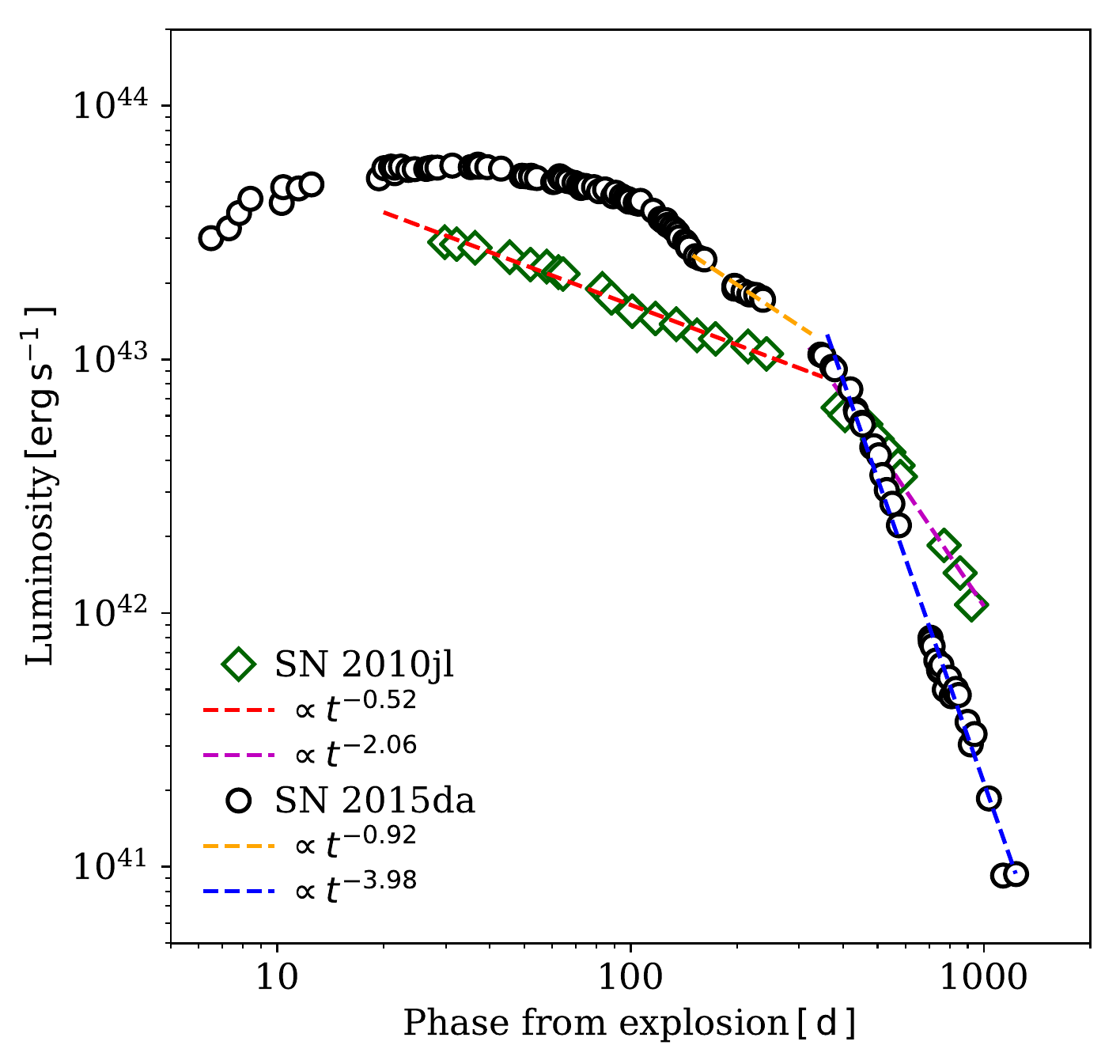}
\caption{Bolometric light curve of \object~compared to that of SN~2010jl in logarithmic units. The best fit parameters for $+150\le t\le+290\,\rm{d}$ and $t>+290\,\rm{d}$ ($+30\le t\le+320\,\rm{d}$ and $t>+320\,\rm{d}$ for SN~2010jl following \citealt{2014ApJ...797..118F} and \citealt{2014ApJ...781...42O}) are reported in the legend. \label{fig:bololog}}
\end{center}
\end{figure}

The resulting bolometric light curve, shown in Figure~\ref{fig:bololog}, has a significantly shorter rise of $\simeq30\,\rm{d}$, with a much brighter peak, $\simeq6.2\times10^{43}$\ergs, more than a factor two brighter than our previous estimate, and more than three times higher than that inferred for SN~2010jl by \citet{2014ApJ...797..118F}. 
The UV contribution for SN 2010jl was, however, estimated from direct observations and the explosion is uncertain, hence its peak may have been missed, resulting in a possible brighter luminosity. 
In any case, this would make \object~one of the brightest Type IIn SNe ever observed.
The total radiated energy in the first $1233\,\rm{d}$, after removing the IR contribution (likely due to a dust echo; see Section~\ref{sec:NIRexcess}), is $\simeq10^{51}\,\rm{erg}$.
This high value and the prominent narrow emission lines suggest that an ejecta-CSM interaction with a highly efficient conversion of kinetic energy into radiation is the main mechanism powering the light curve of \object~\citep[see, e.g.,][]{2014ApJ...797..118F}. 

The luminosity evolution of \object~can then be modeled in the context of an SN explosion within an extended and dense CSM \citep[see][]{2011ApJ...729L...6C,2012ApJ...759..108S,2014ApJ...781...42O}.
After CC, a radiation-mediated shock wave propagates outwards until it breaks through the stellar photosphere.
If the progenitor is surrounded by a dense and extended CSM with an optical depth $\tau_{CSM}>c/V_{sh}=\tau_{sh}$ (where $c$ is the speed of light and $V_{sh}$ the velocity of the shock), the SN shock propagates into the CSM.
As a consequence, the shock breakout signal is delayed, and the reverse-forward shock structure typical of interacting transients forms at the interface between the SN ejecta and the CSM \citep{1994ApJ...420..268C}.
As the photon diffusion time drops below the shock expansion timescale (i.e., when $\tau_{CSM}\simeq \tau_{sh}\simeq c/V_{s}$), the thermalized shock radiation escapes the CSM, releasing energy over timescales that may be much longer than those typically observed in windless breakouts \citep[see][]{2010ApJ...724.1396O,2011ApJ...729L...6C}.
If the density of the CSM above the breakout layer is sufficiently high, the shock becomes collisionless, the photons are no longer trapped \citep{2012ApJ...747..147K}, and the kinetic energy of the ejecta is efficiently converted into radiation at a rate of $\epsilon(\rho_{CSM}V_{sh}^2/2)(4\pi r_{sh}^2V_{sh})$, where $\epsilon$ is the conversion efficiency, $\rho_{CSM}$ the density of the CSM, and $r_{sh}$ and $V_{sh}$ the radius and velocity of the shock, respectively \citep{2012ApJ...759..108S}.

At later times, when the mass of the swept-up CSM is comparable to, or larger than, the ejected mass, the system can either enter a phase of energy (the Sedov-Taylor phase) or momentum conservation \citep["snowplow" phase;][]{2012ApJ...759..108S}, which, in both cases, would result in a steeper decay of the observed luminosity. 
If the shock reaches the outer boundary of the dense CSM before this phase occurs, an even steeper drop in luminosity takes place. 

Assuming spherical symmetry, \citet{1982ApJ...259..302C} showed that the density profiles of the SN ejecta and dense CSM can be described by power laws of the form $\rho_{ej}\propto t^{-3}(r/t)^{-n}$ and $\rho_{CSM}=qr^{-s}$, respectively.
In a wind profile ($s=2$), the density profile of the surrounding material can be expressed by $\rho_w=\dot{M}/(4\pi u_{\rm{w}} r^{2})$ \citep[see, e.g.,][]{2017hsn..book..875C}, and therefore when we know the normalization constant $q,$ it is possible to infer the mass-loss rate of the progenitor star during the last stages of its evolution\footnote{Hereafter we adopt the nomenclature used by \citet{1982ApJ...258..790C}, while in \citet{2014ApJ...781...42O} $q$, $n,$ and $s$ are called $K$, $m,$ and $w$, respectively.}. \\
\begin{figure}
\begin{center}
\includegraphics[width=\linewidth]{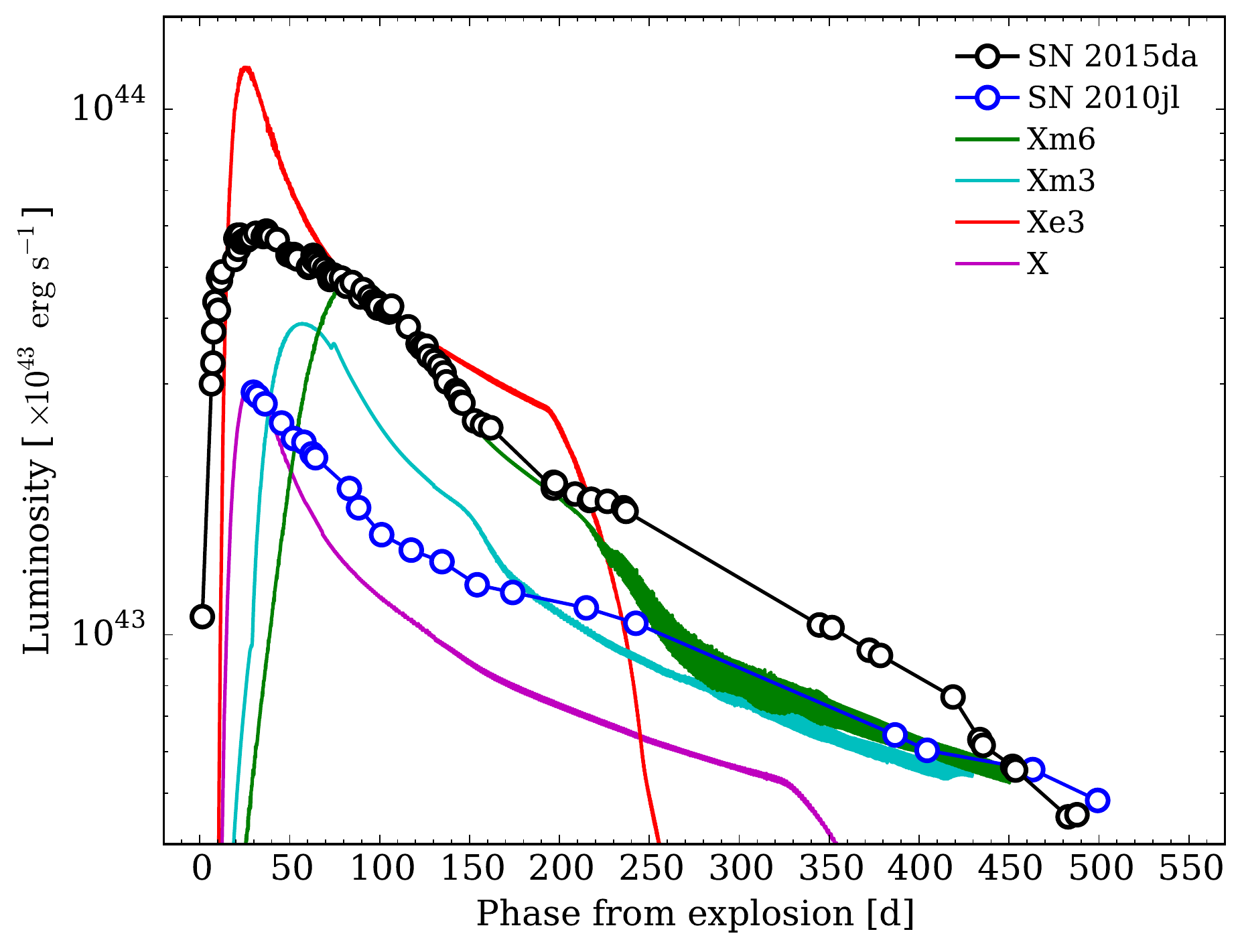}
\caption{Bolometric light curve of \object~compared to models presented in \citet{2015MNRAS.449.4304D} and bolometric light curve of SN~2010jl \citep[computed as in ][]{2014ApJ...797..118F}. Model X has $\dot{M}=0.1\,\rm{M_{\odot}}\,\rm{yr^{-1}}$, model Xm3 $\dot{M}=0.3\,\rm{M_{\odot}}\,\rm{yr^{-1,}}$ and model Xm6 $\dot{M}=0.6\,\rm{M_{\odot}}\,\rm{yr^{-1}}$. In all cases, the total energy is $10^{51}$ergs. Model Xe3 has $\dot{M}=0.1\,\rm{M_{\odot}}\,\rm{yr^{-1}}$ and total energy $3\times10^{51}$ergs. \label{fig:DessartBolom}}
\end{center}
\end{figure}

Assuming that the shock breakout occurs at $\tau\approx c/V$, \citet{2014ApJ...781...42O} showed that the value of $q$ can be estimated using
\begin{equation}
q=\frac{c}{\kappa}\left(s-1\right)\left(\frac{n-s}{n-3}\right)^{s-1}V_{bo}^{s-2}t_{bo}^{s-1,}
\label{eq:massloadpar}
\end{equation}
where $\kappa$ is the opacity ($\kappa=0.34\,\rm{cm^2}\,\rm{g^{-1}}$ for electron scattering in a H--rich gas), $c$ the speed of light, and $V_{bo}$ is the shock velocity at the shock breakout $V_{s}(t_{bo})$. 
Therefore, for $s=2$, $q$ is only a function of $n$ and $t_{bo}$.
Using $L(t)=\epsilon\dot{M}V_{s}(t)^3/(2u_{\rm{w}})$, $V_{s}(t_{bo})$ can be expressed as a function of the energy conversion efficiency $\epsilon$
\begin{equation}
V_{bo}=t_{bo}^{(\alpha-1)/3}\left[2\pi\epsilon\frac{n-s}{n-3}\left(s-1\right)\frac{c}{\kappa L_0}\right]^{-1/3},
\label{eq:vbo}
\end{equation}
with $L_0$ directly derived by fitting the bolometric light curve before the break with a power law.
In a wind profile, $q=\dot{M}/4\pi u_{w}$ and the mass-loss rate of the progenitor star can be derived directly from the density profile of the CSM as a function of the wind expansion velocity $u_w$, as Equation~\ref{eq:massloadpar} loses its dependence on $V_{bo}$.
The mass of the CSM swept up by the shock at the time $t$ can be expressed as a function of $L_0$, $t_{bo}$, $n$ and $s$ (see Equation~21 in \citealt{2014ApJ...781...42O}), which, in a wind profile, can be expressed by:
\begin{equation}
M_{\mathrm{CSM}}=4\pi c\left(\frac{n-2}{n-3}\right)^{5/3}\left(\frac{2\pi c\epsilon}{\kappa^2L_0}\right)^{-1/3}t_{bo}^{2/3}t^{(n-3)/(n-2)}.
\label{eq:CSMmass}
\end{equation}
\begin{figure*}
\begin{center}
\includegraphics[width=0.48\linewidth]{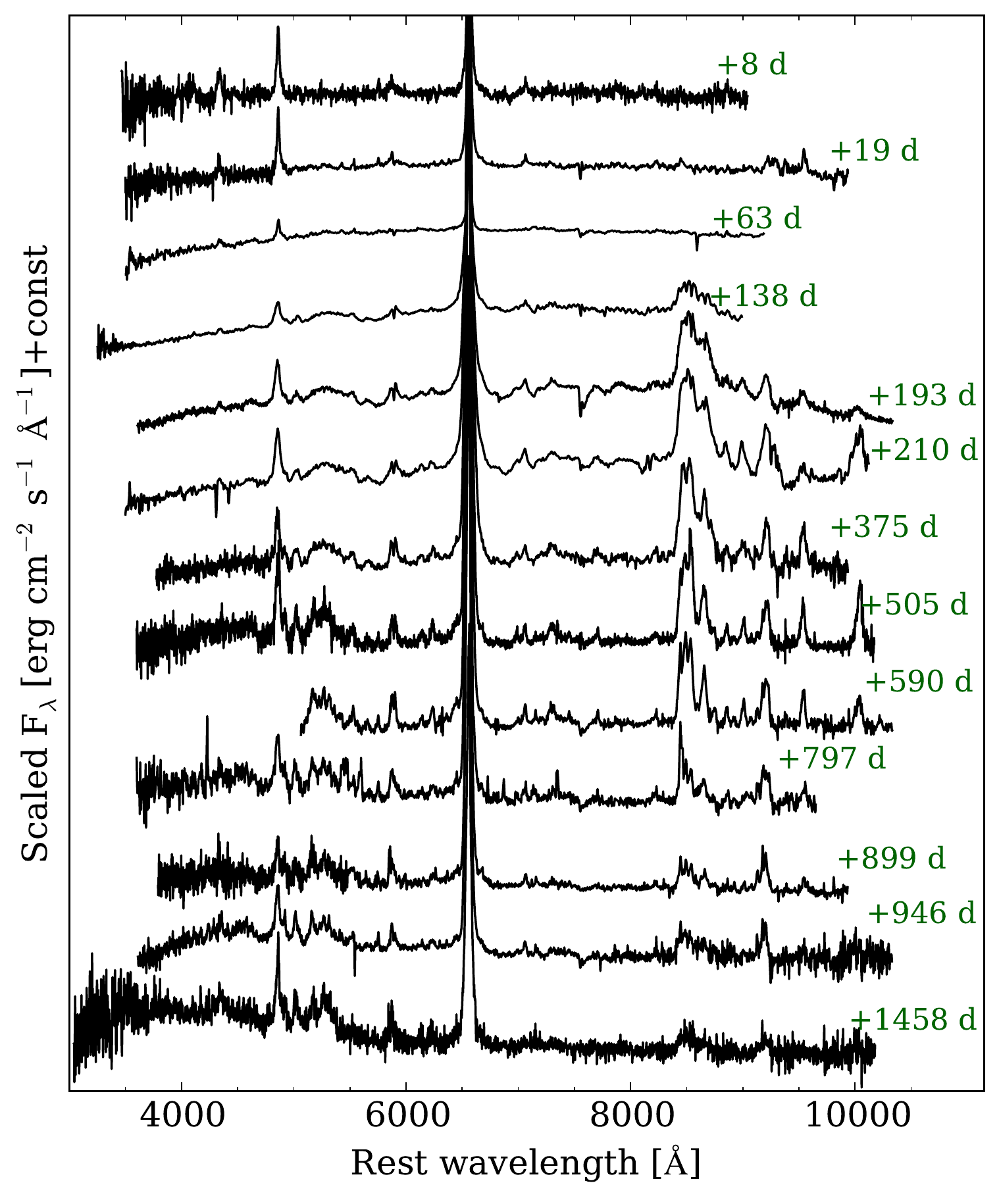}
\includegraphics[width=0.48\linewidth]{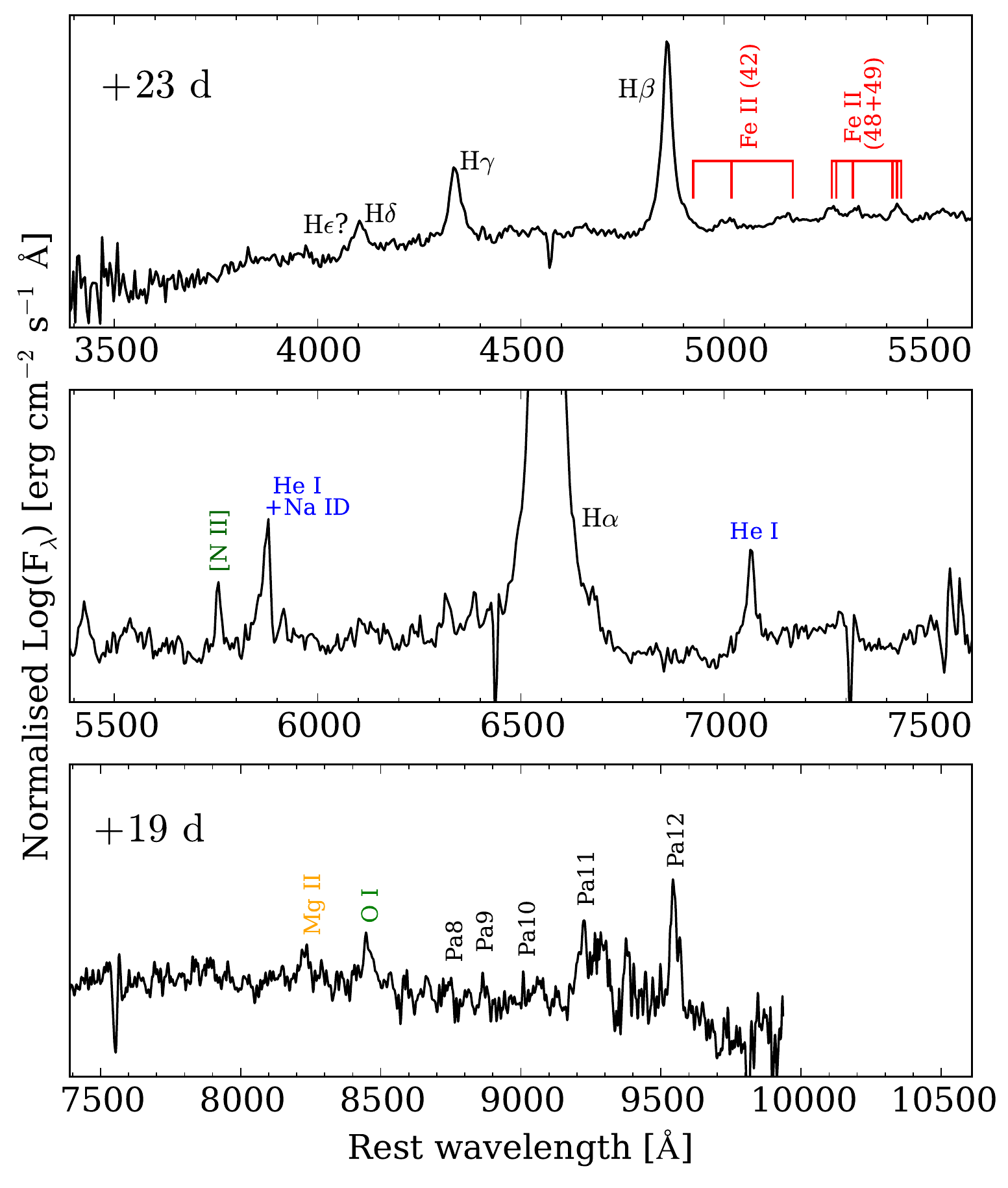}
\caption{{\bf Left:} Selection of optical spectra of \object~with highest SNR, spectral coverage and resolution. Spectra were not corrected for extinction to facilitate the comparison at wavelengths bluer than $\simeq5000$\ang. {\bf Right:} Line identification on the $+23\,\rm{d}$ (first and second panel) and $+19\,\rm{d}$ (bottom panel) spectra. Phases refer to the estimated epoch of the explosion.
\label{fig:Optspec}}
\end{center}
\end{figure*}

The pseudo-bolometric light curve of \object~can be interpreted as the result of a SN shock propagating in a dense wind, with photon diffusion in the dense H-rich CSM being the dominant mechanism until $t\simeq+150\,\rm{d}$.
This epoch corresponds to the onset of the broken power law describing the light curve, when the energy input from the shock directly corresponds to the observed luminosity.
The "shock breakout" through the dense wind can then be estimated from the peak of the light curve at $\simeq30\,\rm{d}$ (see Figure~\ref{fig:bololog}), while the observed break at $t\simeq+290\,\rm{d}$ corresponds to the onset of the snowplow phase. 
On the other hand, as shown by \citet[][]{2014A&A...564A..83M}, this would result in a modest decay rate, and a more likely interpretation is the propagation of the shock into a CSM with a lower density, as also inferred from the narrow P-Cygni in the \ha~profile (see Section~\ref{sec:keckspec}). 

At $t\ge+150\,\rm{d,}$ the bolometric light curve of \object~is well-reproduced by a broken power law, with an early luminosity evolution described by $L=9.59\times10^{49}\,t^{-0.92}\,\rm{erg}\,\rm{s^{-1}}$ followed by a much steeper decline with $L\propto t^{-3.98}\,\rm{erg}\,\rm{s^{-1}}$ until $+1233\,\rm{d}$ (subtracting the contribution of the IR luminosity at $t\ge+433\,\rm{d}$; see Section~\ref{sec:NIRexcess}), while the change in the power law index occurs around $+290\,\rm{d}$.

\cite{2014ApJ...781...42O} showed that the value of $n$ is related to the power law index $\alpha$ through the equation:
\begin{equation}
\alpha\equiv\frac{(2-s)(s-3)+3(s-3)}{n-s},
\label{eq:nexp}
\end{equation}
which, in a wind profile, gives $n=5.26$ for $\alpha=-0.92$.
This value is considerably lower than that inferred for SN~2010jl \citep[$n\simeq7.6$;][]{2014ApJ...797..118F}. 
The relative time interval between the diffusion phase and break in the light curve is, however, considerably shorter for \object, which makes the determination of the power law index more uncertain. 
\begin{figure*}
\begin{center}
\includegraphics[width=\linewidth]{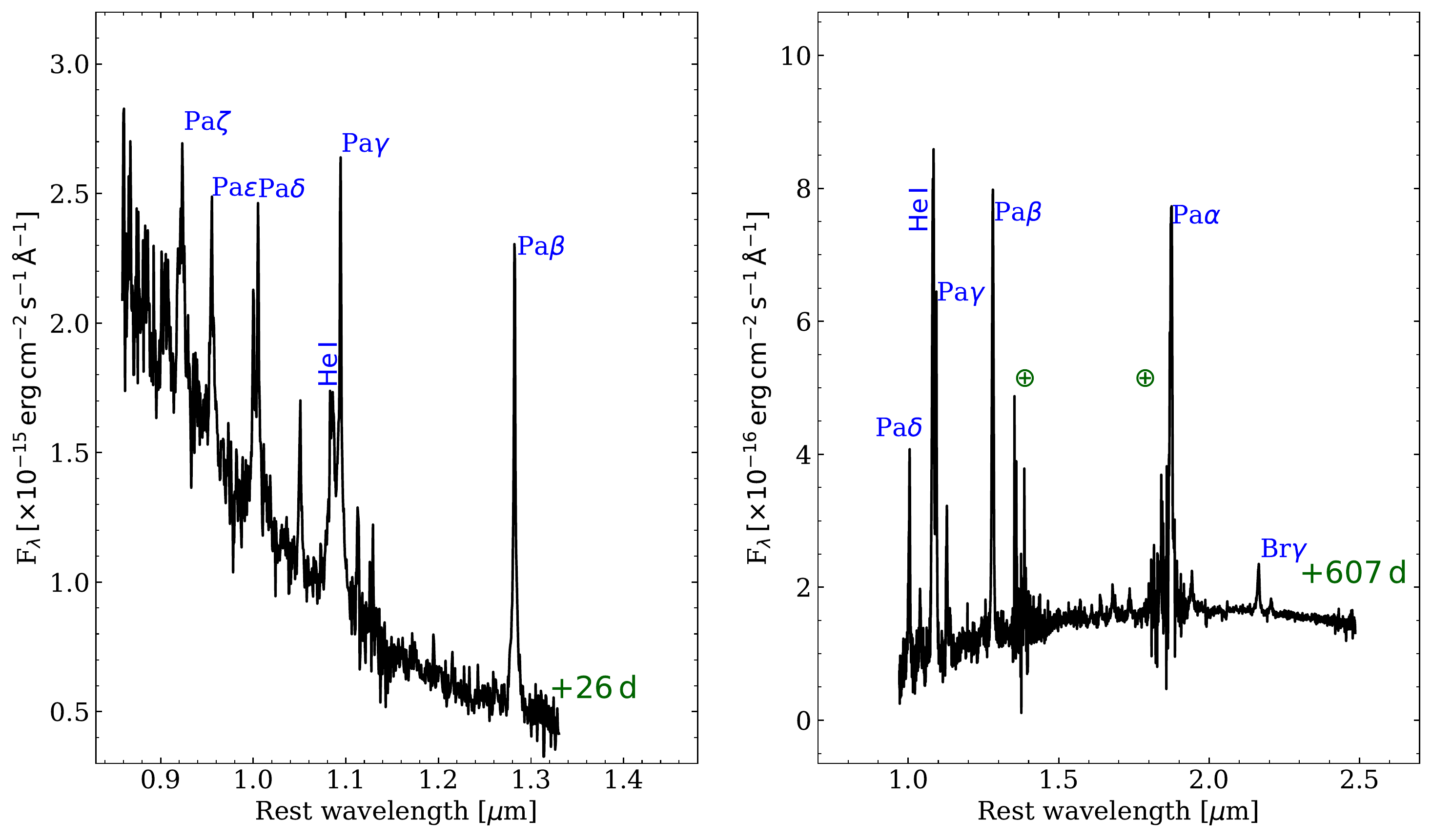}
\caption{NIR spectra of \object~at $+26\,\rm{d}$ ({\bf Left}) and $+607\,\rm{d}$ ({\bf right}). Phases refer to the estimated epoch of the explosion. \label{fig:NIRspec}}
\end{center}
\end{figure*}

Assuming that the time of the shock breakout coincides with the peak in luminosity in the diffusion dominated phase, $t_{bo}\simeq30\,\rm{d,}$ and a wind velocity $u_w\simeq100$\kms~(see Section~\ref{sec:keckspec}), Equation~\ref{eq:massloadpar} gives a mass-loss rate $\dot{M}= 0.66\,\rm{M_{\odot}\,\rm{yr^{-1}}}$ for $n=5.26$.
Equation~\ref{eq:CSMmass} then gives a total mass of $\simeq5.48(\epsilon/0.25)^{-1/3}$\msun~for the CSM swept up by the SN shock at $t=t_{break}=290\,\rm{d}$.
Here, we scaled the efficiency to $\epsilon=0.25$ \citep[][]{2014ApJ...788..154O}. 
Some support for a value of $\epsilon$ in this range comes from two-dimensional radiation hydrodynamics simulations of Type IIn SNe \citep{2016MNRAS.458.1253V}, who found an efficiency of $\sim30\%$ for the conversion of kinetic energy to radiation. 
We emphasize, however, that this value is uncertain.

Alternatively, after the photon diffusion phase, the bolometric light curve can be expressed (for $s=2$) as:
\begin{eqnarray}
L(t)=8.51\times10^{42}\times\epsilon \left(\frac{\dot{M}}{0.1\,\mathrm{M_{\odot}}\,\mathrm{yr^{-1}}}\right)\left(\frac{u_w}{100\,\mathrm{km}\,\mathrm{s^{-1}}}\right)^{-1} \nonumber \\ \times\left(\frac{V_{s,break}}{3000\,\mathrm{km}\,\mathrm{s^{-1}}}\right)^3\left(\frac{t_{break}}{320\mathrm{d}}\right)^{\alpha}\,\mathrm{erg}\,\mathrm{s^{-1}}
\label{eq:boloFransson}
\end{eqnarray}
\citep{2014ApJ...797..118F}.
Equation~\ref{eq:boloFransson} gives the bolometric luminosity at $t_{break}$ as a function of the shock velocity at this epoch, $V_{break}$ (or any other epoch).
The mass of the swept up CSM can then be obtained from the same relation used to get Equation~\ref{eq:CSMmass}:
\begin{equation}
M_{\mathrm{CSM}}=\frac{n-3}{n-2}\dot{M}\frac{V_{s}}{u_w} t_{break}.
\label{eq:massFransson}
\end{equation}

For low mass-loss rates, $V_{bo}$ can be inferred from line profiles often observed from the shocked gas of interacting transients \citep[see, e.g., the cases of SNe~1988Z, 2005ip and 2006jd;][respectively]{1993MNRAS.262..128T,2012ApJ...756..173S}.
However, \object~shows symmetric line profiles with broad wings typical of electron scattering from an outer un-shocked CSM at all epochs (see Section~\ref{sec:spectroscopy}).
Therefore, we could not estimate the velocity of the shock at the breakout directly from the spectra.
Alternatively, the shock velocity at the time of the breakout can be derived using Equation~\ref{eq:vbo} for given values of $\epsilon$, resulting in $\simeq6150(\epsilon/0.25)^{-1/3}$\kms~at $t=t_{b0}=30\,\rm{d}$, and $\simeq1440(\epsilon/0.25)^{-1/3}$\kms~at $t=t_{break}=290\,\rm{d}$.

Assuming a typical velocity $V_s\simeq3000$\kms~at an age of one year, as inferred for other Type IIn SNe such as SNe~2010jl \citep[][]{2014ApJ...797..118F} or 2013L (Taddia et al. in preparation), we can infer independent values for $\dot{M}$ and the mass of the swept up CSM, obtaining (adopting $\epsilon=0.25$) $\dot{M}=0.62\,(V_s/3000\,\mathrm{km}\,\mathrm{s^{-1}})^{-3}\,\rm{M_{\odot}}\,\rm{yr^{-1}}$, in agreement with our former estimate. 
The agreement between these two estimates gives some extra confidence in the choice of this parameter. 
The total CSM mass from Equation~\ref{eq:massFransson} is therefore $M_{\rm{CSM}}\simeq10.34(\epsilon/0.25)^{-1}\,(V_s/3000\,\mathrm{km}\,\mathrm{s^{-1}})^{-2}$\msun.
Again, we point to the considerable uncertainty in this estimate from the assumed values of $V_s$ and $\epsilon$.  
In addition, we did not take into account possible variations in the kinetic energy conversion efficiency $\epsilon$ due to a high density gradient within the shocked regions \citep[see, e.g.,][]{2019ApJ...884...87T}.

It is of some interest to compare our simplified modeling with the detailed simulations by \citet{2015MNRAS.449.4304D}, using the Eulerian radiation-hydrodynamics code {\sc heracles} \citep{2007A&A...464..429G}. 
Although these were designed to model SN~2010jl and are one-dimensional simulations, they are of interest for understanding the effects of photon diffusion on the observed light curve through a qualitatively visual comparison.

Figure~\ref{fig:DessartBolom} shows that models X ($\dot{M}=0.1\,\rm{M_{\odot}}\,\rm{yr^{-1}}$) and Xm3 ($\dot{M}=0.3\,\rm{M_{\odot}}\,\rm{yr^{-1}}$) are able to roughly reproduce the early evolution of \object, with comparable rise times, although with lower luminosities. 
Model Xm6, with a higher mass-loss rate ($0.6\,\rm{M_{\odot}}\,\rm{yr^{-1}}$), gives roughly the correct luminosity, but with a slower rise time, while Xe3 ($\dot{M}=0.1\,\rm{M_{\odot}}\,\rm{yr^{-1}}$ and total energy of $3\times10^{51}\,\rm{erg}$) seems to reproduce the rise time and the early (up to $\simeq150\,\rm{d}$ after explosion) part of the light curve after maximum well, although with a much higher peak luminosity. 
However, we note that the early light curve is uncertain due to the large bolometric correction from the assumed spectral shape in the UV.  
All these models correspond to a 9.8\msun~inner ejecta breaking through a dense CSM (with a total mass of 2.89\msun) with a wind density profile, $u_w=100$\kms, a constant temperature of $2000\,\rm{K,}$ and total kinetic energy of $10^{51}\,\rm{erg}$.  
While a more detailed modeling tuned to the observations of \object~would be of obvious interest, a rough comparison between radiative transfer models and our estimates provides a comparable mass loss, although with a higher total CSM mass.

To account for these high values, we have to assume that a large fraction of the surrounding CSM was expelled by the progenitor star through repeated massive eruptive events like those occasionally experienced by LBV stars \citep[see the case of the Great Eruption of $\eta$ Car, e.g.,][]{1999Natur.402..502M,2003AJ....125.1458S}.
On the other hand, the mass-loss rate and CSM mass inferred through Equations~\ref{eq:massloadpar} and \ref{eq:CSMmass} are strongly dependent on the assumed value of $t_{bo}$ and $s$.
\citet{2014ApJ...788..154O} showed that for $s<2,$ the shock is expected to break through near the edge of the CSM, and the model would not give a light curve with a power law decay lasting long enough to reproduce the light curve of SN~2010jl.
We note, however, that \object~shows a somewhat different behavior at $t\le t_{break}$, with a significantly shorter first decline than that observed in SN~2010jl (see Figure~\ref{fig:bololog}).

\subsection{Spectroscopy} \label{sec:spectroscopy}
\begin{figure*}
\begin{center}
\includegraphics[width=0.45\linewidth]{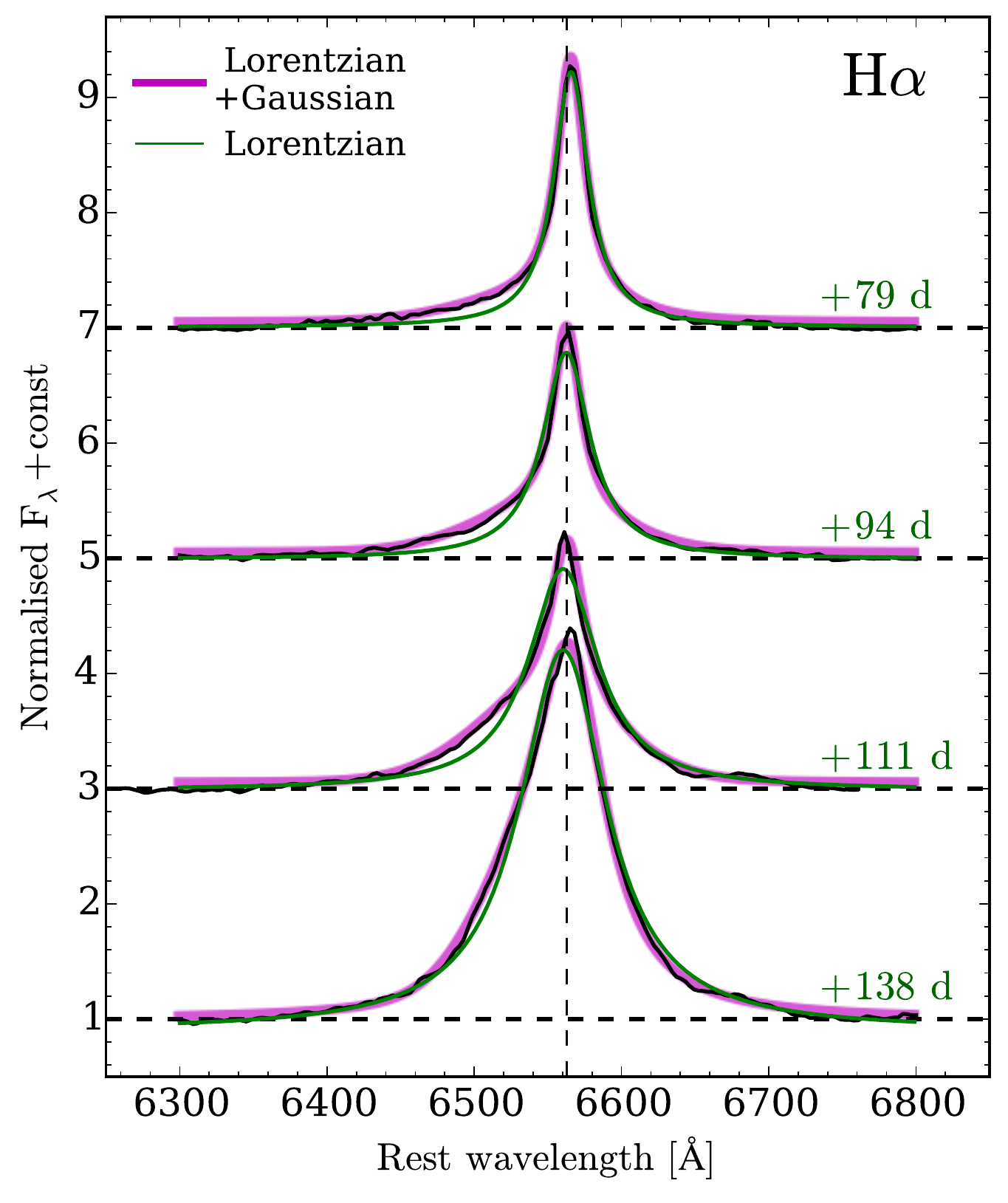}
\includegraphics[width=0.47\linewidth]{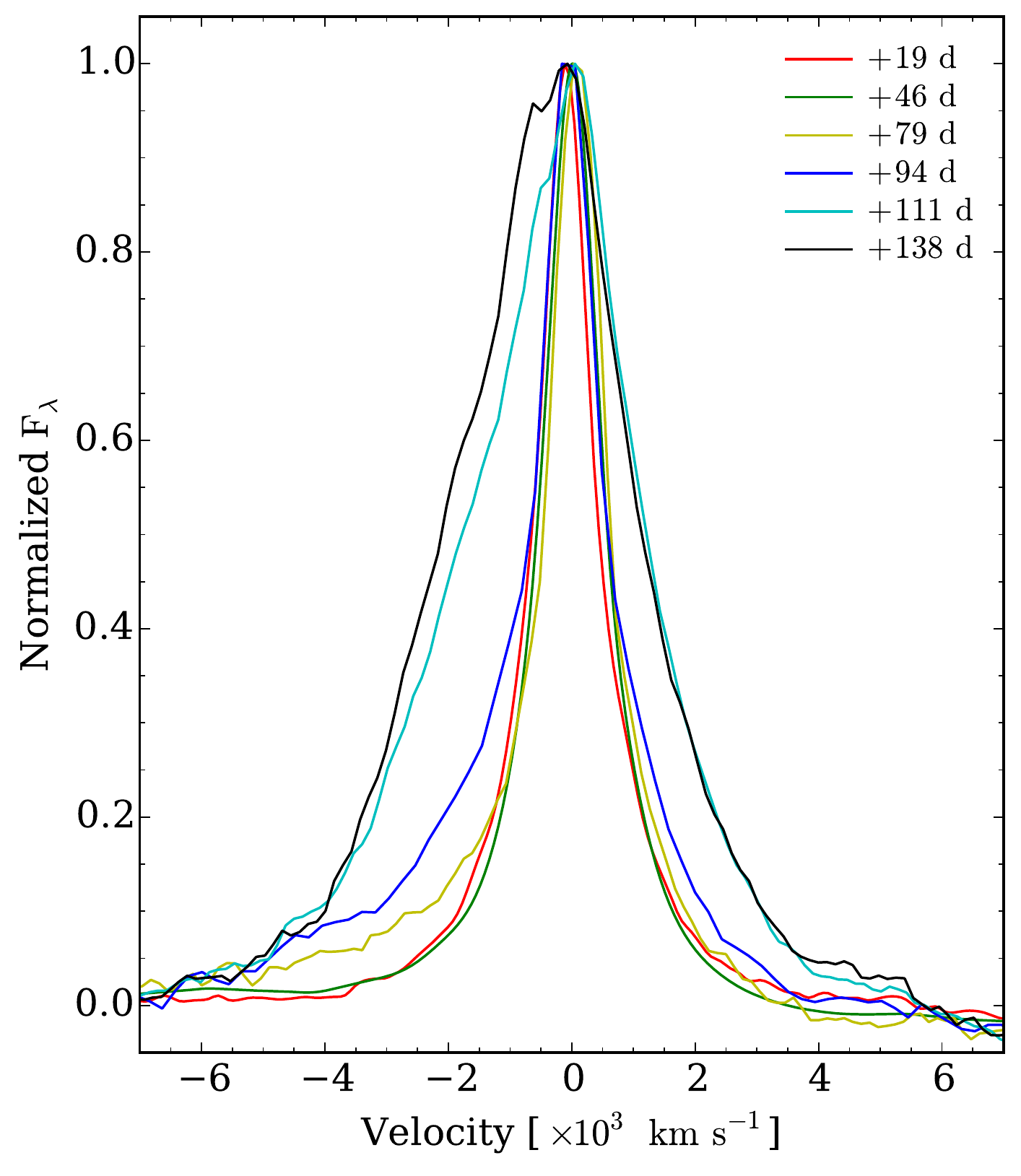}
\caption{{\bf Left:} Fit to the \ha~line profiles on the $+79\,\rm{d}$, $+94\,\rm{d}$, $+111\,\rm{d}$ and $+138\,\rm{d}$ spectra using a combination of Lorentzian and Gaussian functions. {\bf Right:} Evolution of the continuum-subtracted \ha~profile over a selection of phases. The line profiles were normalized to their peak values in order to highlight their evolution. Phases refer to the estimated epoch of the explosion.\label{fig:Hasymmetry}}
\end{center}
\end{figure*}

The optical spectral evolution of \object~is shown in Figure~\ref{fig:Optspec}, which includes a selection of spectra with the highest signal--to--noise ratios (SNRs) and resolutions.
The entire set of spectroscopic data is available, along with the photometric tables, through the Weizmann Interactive Supernova data REPository \citep[WISEREP\footnote{\url{https://wiserep.weizmann.ac.il/}};][]{2012PASP..124..668Y}.
Our 2 NIR spectra are shown in Figure~\ref{fig:NIRspec}, while a complete log of the spectroscopic observations is reported in Table~\ref{table:speclog} and shown in Figure~\ref{fig:allSpectra} in Section~\ref{sec:spectra}.

At early phases, the spectra show classical features of Type IIn SNe, such as a blue continuum with prominent narrow \ion{H}{I} (\ha~to \heps) and \ion{He}{I} ($\lambda5875$ and $\lambda7065$) lines in emission.
The strong \ion{Na}{I\,D} doublet ($\lambda\lambda5890,5896$) is clearly visible until $+193\,\rm{d}$, suggesting a highly extinguished local environment (see Section~\ref{sec:host}).
At later phases, the spectral continuum fades and the \ion{He}{I} $\lambda5875$ line dominates the spectral flux at these wavelengths.
From the positions of the minima of the \ion{Na}{I\,D} features observed in the DEIMOS spectrum we inferred a heliocentric recessional velocity of $\simeq2000$\kms~that we adopt to set the observed spectra at rest wavelengths. 

Early spectra ($t\simeq+8\,\rm{d}$) also show narrow circumstellar \ion{[N}{II]} $\lambda5755$, broad \ion{O}{I} $\lambda8446$ and narrow NIR \ion{H}{I} features (Pa~8 to Pa~12), while \ion{Fe}{II} lines (multiplets 42, 48 and 49) are visible, although faint, already at $+23\,\rm{d}$, becoming more evident at later phases ($t\geq+110\,\rm{d}$), where they start to contribute significantly to the shape of the blue pseudo-continuum (at $\lambda\lesssim5800$\ang).
At $+19\,\rm{d}$ we also identify \ion{Mg}{II} $\lambda\lambda7877,\,7896$, $\lambda\lambda8214,\,8235,$ and $\lambda\lambda9218,\,9244$.
The NIR \ion{Ca}{II} triplet starts to dominate the red part of the optical spectra from $t\simeq+138\,\rm{d}$, becoming most prominent at $\simeq505\,\rm{d}$, when it starts to fade with respect to the spectral continuum.

The NIR \ion{H}{I} lines become progressively stronger up to $+590\,\rm{d}$ whereafter they slowly decrease in strength.
At $+35\,\rm{d,}$ the NIR spectrum is dominated by narrow \ion{H}{I} ($\rm{Pa}\beta$ to $\rm{Pa}\zeta$) and \ion{He}{I} ($\lambda10830$) lines.

Fitting a BB to the spectral continuum, we inferred the temperature evolution of the pseudo photosphere, slowly declining from $\simeq13430\pm355\,\rm{K}$ to $\simeq7180\pm510\,\rm{K}$ during the first $\simeq240\,\rm{d}$ after explosion, in agreement with the one inferred from the SEDs (see Section~\ref{sec:NIRexcess}).
From $t\simeq193\,\rm{d}$, the spectra show a blue excess at wavelengths shorter than $\simeq5500$\ang, becoming more evident at later epochs, observed throughout the remaining $\sim1300\,\rm{d}$ of the spectroscopic monitoring.
The source of the excess is likely not thermal, since spectra at these wavelengths ($\lambda<5500-5800$\ang) are significantly affected by strong \ion{Fe}{II} blends.
On the other hand, late ($t\gtrsim100\,\rm{d}$) blue excesses are common in interacting transients (see the cases of SNe~2006jc; \citealt{2007ApJ...657L.105F,2007Natur.447..829P} and 2005ip; \citealt{2009ApJ...695.1334S}) and are likely due to the contribution of fluorescence from a number of blended Fe lines or to a "revitalized" late--time ejecta--CSM interaction with a low energy conversion efficiency \citep[see, e.g.,][]{2009ApJ...695.1334S}.
\begin{figure}
\begin{center}
\includegraphics[width=0.98\linewidth]{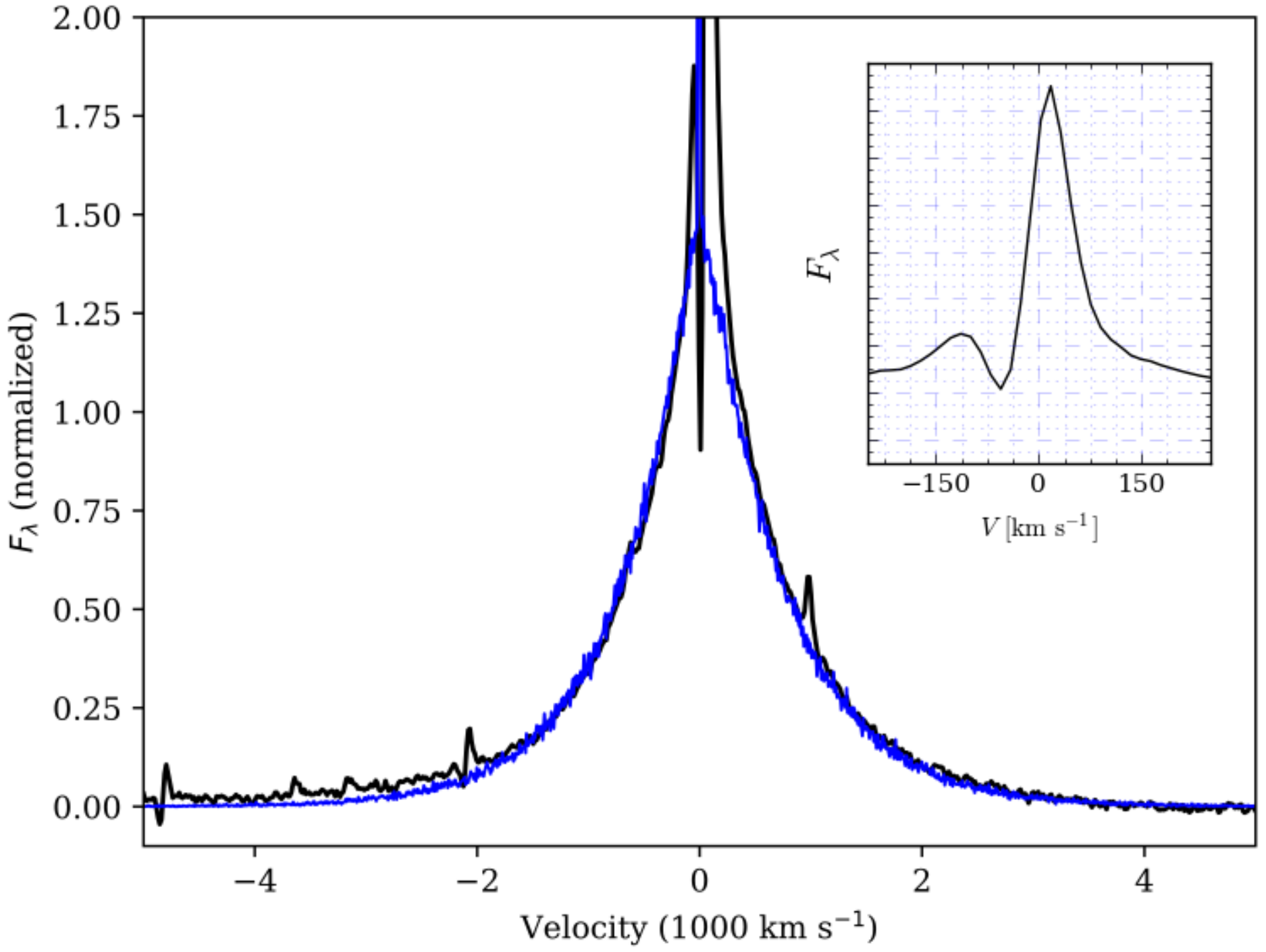}
\caption{The \ha~line profile observed in the $+46\,\rm{d}$ Keck DEIMOS spectrum of \object~along with a fit to the line profile from electron scattering. The line profile below $\pm200$\kms, in the model from un-scattered photons, is affected by the resonance line scattering in the \ha~transition and was not modeled. \label{fig:keckHa}}
\end{center}
\end{figure}

\subsubsection{The DEIMOS spectrum} \label{sec:keckspec}
A moderate-resolution spectrum ($R\simeq3600$ from the \ion{[O}{I]} sky lines) was obtained on 2015 February 23.66~UT ($\rm{JD}=2457077.16$, corresponding to $t=+46\,\rm{d}$; Program ID U063D, PI Filippenko), with the DEep Imaging Multi--Object Spectrograph \citep[DEIMOS;][]{2003SPIE.4841.1657F} mounted at the $8\,\rm{m}$ Keck II telescope at Mauna Kea.
These data are available in the public section of the Keck Observatory Archive (KOA\footnote{\url{https://www2.keck.hawaii.edu/koa/public/koa.php}}).

\begin{table}
\centering
\caption[]{Main parameters of the BB fit to the observed SED evolution of \object.}
\smallskip
\label{table:BalmerLines}
\scriptsize
\begin{tabular}{cccccc}
\hline\hline
\noalign{\smallskip}
JD & Phase & Luminosity (\ha/\hb) & EW (\ha/\hb) & FWHM(\ha/\hb) \\ 
   & (d)   & \ergs                & (\AA)        & (\kms)        \\
\noalign{\smallskip}
\hline
2457039.37 &   $+8$  & 5.78/3.62  & 370/110  & 1425/1970           \\
2457041.37 &  $+10$  & 6.45/3.43  & 385/80   & 1630/2400           \\
2457042.38 &  $+11$  & 6.11/3.43  & 320/85   & 1575/2025           \\
2457049.99 &  $+19$  & 7.55/4.15  & 240/85   & 1350/1525           \\
2457050.37 &  $+19$  & 6.55/4.06  & 240/85   & 1410/2100           \\
2457053.67 &  $+23$  & 7.00/3.93  & 275/75   & 1335/2060           \\
2457057.34 &  $+26$  & 6.60/2.53  & 230/65   & 1415/1960           \\
2457063.55 &  $+33$  & 5.60/3.27  & 175/55   & 1275/2370           \\
2457067.35 &  $+36$  & 5.45/2.94  & 165/50   & 1440/2235           \\
2457077.16 &  $+46$  & 3.95/1.27  & 105/20   & 540/1060\textdagger \\
2457093.52 &  $+63$  & 3.05/1.32  & 90/20    & 1135/1930           \\
2457109.59 &  $+79$  & 3.80/0.92  & 90/15    & 1450/1980           \\
2457119.83 &  $+89$  & 3.15/1.20  & 65/20    & 1590/2295           \\
2457124.48 &  $+94$  & 3.80/0.94  & 85/15    & 2110/2230           \\
2457141.62 & $+111$  & 5.20/1.58  & 1450/25  & 3120/3045           \\
2457168.64 & $+138$  & 8.75/2.89  & 265/50   & 3770/3230           \\
2457185.64 & $+155$  & 10.85/3.15 & 370/65   & 3710/3165           \\
2457224.39 & $+193$  & 11.05/3.00 & 505/80   & 3480/3020           \\
2457224.43 & $+193$  & 12.80/2.40 & 655/85   & 3820/2530           \\
2457241.32 & $+210$  & 13.75/3.10 & 715/105  & 3700/2890           \\
2457259.39 & $+228$  & 13.15/3.10 & 825/80   & 3560/3010           \\
2457273.30 & $+242$  & 13.85/2.95 & 880/105  & 3700/3030           \\
2457359.68 & $+329$  & 12.00/2.10 & 1250/115 & 3470/2560           \\
2457406.02 & $+375$  & 12.35/2.00 & 1040/110 & 3200/2230           \\
2457465.60 & $+435$  & 3.60/0.50  & 1940/145 & 2990/2350           \\
2457535.90 & $+505$  & 6.60/1.20  & 1190/85  & 2600/2300           \\
2457562.44 & $+531$  & 1.35/0.30  & 1095/160 & 2430/3060           \\
2457603.68 & $+573$  & 6.10/0.90  & 1550/100 & 2370/2130           \\
2457621.38 & $+590$  & 6.00/-\textdagger     & 1735/-\textdagger   & 2335/-\textdagger \\
2457626.64 & $+596$  & 4.25/0.70  & 1235/80  & 2285/2180           \\
2457636.66 & $+606$  & 7.70/0.45  & 2265/110 & 2295/2430           \\
2457644.36 & $+613$  & 3.20/0.35  & 1150/65  & 2205/2780           \\
2457728.75 & $+698$  & 2.25/0.40  & 945/75   & 2080/2625           \\
2457736.74 & $+706$  & 1.95/0.30  & 1190/60  & 2130/2940           \\
2457780.98 & $+750$  & 2.70/0.35  & 1135/80  & 2110/2150           \\
2457806.73 & $+776$  & 1.85/0.40  & 1145/75  & 2030/2715           \\
2457827.57 & $+797$  & 1.90/0.35  & 1160/65  & 1995/2950           \\
2457841.40 & $+810$  & 1.80/0.35  & 1160/75  & 2010/2365           \\
2457863.55 & $+833$  & 2.10/0.40  & 1490/110 & 2040/2315           \\
2457867.59 & $+837$  & 1.60/0.30  & 1070/80  & 2000/2445           \\
2457875.54 & $+845$  & 1.65/0.30  & 945/70   & 2010/2940           \\
2457899.58 & $+869$  & 1.90/0.30  & 1125/85  & 1965/2765           \\
2457914.52 & $+884$  & 1.30/0.25  & 940/50   & 2030/2620           \\
2457929.81 & $+899$  & 2.65/0.25  & 1360/60  & 2070/2250           \\
2457951.44 & $+920$  & 2.00/0.35  & 1155/60  & 2030/2235           \\
2457961.69 & $+931$  & 2.10/0.40  & 1285/50  & 2000/2605           \\
2457968.40 & $+937$  & 1.50/0.25  & 1300/65  & 2085/3135           \\
2457977.41 & $+946$  & 1.35/0.30  & 1030/50  & 2020/2330           \\
2458121.74 & $+1091$ & 1.75/-\textdagger     & 1025/-\textdagger   & 2140/-\textdagger \\
2458489.12 & $+1458$ & 1.45/0.50  & 1755/95  & 2340/2255           \\
\hline
\noalign{\smallskip}
\end{tabular}
\tablefoot{\textdagger Uncertain or missing values due to the limited spectral coverage of the spectrum.}
\end{table}

The spectrum shows many marginally resolved \ion{Fe}{II} lines with P-Cygni absorption features (multiplets 40, 42, 46, 48, 49, 74) or purely in emission (200, 40 apart from $\lambda6516$ and 49, with the possible exception of $\lambda6113$).
We also identified a number of narrow lines purely in emission corresponding to other transitions, namely \ion{[N}{II]} $\lambda5755$, \ion{He}{I} $\lambda5875$ and \ion{Si}{II} $\lambda\lambda6347$, 6371 as well as a few other unidentified lines at 5568, 5587, 6318, 6332, and 6384/6385\ang.
The \ion{Fe}{II} $\lambda5169$ line, typically used to infer the photospheric expansion velocity in SNe, shows a faint narrow emission with a structured absorption component, a minimum of $\simeq10$\kms, and a blue wing, possibly contaminated by a second component, extending up to $\simeq110$\kms.
\ion{Fe}{II} lines are blueshifted\footnote{With the exception of multiplet 74, which shows peaks almost at rest wavelengths.} by $\simeq40-50$\kms, from which we infer minima of $\simeq50$\kms~with a terminal velocity of $100-110$\kms.

In Figure~\ref{fig:keckHa}, we zoom in on the DEIMOS spectrum in the region of \ha~showing a narrow P-Cygni feature with a minimum blueshifted by $\simeq55$\kms, and a terminal velocity of $\simeq110$\kms~on top of a broader profile, typical of electron scattering, with wings showing a full-width-at-zero-intensity (FWZI) of $\simeq3\times10^3$\kms. 

To illustrate the dominance of electron scattering in the formation of the line profiles, we used the same Monte Carlo code as in \citet{2014ApJ...797..118F}. 
The input photons from recombination and collisions are emitted from the ionized region of the slowly moving "precursor shock" \citep{2017ApJS..229...34S} and then undergo electron scattering in the same region, although most of them are emitted close to the shock. 
The main parameters of the fit are the optical depth to electron scattering, $\tau_{\rm{e}}$ and the electron temperature, $T_{\rm e}$. 
Since the change in frequency in each scattering is $\propto T_{\rm{e}}^{1/2}$, to obtain a given FWHM, we need $\tau_{\rm{e}}\propto T_{\rm{e}}^{-1/2}$. 
The two parameters are therefore degenerate. 
In our calculations, we assumed $T_{\rm e}=10^4\,\rm{K,}$ and we did not attempt to model the resonance scattering by the \ha~line giving rise to the narrow P-Cygni profile. 

In Figure~\ref{fig:keckHa}, we show the resulting fit, where the broad wings are well-reproduced by an exponential profile typical of electron scattering \citep{2018MNRAS.475.1261H}.
To obtain the observed FWHM at this epoch, we need $\tau_{\rm{e}}\approx6.0(T_{\rm{e}}/10^4\,\rm{K})^{-1/2}$. 
This value is similar to the one obtained for SN~2010jl \citep{2014ApJ...797..118F}, and it shows that the gas is optically thick to electron scattering. 
The narrow line in the model is due to un-scattered photons at zero velocity. 
These photons are scattered by the \ha~line itself and form part of the P-Cygni profile below $\sim200$\kms. 
The fact that this results in a P-Cygni profile means that the emission must come from a more extended region, producing both the absorption and emission component.

An immediate conclusion is that the emission from the inner parts of the ejecta (with respect to the forward shock), have an even higher optical depth. 
Emission lines from this region would therefore be washed out into a continuum, explaining why we do not observe broad lines from the expanding ejecta or post-shock gas. 

\subsubsection{Evolution of the H lines} \label{sec:HEv}
Physical quantities inferred from the main Balmer lines (i.e., \ha~and \hb) and described in this Section are reported in Table~\ref{table:BalmerLines}, as obtained through the {\sc iraf} task {\sc splot}, including FWHM velocities and equivalent widths (EWs).
EWs for \ha~and \hb~show average values of $\sim880$\AA~and $\sim75$\AA, respectively, in agreement with the distribution of values inferred for the sample of SNe IIn presented in \citet{2013ApJS..207....3S}, who also proposed weaker \hb~lines (with $EW\simeq6$\AA, as well as the absence of \ion{He}{I} $\lambda5876$ features) as a hallmark feature of Ia-CSM SNe. 
The values inferred for SN 2015da argue against a Type Ia-CSM interpretation for this object.

At $t<+63\,\rm{d}$, Balmer lines show narrow profiles purely in emission, with roughly constant FWHM velocities of $\simeq10^3$\kms~and $\simeq1.7\times10^3$\kms~for \ha~and \hb, respectively.
All Balmer lines are well reproduced using a Lorentzian profile, and we do not see any trace of the narrow component observed in the higher resolution DEIMOS spectrum obtained at $+46\,\rm{d}$ (see Section~\ref{sec:keckspec}).
This is most likely due to an effect of resolution, as suggested by the appearance of the narrow component in the moderate-resolution spectra obtained at $+505\le t\le+606\,\rm{d}$, all having resolutions $\lesssim10$\ang~in the $6300-6800$\ang~region.

From $+79\,\rm{d}$, we note small deviations from a single Lorentzian profile in the blue wing of \ha~(see Figure~\ref{fig:Hasymmetry}) and a second Gaussian component is required to fit the entire profile.
The \hb~line profile, on the other hand, is well-reproduced by a single Lorentzian component at all phases.
A second component is required also at later phases, although at $+138\,\rm{d}$ \ha~does not show significant asymmetries, and a single Lorentzian component is again sufficient to fit the entire profile (see Figure~\ref{fig:Hasymmetry}).
In the early NIR spectrum ($+26\,\rm{d}$), Paschen lines are marginally resolved ($\rm{FWHM}\simeq600$\kms) and show symmetric profiles centered at the corresponding rest wavelengths.
At later phases ($+607\,\rm{d}$), we notice a broadening ($\rm{FWHM}\simeq1100-1300$\kms) in the $\rm{Pa}\gamma$ and $\rm{Pa}\beta$ lines, with slightly blueshifted peaks of $\simeq100$\kms~($\simeq40$\kms~for $\rm{Pa}\gamma$, which is strongly contaminated by the prominent \ion{He}{I} $\lambda10830$ line), although the overall profiles are still well-reproduced by single symmetric Lorentzian (or Gaussian, for marginally resolved lines) profiles.

We note a similar evolution in \ha, with the centroid of the line progressively shifting toward bluer wavelengths with time, up to $300-500$\kms~at $t\gtrsim833\,\rm{d}$, where the uncertainty is due to the different SNR and resolutions of the spectra.
This is highlighted in Figure~\ref{fig:Hasymmetry} (right panel), showing the evolution of \ha~over selected phases.
The \ha~profile shows a broadening at $t\gtrsim100\,\rm{d}$, which might be due to a gradual emergence of the emission from the shock, while the overall profile remains symmetric (see below).
\begin{figure}
\begin{center}
\includegraphics[width=0.95\linewidth]{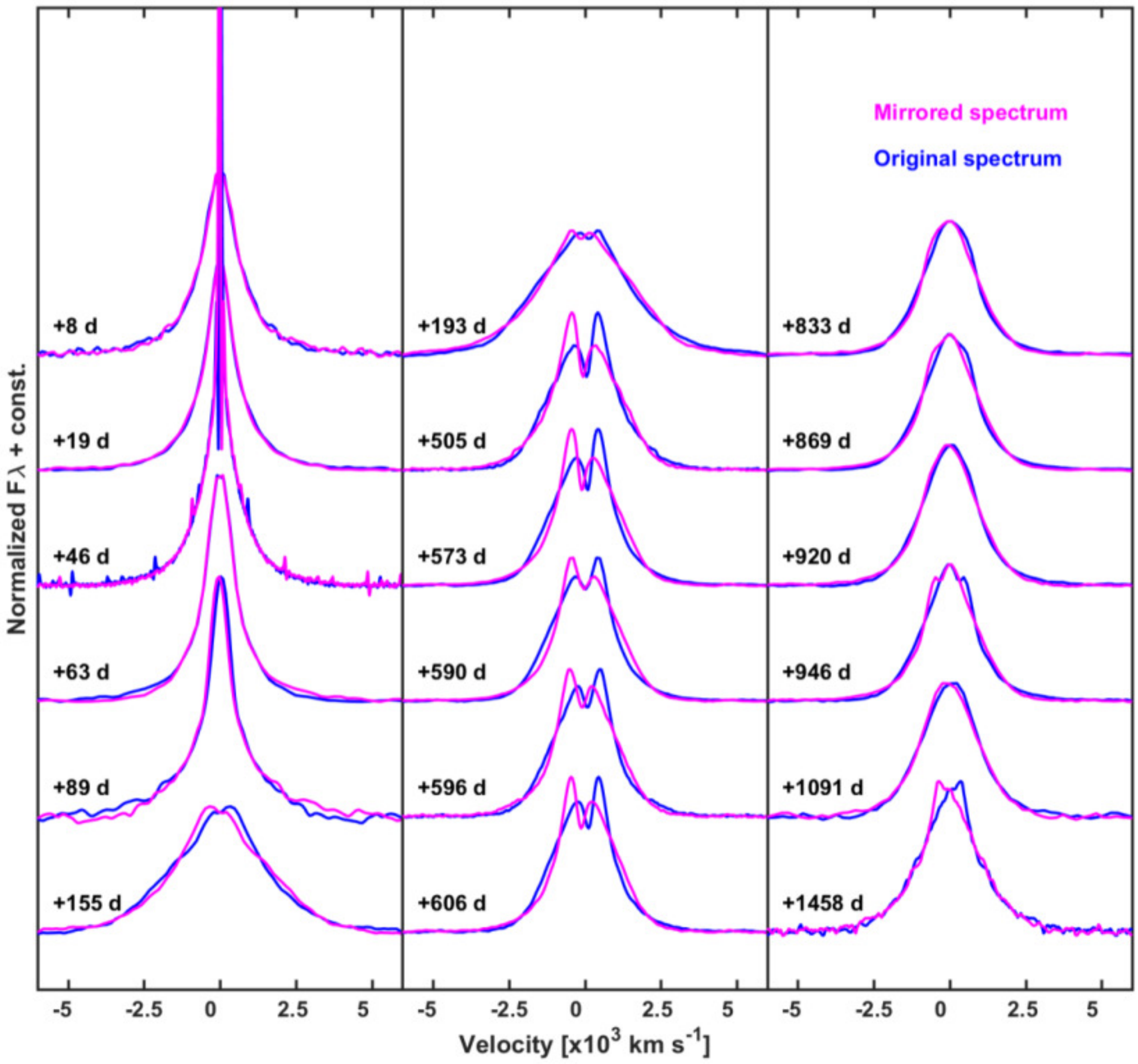}
\caption{\ha~profiles at selected epochs, redshifted to the line rest wavelength (blue) and mirrored with respect to the computed centroids (magenta). 
Lines show highly symmetric profiles, with blue and red wings overlapping almost perfectly at all phases. Small asymmetries in the top part of the lines are due to the presence of the narrow P--Cygni features, more evident in the higher resolution spectra. Phases refer to the estimated epoch of the explosion.
\label{fig:Hamirrored}}
\end{center}
\end{figure}

In SN~2010jl, wavelength-dependent asymmetries and the apparent dimming of the red wings of emission components was used by \citet{2012AJ....143...17S} and \citet{2014Natur.511..326G} to suggest rapid dust formation in the SN ejecta as the possible cause of the IR excess.
However, \citet{2014ApJ...797..118F} showed that the profile of \ha~remains symmetric with respect to a shifted centroid and attributed this shift to a bulk velocity of the emitting shell or to acceleration of the un-shocked CSM by the radiation field generated in the inner shocked regions.
Following \citet{2014ApJ...797..118F}, we therefore mirrored the red wing of \ha~with respect to the computed centroid of the line profile at each epoch.
The resulting profiles are shown in Figure~\ref{fig:Hamirrored} for a selection of spectra with high SNR and good resolution.
As in SN~2010jl, no sign of asymmetries is seen at any epoch, suggesting that a macroscopic velocity is the most likely reason for the blueshift of the \ha~profile of \object.

\begin{figure}
\begin{center}
\includegraphics[width=\linewidth]{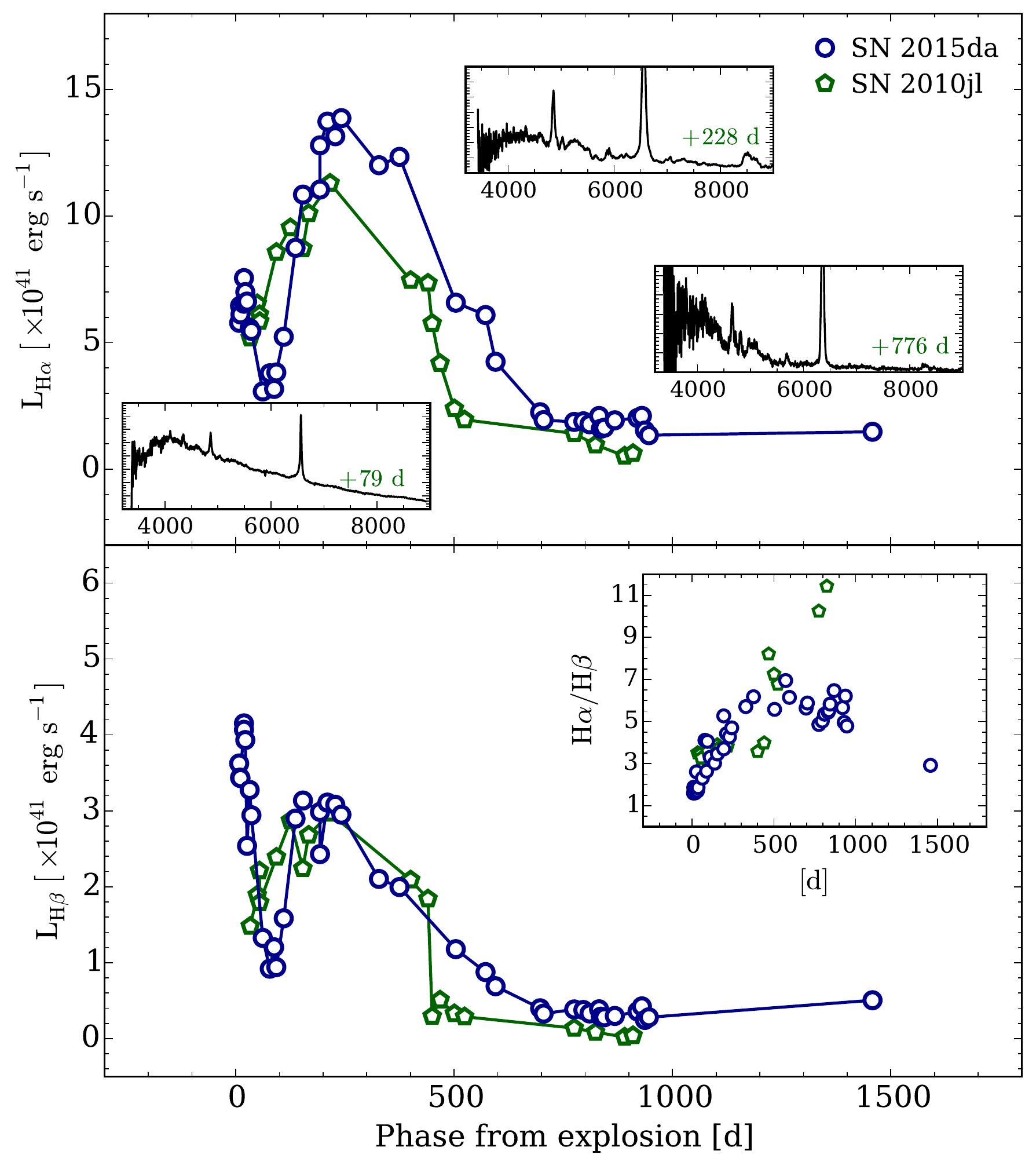}
\caption{Evolution of the integrated luminosity of the most prominent Balmer lines (\ha~and \hb) compared to those inferred for SN~2010jl. Insets in the upper panel show representative spectra during the different phases of the evolution of \object. The inset in the bottom panel shows the evolution of the \ha/\hb~ratio compared to that computed for SN~2010jl. \label{fig:BalmerLum}}
\end{center}
\end{figure}

The evolution of the \ha~and \hb~integrated luminosity is shown in Figure~\ref{fig:BalmerLum}.
We note a rapid decline for both lines during the first $63\,\rm{d}$, with the luminosity evolving from $\simeq6.58/3.60\times10^{41}$\ergs~to $5.46/2.94\times10^{41}$\ergs~for \ha/\hb, respectively. 
The luminosity shows a subsequent re-brightening up to $1.39/0.29\times10^{42}$\ergs~during the following $\simeq180\,\rm{d}$, with an offset of $\simeq+30\,\rm{d}$ with respect to the onset of the re-brightening observed in SN~2010jl \citep[assuming $\rm{JD}=2455479$ as the explosion epoch for SN~2010jl;][]{2014ApJ...797..118F}.
The integrated luminosities show a further decline at later phases, until $+706\,\rm{d}$, when it sets at $1.94/0.33\times10^{41}$, staying roughly constant for the remaining $\simeq750\,\rm{d}$.
In \object, the \ha/\hb~ratio increases monotonically up to $\simeq+560\,\rm{d}$, when it settles at a roughly constant value of $\simeq5.7$, while at $+1458\,\rm{d,}$ it drops again to $2.92$, showing a quite different evolution with respect to that of SN~2010jl.
This different behavior might be attributed to the blue pseudo continuum contamination of the spectra at $t\gtrsim228\,\rm{d}$, which can bias the integrated luminosity inferred for \hb~(see Section~\ref{sec:spectroscopy} and Figure~\ref{fig:BalmerLum}, upper panel).

\subsection{The progenitor star} \label{sec:progenitor}
The field of \object~was monitored by the Palomar Transient Factory (PTF\footnote{\url{https://www.ptf.caltech.edu/}}), which imaged its host galaxy with a roughly constant cadence from 2009 March 17 to 2014 May 28 (see Table A.2).
Frames were recovered through the NASA/IPAC Infrared Science Archive \footnote{\url{https://irsa.ipac.caltech.edu/}}.
We could not detect traces of pre-SN variability down to $g\simeq-16.1\,\rm{mag}$ or $-15\,\rm{mag}$, assuming a distance of $53.2\,\rm{Mpc}$ and the total extinction reported in Section~\ref{sec:host}.
LBVs are among the most luminous stars known, and typically have absolute magnitudes $M_R\simeq-9\,\rm{mag}$ in quiescence.
On the other hand, they occasionally experience nonterminal major eruptive events, like the ones observed in $\eta-$Car in the 19th century \citep[e.g.,][]{2011MNRAS.415.2009S}, producing optical transients that mimic the behavior of SNe IIn \citep[hence dubbed "SN impostors";][]{2000PASP..112.1532V,2006MNRAS.369..390M}, although with fainter absolute magnitudes \citep[$M_{peak}\simeq-14\,\rm{mag}$, e.g.,][]{2015MNRAS.447..117T,2016ApJ...823L..23T}.
In addition, we cannot rule out a larger value of reddening in the environment of the progenitor star before explosion.
An SN explosion in a dusty environment is expected to produce a dust-free cavity within a radius directly dependent on the peak luminosity of the transient.
\citet{1983ApJ...274..175D} showed that an SN with $L_{peak}=10^{10}$\lsun~can produce a cavity of $5.1\times10^{17}\,\rm{cm}$, depending on the chemical composition, size of the dust gains, density and optical depth of the shell. Therefore, a larger amount of dust could survive less luminous outbursts.
It is therefore possible that the local extinction in the environment of \object~was higher during the pre-SN stages than was estimated in Section~\ref{sec:host}, possibly masking multiple LBV--like outbursts.
In this scenario, the available observations would not put sensible constraints on the pre-explosion variability of the precursor.

The prominent narrow \ion{[N}{II]} line visible in the spectra almost at all phases is indicative of a possible nitrogen enrichment of the narrow-line region of the CSM, which in turn might indicate a large enrichment of CNO processed gas.
Significant CNO enrichments have been observed in SN ejecta or CSM for a small number of CC SNe \citep[see, e.g.,][and references therein]{2005ApJ...622..991F}, as well as evolved massive stars such as LBVs.
Unfortunately, most of the diagnostic lines commonly used for CNO analyses are out of the observed spectral range \citep[e.g., \ion{C}{III}, \ion{C}{IV}, \ion{N}{III]}, \ion{N}{IV],} and \ion{O}{III]}, all at $\lambda<2000$\ang;][]{2014ApJ...797..118F}, and we therefore did not have the sufficient spectroscopic coverage or resolution to perform a detailed analysis. 

On the other hand, from the narrow \ha~P--Cygni absorption observed in the higher resolution DEIMOS spectrum at $+46\,\rm{d}$, we inferred a wind velocity of $\simeq100$\kms~(similar to that of SN~2010jl), suggesting a very massive star as a viable progenitor for \object.
This corresponds to the lower end of the range of typical velocities inferred for LBVs \citep[$10^2-10^3$; e.g.][]{2011MNRAS.415..773S}, which, on the other hand, is significantly higher than those displayed by red supergiants \citep[RSG, $10-50$\kms; e.g.][]{1990ApJS...73..769J}.
In addition, the total mass of the CSM inferred from our analysis of the pseudo-bolometric light curve ($\gtrsim5.5$\msun; see Section~\ref{sec:boloLC}) is more than a factor of five higher than that expected to be lost by a lower mass star during the RSG phase \citep[$<1$\msun~for a star with an initial mass of 16\msun;][]{2018MNRAS.475...55B}.
As a consequence, there are a number of consistent, although not compelling, indications suggesting that the progenitor of \object~was an LBV star.

\section{Summary and conclusions} \label{sec:conclusions}
In this paper, we reported the main results of the photometric and spectroscopic follow-up of the Type IIn \object, exploded in the relatively nearby ($D\simeq53.2\,\rm{Mpc}$) host galaxy NGC~5337.

\object~is a long-lasting Type IIn SN discovered soon after explosion and with well-sampled photometric and spectroscopic follow-up from the optical to MIR wavelengths.
This makes it one of the best followed SN IIn, with, to date, more than four years of collected data.
Follow-up campaigns of this nearby optical transient are still ongoing both in the optical and NIR and MIR domains, and additional data could help us to better constrain the physical properties of the explosion, its dusty environment, and ultimately the nature of its progenitor star.

The transient exploded in a highly obscured environment, contributing a reddening of $E(B-V)=0.97\,\rm{mag}$ ($A_V\simeq3\,\rm{mag}$ assuming a canonical extinction law with $R_V=3.1$) in the direction of \object.
The IR excess observed from $+433\,\rm{d}$ suggests that the IR luminosity is most likely produced by radiatively heated dust.
This conclusion is supported by the shape of the \ha~line profiles, which show symmetric wings with respect to their centroids at all phases, while alternative explanations, like rapid formation of large dust grains at the interface between the forward and reverse shock, would result in strong asymmetries in the line shape.
On the other hand, given the simplicity of the models adopted here and the assumptions made, we cannot definitely rule out any of the possible emission mechanisms.

The analysis of the pseudo-bolometric light curve of SN 2015da revealed an extended CSM with a total mass $5.5-10.3$\msun~(assuming an energy conversion efficiency $\epsilon=0.25$), while wind velocities suggest a very massive precursor, possibly an LBV, as a viable progenitor for \object.
This conclusion is supported by the very high mass-loss rate inferred from the evolution of the pseudo-bolometric luminosity ($\dot{M}\simeq0.6-0.7\,\rm{M_{\odot}}\,\rm{yr^{-1}}$), which indicates that multiple outbursts during a long-lasting eruptive phase, similar to those observed in other SN impostors such as SN 2000ch \citep{2004PASP..116..326W,2010MNRAS.408..181P} are responsible for the dense CSM surrounding the progenitor star at the time of its explosion.
We note, however, that the upper CSM mass limit would require a progenitor that retained a massive envelope until the very last phases of its evolution. 
According to the Geneva stellar evolutionary models \citep{2004A&A...425..649H}, this is problematic at solar metallicity. 
For example, a nonrotating 25\msun~star has an envelope mass of $\simeq10$\msun~at the point of explosion, while a more massive (or rotating) star would lose even more mass to then explode as a stripped-envelope SN. 
While this envelope mass is larger than the ejected CSM, it would require that the "entire" envelope was ejected just before explosion. 
If we assume that only some fraction of the envelope can be ejected immediately prior to CC, we must turn to lower metallicity models, which, on the other hand, have commensurately smaller $\dot{M}$. 
On the other hand, even at SMC metallicity, it is hard to form a H-rich star that has an envelope mass exceeding 10\msun~at the time of CC.
At even lower metallicity, for example, $Z=10^{-5}$, wind-driven mass loss is much smaller, and it is possible to form a progenitor with a ZAMS mass of 60\msun~that will explode as a Type II SN \citep{2002A&A...390..561M}. 
However, while this was plausible for SN~2010jl that exploded in a faint irregular galaxy, NGC~5337 is a normal spiral, with a roughly solar metallicity (see Section~\ref{sec:host}). 
One possibility is that \object~is hosted in a faint dwarf satellite of NGC~5337, which is by chance projected on the sky towards towards the main galaxy. 
Future spectroscopy at the site of \object~after the SN has faded may allow us to measure the local metallicity, or, alternatively, to identify a kinematically distinct dwarf host. 
Without this information, the contradiction between the nearly solar metallicity host and the requirement for a low metallicity progenitor remains a puzzle.

\begin{acknowledgements}
We thank Marco Berton, Sina Chen, Fabio Briganti, Fabio Martinelli, Emmanuel Conseil, Stan Howerton, Masanori Mizutani and Kunihiro Shima for their help with the observations of \object. We are also grateful to our late friend Alex Dimai, whose observations have been used in this study.

We gratefully acknowledge support from the Knut and Alice Wallenberg Foundation. \\
The Oskar Klein Centre is funded by the Swedish Research Council. \\
We acknowledge the support of the staff of the Xinglong 2.16m telescope. \\ 
This work was partially supported by the Open Project Program of the Key Laboratory of Optical Astronomy, National Astronomical Observatories, Chinese Academy of Sciences. \\

M.~F. is supported by a Royal Society - Science Foundation Ireland University Research Fellowship. \\
J.~H. acknowledges financial support from the Finnish Cultural Foundation and the Vilho, Yrj\"o and Kalle V\"ais\"al\"a Foundation of the Finnish Academy of Science and Letters. \\
Research by D. J. S. is supported by NSF grants AST-1821987, AST-1821967, AST-1813708, AST-1813466 and AST-1908972. \\
S.~B., L.~Tomasella and M.~T. are partially supported by the PRIN-INAF 2016 with the project "Towards the SKA and CTA era: discovery, localisation, and physics of transient sources" (P.I. M.~Giroletti). \\
N.~E.-R. acknowledges support from the Spanish MICINN grant ESP2017--82674--R and FEDER funds.
D.~A.~H., C.~M., and G.~H. were supported by NSF AST-1313484 \\
The work of X.~W. is supported by the National Natural Science Foundation of China (NSFC grants 11325313, 11633002, and 11761141001), and the National Program on Key Research and Development Project (grant no. 2016YFA0400803). \\
Research by S.~V. is supported by NSF grant AST-1813176. \\
 J.~Zhang is supported by the National Natural Science Foundation of China (NSFC, grants 11773067, 11403096), the Youth Innovation Promotion Association of the CAS (grants 2018081), and the Western Light Youth Project. \\

Based on observations collected at: \\
ESO La Silla Observatory under program "Optical \& NIR monitoring of bright supernovae with REM" during AOT30. \\ 
The Gemini Observatory, under program GN--2016B-Q-57, which is operated by the Association of Universities for Research in Astronomy, Inc., under a cooperative agreement with the NSF on behalf of the Gemini partnership: the National Science Foundation (United States), the National Research Council (Canada), CONICYT (Chile), Ministerio de Ciencia, Tecnolog\'{i}a e Innovaci\'{o}n Productiva (Argentina), and Minist\'{e}rio da Ci\^{e}ncia, Tecnologia e Inova\c{c}\~{a}o (Brazil). \\
Tthe Nordic Optical Telescope, operated by the Nordic Optical Telescope Scientific Association and the Gran Telescopio Canarias (GTC), both installed at the Spanish Observatorio del Roque de los Muchachos of the Instituto de Astrof\`{i}sica de Canarias, on the island of La Palma (Spain). \\
The Copernico telescope (Asiago, Italy) operated by INAF -- Osservatorio Astronomico di Padova. \\
The $3\,\rm{m}$ Shane Reflector, located at the Lick Observatory (7281 Mt Hamilton Rd, Mt Hamilton, CA 95140, U.S.A.) owned and operated by the University of California. \\
This work makes use of observations from the Las Cumbres Observatory network of telescopes. \\
We acknowledge the support of the staff of the Li--Jiang $2.4\,\rm{m}$ telescope (LJT). \\ Funding for the LJT. has been provided by the Chinese Academy of Sciences (CAS) and the People's Government of Yunnan Province. \\
The LJT is jointly operated and administrated by Yunnan Observatories and Center for Astronomical Mega--Science, CAS. \\

This research has made use of the Keck Observatory Archive (KOA), which is operated by the W. M. Keck Observatory and the NASA Exoplanet Science Institute (NExScI), under contract with the National Aeronautics and Space Administration. \\
Some of the data presented herein were obtained at the W. M. Keck Observatory, which is operated as a scientific partnership among the California Institute of Technology, the University of California and the National Aeronautics and Space Administration. The Observatory was made possible by the generous financial support of the W. M. Keck Foundation. 
The authors wish to recognize and acknowledge the very significant cultural role and reverence that the summit of Mauna Kea has always had within the indigenous Hawaiian community. We are most fortunate to have the opportunity to conduct observations from this mountain. \\
This publication makes use of data products from NEOWISE, which is a project of the Jet Propulsion Laboratory/California Institute of Technology, founded by the Planetary Science Division of the National Aeronautics and Space Administration. \\

This research has made use of the Keck Observatory Archive (KOA), which is operated by the W. M. Keck Observatory and the NASA Exoplanet Science Institute (NExScI), under contract with the National Aeronautics and Space Administration. \\
This research has made use of the NASA/IPAC Extragalactic Database (NED) which is operated by the Jet Propulsion Laboratory, California Institute of Technology, under contract with the National Aeronautics and Space Administration. \\
This research has made use of the NASA/ IPAC Infrared Science Archive, which is operated by the Jet Propulsion Laboratory, California Institute of Technology, under contract with the National Aeronautics and Space Administration. \\
This publication makes use of data products from the Two Micron All Sky Survey, which is a joint project of the University of Massachusetts and the Infrared Processing and Analysis Center/California Institute of Technology, funded by the National Aeronautics and Space Administration and the National Science Foundation. \\
Funding for the Sloan Digital Sky Survey (SDSS) has been provided by the Alfred P. Sloan Foundation, the Participating Institutions, the National Aeronautics and Space Administration, the National Science Foundation, the U.S. Department of Energy, the Japanese Monbukagakusho, and the Max Planck Society. The SDSS Web site is \url{http://www.sdss.org/}. \\

This publication makes use of data products from the Wide-field Infrared Survey Explorer, which is a joint project of the University of California, Los Angeles, and the Jet Propulsion Laboratory/California Institute of Technology, funded by the National Aeronautics and Space Administration. \\

The SDSS is managed by the Astrophysical Research Consortium (ARC) for the Participating Institutions. The Participating Institutions are The University of Chicago, Fermilab, the Institute for Advanced Study, the Japan Participation Group, The Johns Hopkins University, Los Alamos National Laboratory, the Max-Planck-Institute for Astronomy (MPIA), the Max-Planck-Institute for Astrophysics (MPA), New Mexico State University, University of Pittsburgh, Princeton University, the United States Naval Observatory, and the University of Washington. \\

The intermediate Palomar Transient Factory project is a scientific collaboration among the California Institute of Technology, Los Alamos National Laboratory, the University of Wisconsin, Milwaukee, the Oskar Klein Center, the Weizmann Institute of Science, the TANGO Program of the University System of Taiwan, and the Kavli Institute for the Physics and Mathematics of the Universe. 

{\sc iraf} is distributed by the National Optical Astronomy Observatory, which is operated by the Association of Universities for Research in Astronomy (AURA) under a cooperative agreement with the National Science Foundation. \\

{\sc SNOoPy} is a package for SN photometry using PSF fitting and/or template subtraction developed by E.~Cappellaro. A package description can be found at \url{http://sngroup.oapd.inaf.it/snoopy.html}. \\

{\sc foscgui} is a graphic user interface aimed at extracting SN spectroscopy and photometry obtained with FOSC-like instruments. It was developed by E.~Cappellaro. A package description can be found at \url{http://sngroup.oapd.inaf.it/foscgui.html}.
\end{acknowledgements}

\begin{appendix}
\section{Observations and data reduction} \label{sec:obsredu}
The follow-up campaign of \object~spanned a period of more than $1500\,\rm{d}$ after the SN explosion and involved a number of collaborations and facilities.
The names of the telescopes and instruments used are reported in Tables~A.1-A.4 (available at the CDS), and \ref{table:speclog}.

\subsection{Photometric data} \label{sec:lightcurves}
Optical photometric data were mainly obtained using the telescopes of the Las Cumbres Observatory\footnote{\url{https://lco.global/}} network \citep{2013PASP..125.1031B} within the Supernova Key Project, while most of the NIR data and additional optical photometry were provided by the NUTS collaboration\footnote{\url{http://csp2.lco.cl/not/}}, using the $2.56\,\rm{m}$ Nordic Optical Telescope (NOT, at the Observatorio del Roque de los Muchacos, La Palma, Spain) with NOTCam and ALFOSC.
Optical data were also collected using the $1.82\,\rm{m}$ Copernico Telescope (at the INAF Osservatorio Astronomico di Asiago, Italy) with AFOSC and iPTF, the automated Mount Palomar $60\,\rm{inch}$ (P60) and the $48\,\rm{inch}$ Samuel Oschin (P48) telescopes. 
The iPTF survey did not monitor this part of the sky at all in 2015, so the transient was only detected by iPTF in March 2016, which is the reason why it was internally dubbed iPTF16tu. iPTF then followed the object for a full year until the end of the survey.
Early data covering the rise to maximum light and a few additional epochs were provided by amateurs and calibrated to the $R$ band. These data helped constrain the explosion epoch and the rise of \object~(see Section~\ref{sec:photometry}).

NIR photometry was almost entirely provided by NUTS using the NOT with NOTCam, while early data around maximum were provided by the ESO $0.6\,\rm{m}$ Rapid Eye Mount telescope in La Silla (Chile).
One additional NIR epoch was obtained using the Near Infrared Camera Spectrometer (NICS\footnote{\url{http://www.tng.iac.es/instruments/nics/}}) mounted at the $3.58\,\rm{m}$ Telescopio Nazionale Galileo (TNG) located at the Observatorio del Roque de los Muchachos in La Palma (Spain).

Optical and NIR pre-reduction steps were performed using standard {\sc iraf} tasks.
NOTCam frames pre-reduced using an adapted version of the external {\sc iraf} package {\sc NOTCam}\footnote{v.2.5; \url{http://www.not.iac.es/instruments/notcam/guide/observe.html\#reductions}}, using bad pixel masking, differential flat-fielding method, sky subtraction, distortion correction, and stacking of dithered images.
Final magnitudes were mostly obtained using the SuperNOva PhotometrY ({\sc SNOoPY}\footnote{\url{http://sngroup.oapd.inaf.it/snoopy.html.}}) pipeline and calibrated on a local sequence of stars obtained through the Sloan Digital Sky Survey (SDSS\footnote{\url{http://www.sdss.org/}}) (for the optical frames) and the Two Micron All Sky Survey (2MASS\footnote{\url{https://www.ipac.caltech.edu/2mass/}}) catalogs.
$UBVRI$ magnitudes of the reference stars were obtained transforming the SDSS magnitudes following \citet{2008AJ....135..264C}. 
Magnitudes of the local standard stars are provided in Table~\ref{table:localstandards}.
P60 data were reduced using the dedicated pipeline described in \citet{2016A&A...593A..68F}.

MIR magnitudes were obtained as detailed below.
For every pass, after a quick check to verify that the SN did not show rapid variations, all high-quality images (obtained typically within $3-5\,\rm{d}$) available in the NEOWISE 2019 Data Release have been co-added\footnote{With the WISE/NEOWISE Co--adder \\ (\url{https://irsa.ipac.caltech.edu/applications/ICORE/)}} and retrieved. 
The SN exploded in a region in which the galaxy background was still significant, while the spatial resolution was relatively poor ($\geq6\,\rm{arcsec}$). 
We therefore decided to apply the galaxy background subtraction method.
To this aim, we use the averages of all images acquired during the year 2014, meaning before explosion, as background in both bands. 
After accurate re-entering, the images were ready to be subtracted, satisfactorily removing the galaxy, and leaving the SN alone.
On the subtraction image we performed aperture photometry with aperture of eight and nine pixels.
We then applied the aperture correction determined by applying the same aperture on a number of nearby, isolated reference stars of the same magnitude of the SN, present in the AllWISE Source Catalog.
Following these prescriptions, the $W_1$ and $W_2$ magnitudes of \object~reported in Table A.4 are on the same scale as the AllWISE catalog.

\longtab[5]{
\scriptsize
\begin{landscape}
\begin{longtable}{ccccccccccccccc}
\caption{Local sequence of stars used for the optical and NIR photometric calibration. \label{table:localstandards}} \\
\hline\hline
\noalign{\smallskip}
RA [J2000] & Dec [J2000] & $u$ (err) & $g$ (err) & $r$ (err) & $i$ (err) & $z$ (err) & $U$ (err) & $B$ (err) & $V$ (err) & $R$ (err) & $I$ (err) & $J$ (err) & $H$ (err) & $K$ (err) \\ 
(hh:mm:ss)            & (dd:mm:ss)  & (mag) & (mag) & (mag) & (mag) & (mag) & (mag) & (mag) & (mag) & (mag) & (mag) & (mag) & (mag)     & (mag) \\
\noalign{\smallskip}
\hline
\endfirsthead
\caption{continued.} \\
\noalign{\smallskip}
RA [J2000] & Dec [J2000] & $u$ (err) & $g$ (err) & $r$ (err) & $i$ (err) & $z$ (err) & $U$ (err) & $B$ (err) & $V$ (err) & $R$ (err) & $I$ (err) & $J$ (err) & $H$ (err) & $K$ (err) \\ 
(hh:mm:ss)            & (dd:mm:ss)  & (mag) & (mag) & (mag) & (mag) & (mag) & (mag) & (mag) & (mag) & (mag) & (mag) & (mag) & (mag)     & (mag) \\
\noalign{\smallskip}
\endhead
\hline
\endfoot
\noalign{\smallskip}
13:52:08.58 & $+39$:43:33.78 & --     & --     & --       & --       & --     & --     & --     & --      & --     & --     & 16.03(0.07) & 15.17(0.09) & 14.91(0.10) \\
13:52:09.85 & $+39$:38:56.11 & --     & --     & --       & --       & --     & --     & --     & --      & --     & --     & 15.44(0.05) & 14.91(0.06) & 14.75(0.09) \\
13:52:12.32 & $+39$:42:37.14 & 21.15(0.07) & 18.67(0.02) & 17.51(0.02)   & 17.02(0.02)   & 16.71(0.02) & 20.30(0.07) & 19.26(0.07) & 17.98(0.04) & 17.22(0.05) & 16.49(0.11) & 15.44(0.04) & 14.82(0.06) & 14.66(0.09) \\
13:52:15.65 & $+39$:40:02.69 & 17.39(0.02) & 16.47(0.02) & 16.26(0.01)   & 16.17(0.02)   & 16.17(0.02) & 16.53(0.02) & 16.76(0.04) & 16.34(0.03) & 16.08(0.03) & 15.77(0.05) & 15.40(0.04) & 15.19(0.08) & 14.88(0.10) \\
13:52:17.43 & $+39$:40:56.13 & 16.68(0.02) & 14.68(0.02) & 13.80(0.01)   & 13.411(0.001) & 13.22(0.02) & 15.83(0.02) & 15.18(0.05) & 14.15(0.03) & 13.54(0.04) & 12.91(0.09) & 12.15(0.02) & 11.55(0.02) & 11.46(0.02) \\
13:52:20.88 & $+39$:44:19.68 & --     & --     & --       & --       & --     & --     & --     & --      & --     & --     & 12.78(0.02) & 12.24(0.03) & 12.13(0.03) \\
13:52:26.97 & $+39$:39:55.39 & 21.02(0.08) & 19.19(0.02) & 18.45(0.02)   & 18.14(0.02)   & 18.01(0.02)   & 20.17(0.08) & 19.65(0.05) & 18.75(0.03) & 18.21(0.04) & 17.67(0.07) & --     & --     & --     \\
13:52:28.08 & $+39$:39:14.57 & 21.21(0.09) & 18.85(0.01) & 17.61(0.02)   & 17.10(0.02)  & 16.81(0.02) & 20.35(0.09) & 19.47(0.07) & 18.11(0.04) & 17.31(0.05) & 16.55(0.11) & 15.69(0.05) & 15.03(0.05) & 14.86(0.09) \\
13:52:29.22 & $+39$:41:02.24 & 19.99(0.04) & 17.66(0.01) & 16.60(0.02)   & 16.19(0.02)   & 15.94(0.02) & 19.13(0.04) & 18.22(0.06) & 17.03(0.03) & 16.33(0.05) & 15.68(0.09) & 14.83(0.03) & 14.19(0.04) & 14.12(0.05) \\
13:52:29.52 & $+39$:41:24.69 & 17.86(0.02) & 16.35(0.02) & 15.83(0.02)   & 15.62(0.02)   & 15.58(0.02) & 17.01(0.02) & 16.74(0.04) & 16.03(0.03) & 15.61(0.03) & 15.18(0.06) & 14.74(0.03) & 14.44(0.04) & 14.30(0.06) \\
13:52:29.64 & $+39$:39:56.96 & 15.28(0.02) & 13.97(0.01) & 14.286(0.001) & 13.49(0.02)   & 13.49(0.01) & 14.42(0.02) & 14.08(0.03) & 14.14(0.02) & 13.91(0.08) & 12.86(0.16) & 12.70(0.02) & 12.40(0.02) & 12.37(0.02) \\
13:52:32.03 & $+39$:39:17.95 & --     & --     & --       & --       & --     & --     & --     & --      & --     & --     & 15.76(0.06) & 15.04(0.07) & 14.81(0.09) \\
13:52:35.03 & $+39$:41:22.91 & 21.52(0.08) & 19.59(0.02) & 18.48(0.02)   & 18.13(0.02)   & 17.66(0.02) & 20.67(0.08) & 20.17(0.06) & 18.93(0.04) & 18.23(0.04) & 17.64(0.08) & 15.72(0.08) & 14.96(0.08) & 14.38(0.08) \\
13:52:39.33 & $+39$:43:43.11 & 18.14(0.02) & 16.86(0.02) & 16.41(0.02)   & 16.26(0.02)   & 16.23(0.02) & 17.29(0.02) & 17.23(0.04) & 16.59(0.03) & 16.21(0.03) & 15.84(0.06) & --     & --     & --    \\
\noalign{\smallskip}
\end{longtable}
\tablefoot{$UBVRI$, $JHK$ and $W_1W_2$ magnitudes are calibrated in the Vega photometric system, $ugriz$ in the AB photometric system using values reported in the SDSS database.}
\end{landscape}
}

\subsection{Spectroscopic data} \label{sec:spectra}
\begin{figure*}
\begin{center}
\includegraphics[width=\linewidth]{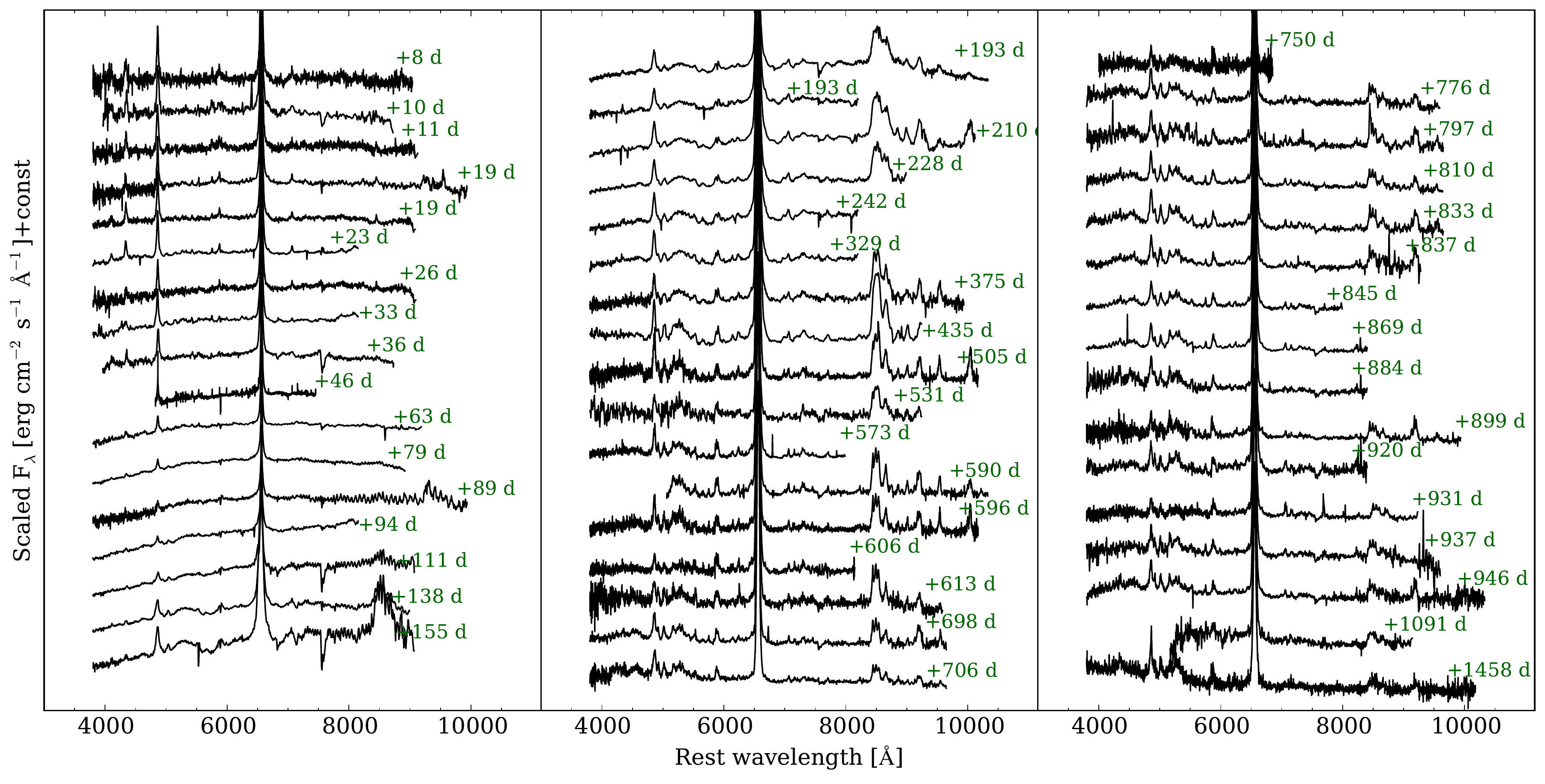}
\caption{Complete set of optical spectroscopic observations of \object. Spectra were not corrected for extinction to facilitate the comparison at wavelengths bluer than $\simeq5000$\ang. Phases refer to the estimated epoch of the explosion.\label{fig:allSpectra}}
\end{center}
\end{figure*}

The spectroscopic follow-up campaign was carried out mostly using the NOT with ALFOSC and the $1.82\,\rm{m}$ Copernico Telescope with AFOSC, and the data were reduced using the dedicated pipeline {\sc foscgui}\footnote{\url{http://sngroup.oapd.inaf.it/foscgui.html}}.
Three additional spectra were obtained using the $2\,\rm{m}$ Faulkes Telescope North (FTN) telescope with FLOYDS of the Las Cumbres Observatory network.
Spectra at different phases were provided by the TNG with DOLORes, the $10.4\,\rm{m}$ Gran Telescopio Canarias (GTC, located at the Observatorio del Roque de los Muchachos in La Palma) with OSIRIS, the $3\,\rm{m}$ Donald Shane Telescope (at the Lick Observatory in San Jose, California, U.S.A.) with KAST and the $200\,\rm{inch}$ Hale telescope with DBSP at the Mount Palomar Observatory (San Diego, California, U.S.A.).
All these spectra were reduced using standard {\sc iraf} tasks.
The classification spectrum was obtained using the Lijiang $2.4\,\rm{m}$ telescope (LJT) at the Lijiang Observatory of Yunnan Observatories with YFOSC, which also provided three early-phase spectra ($+11$, $+19$ and $+26\,\rm{d}$ after explosion).
Two additional early-phase spectra ($+10$ and $+36\,\rm{d}$) were obtained using the ZTA $2.6\,\rm{m}$ telescope located at the Byurakan Astrophysical Observatory (BAO, Armenia) with XLT.
All these spectra were reduced using standard \textsc{iraf} tasks.
NIR spectra were provided by the TNG using NICS and reduced using {\sc iraf} tasks, and the $8.1\,\rm{m}$ Gemini North telescope (located at the Mauna Kea Observatory, Hawaii, U.S.A.) using the Gemini Near-InfraRed Spectrograph (GNIRS), reduced as in \citet{2019PASP..131a4002H}. 
A complete set of optical spectra is shown in Figure~\ref{fig:allSpectra}.

\longtab[6]{
\small
\begin{longtable}{cccccccc}
\caption{\label{table:speclog} Log of the optical spectroscopic observations of \object.} \\
\hline\hline
\noalign{\smallskip}
Date & JD & Phase & Instrumental setup & Grism/Grating & Spectral range & Resolution & Exposure time \\ 
 &  & (d) &  &  & (\AA) & (\AA) & (s) \\
\noalign{\smallskip}
\hline
\endfirsthead
\caption{continued.}\\
\hline\hline
\noalign{\smallskip}
Date & JD & Phase & Instrumental setup & Grism/Grating & Spectral range & Resolution & Exposure time \\ 
 &  & (d) &  &  & (\AA) & (\AA) & (s) \\
\noalign{\smallskip}
\endhead
\hline
\endfoot
\noalign{\smallskip}
20150116 & 2457039.37 &   $+8$ & GMG24+YFOSC/YSU           & gr3              & $3470-9110$  & 15.95  & 1560                       \\
20150118 & 2457041.37 &  $+10$ & BAO $2.16\,\rm{m}$+XLT    & 14.00                & $3970-8720$  & 15.00      & 3600                       \\
20150119 & 2457042.38 &  $+11$ & GMG24+YFOSC/YSU           & gr3              & $3470-9130$  & 20.86  & 1800                       \\
20150127 & 2457049.99 &  $+19$ & FTN+FLOYDS                & $235\,\rm{l/mm}$ & $3180-9940$  & 14.08  & 3600                       \\
20150127 & 2457050.37 &  $+19$ & GMG24+YFOSC/YSU           & gr3              & $3470-9080$  & 15.97  & 2100                       \\
20150131 & 2457053.67 &  $+23$ & Ekar $1.82\,\rm{m}$+AFOSC & gr4              & $3330-8150$  & 15.20  & 1200                       \\
20150203 & 2457056.79 & $+26$ & TNG+NICS           & IJ & $9000-14000$ & $26.86\star\star$ & $4\times150$ \\
20150203 & 2457057.34 &  $+26$ & GMG24+YFOSC/YSU           & gr3              & $3480-9100$  & 15.20  & 1800                       \\
20150210 & 2457063.55 &  $+33$ & Ekar182+AFOSC             & gr4              & $3330-8150$  & 14.83  & 1800                       \\
20150213 & 2457067.35 &  $+36$ & BAO $2.16\,\rm{m}$+XLT    & 14.00                & $3960-8730$  & 15.00      & 2700                       \\
20150223 & 2457077.16 &  $+46$ & Keck+DEIMOS               & 1200G/6170       & $4830-7460$  &  1.56  & $300+300$                  \\
20150312 & 2457093.52 &  $+63$ & Ekar182+AFOSC             & gr4+VPH6         & $3340-9190$  & 16.44  & $2400+2400$                \\
20150328 & 2457109.59 &  $+79$ & NOT+ALFOSC                & gr4              & $3360-8920$  & 17.82  & 2400                       \\
20150407 & 2457119.83 &  $+89$ & FTN+FLOYDS                & $235\,\rm{l/mm}$ & $3180-9940$  & 15.17  & 3600                       \\
20150411 & 2457124.48 &  $+94$ & Ekar $1.82\,\rm{m}$+AFOSC & gr4              & $3340-8150$  & 13.93  & 1800                       \\
20150428 & 2457141.62 & $+111$ & NOT+ALFOSC                & gr4              & $3250-9070$  & 14.74  & 1800                       \\
20150525 & 2457168.64 & $+138$ & NOT+ALFOSC                & gr4              & $3250-8990$  & 14.59  & 2400                       \\
20150612 & 2457185.64 & $+155$ & NOT+ALFOSC                & gr4              & $3490-9070$  & 16.36  & 1800                       \\
20150720 & 2457224.39 & $+193$ & GTC+OSIRIS                & R1000B+R1000R    & $3610-10330$ &  7.98  & $900+900$                  \\
20150720 & 2457224.43 & $+193$ & Ekar $1.82\,\rm{m}$+AFOSC & gr4              & $3380-8200$  & 16.34  & 1800                       \\
20150806 & 2457241.32 & $+210$ & Ekar $1.82\,\rm{m}$+AFOSC & gr4              & $3380-10120$ & 14.52  & 2400                       \\
20150824 & 2457259.39 & $+228$ & NOT+ALFOSC                & gr4              & $3490-8990$  & 14.00  & 1800                       \\
20150907 & 2457273.30 & $+242$ & Ekar $1.82\,\rm{m}$+AFOSC & gr4              & $3380-8200$  & 14.06  & 1800                       \\
20151203 & 2457359.68 & $+329$ & Ekar $1.82\,\rm{m}$+AFOSC & gr4              & $3380-8200$  & 13.25  & 1800                       \\
20160118 & 2457406.02 & $+375$ & FTN+FLOYDS                & $235\,\rm{l/mm}$ & $3760-9940$  & 14.89  & 3600                       \\
20160318 & 2457465.60 & $+435$ & Ekar $1.82\,\rm{m}$+AFOSC & VPH6             & $3305-9240$  & 16.21  & 2700                       \\
20160527 & 2457535.90 & $+505$ & P200+DBSP                 & 316/7500         & $3180-10170$ & $10.30\star$ & 840                        \\
20160622 & 2457562.44 & $+531$ & Ekar $1.82\,\rm{m}$+AFOSC & VPH6+VPH7        & $3600-9240$  & 15.61  & $2400+2400$                \\
20160803 & 2457603.68 & $+573$ & SHANE+KAST                & 600/431+300/7500 & $3420-7990$  &  6.78  & $2\times2700+2\times2700$  \\
20160820 & 2457621.38 & $+590$ & GTC+OSIRIS                & R1000R           & $5060-10330$ &  8.41  & $2\times1800$              \\
20160826 & 2457626.64 & $+596$ & P200+DBSP                 & 600/4000         & $3300-10170$ & $10.30\star$ &  600                       \\
20160905 & 2457636.66 & $+606$ & SHANE+KAST                & 600/431+300/7500 & $3410-8150$  &  7.25  & 1640                       \\
20160906 & 2457637.50 & $+607$ & Gemini N+GNIRS & HK & $8000-25000$ & $12.57\star\star$ & $12\times30$ \\
20160912 & 2457644.36 & $+613$ & NOT+ALFOSC                & gr4              & $3380-9160$  & 18.04  & 1800                       \\
20161206 & 2457728.75 & $+698$ & NOT+ALFOSC                & gr4              & $3580-9650$  & 14.15  & 2700                       \\
20161214 & 2457736.74 & $+706$ & NOT+ALFOSC                & gr4              & $3380-9650$  & 14.08  & 2400                       \\
20170127 & 2457780.98 & $+750$ & SHANE+KAST                & 600/431          & $4000-6850$  & 11.53  & 2700                       \\
20170222 & 2457806.73 & $+776$ & NOT+ALFOSC                & gr4              & $3380-9590$  & 18.08  & 2400                       \\
20170315 & 2457827.57 & $+797$ & NOT+ALFOSC                & gr4              & $3550-9640$  & 14.58  & $2\times2300$              \\
20170328 & 2457841.40 & $+810$ & NOT+ALFOSC                & gr4              & $3380-9640$  & 14.23  & 2700                       \\
20170420 & 2457863.55 & $+833$ & NOT+ALFOSC                & gr4              & $3630-9650$  & 14.29  & $2\times2300$              \\
20170424 & 2457867.59 & $+837$ & TNG+DOLoRes               & LRB+LRR          & $3480-9280$  & 17.28  & $2400+2400$                \\
20170502 & 2457875.54 & $+845$ & TNG+DOLoRes               & LRB              & $3300-7990$  & 11.83  & 2400                       \\
20170526 & 2457899.58 & $+869$ & TNG+DOLoRes               & LRB+LRR          & $3300-10400$ & 15.78  & $2400+2400$                \\
20170610 & 2457914.52 & $+884$ & TNG+DOLoRes               & LRB+LRR          & $3480-9240$  & 11.58  & $2400+2400$                \\
20170625 & 2457929.81 & $+899$ & SHANE+KAST                & 600/431+300/7500 & $3400-9930$  &  7.53  & 1200                       \\
20170716 & 2457951.44 & $+920$ & TNG+DOLoRes               & LRR              & $3400-10410$ & 19.14  & 2700                       \\
20170727 & 2457961.69 & $+931$ & SHANE+KAST                & 600/431+300/7500 & $3460-9230$  & 7.08   & 1200                       \\
20170802 & 2457968.40 & $+937$ & TNG+DOLoRes               & LRB              & $3580-10400$ & 10.76  & 2700                       \\
20170811 & 2457977.41 & $+946$ & GTC+OSIRIS                & R1000B+R1000R    & $3610-10330$ & 8.81   & $2\times1800+2\times1800$  \\
20180102 & 2458121.74 & $+1091$ & GTC+OSIRIS               & R500R            & $5170-9140$  & 14.64  & $2\times1800$              \\     
20190105 & 2458489.12 & $+1458$ & KECK+LRIS                     & 400/3400 & $3000-10200$ & 7.44 & 900 \\
\noalign{\smallskip}
\end{longtable}
\tablefoot{NOT: $2.56\,\rm{m}$ Nordic Optical Telescope with ALFOSC; GTC: $10.4\,\rm{m}$ Gran Telescopio Canarias with OSIRIS; TNG: $3.56\,\rm{m}$ Telescopio Nazionale Galileo with DOLoRes (all located at Roque de Los Muchachos, La Palma, Spain); Ekar $1.82\,\rm{m}$: Ekar $1.82\,\rm{m}$ Copernico telescope with AFOSC (Mt. Ekar, Asiago, Italy); GMG24: Lijiang $2.4\,\rm{m}$ telescope with YFOSC (Lijiang Gaomeigu Station of Yunnan Observatories, Yunnan, China); BAO: ZTA $2.6\,\rm{m}$ telescope with XLT (Byurakan Astrophysical Observatory, Mt. Aragats, Armenia); SHANE: $3\,\rm{m}$ Donald Shane Telescope with KAST (Lick Observatory, San Jose, California - U.S.A.); P200: $200\,\rm{inch}$ Hale telescope with DBSP (Mt. Palomar Observatory, San Diego, California - U.S.A.); KECK: $\,\rm{m}$ Keck II telescope with DEIMOS and LRIS (Mauna Kea Observatory, Hawaii - U.S.A.); FTN: $2\,\rm{m}$ Faulkes Telescope North with FLOYDS, Las Cumbres Observatory node at the Haleakala Observatory, Hawaii; Gemini N: $8.1\,\rm{m}$ Gemini North telescope with GNIRS (Mauna Kea Observatory, Hawaii - U.S.A). \\ $\star$ Nominal resolution at 6562.8\ang. \\ $\star\star$ Resolution computed around $\rm{Pa}\beta$.}
}
\end{appendix}
\end{document}